\documentclass[paper=a4, fontsize=11pt]{article}
\usepackage{amsmath}
\usepackage{amsthm}
\usepackage{enumitem}
\usepackage{cancel}
\usepackage{appendix}
\usepackage{graphicx}
\usepackage{geometry}
\usepackage{url}
\usepackage[colorlinks=true,linkcolor=blue]{hyperref}
\usepackage{sectsty}
\usepackage[utf8]{inputenc}
\usepackage[T1]{fontenc}
\usepackage{amssymb}
\allsectionsfont{ \normalfont\scshape \textbf}
\newcommand{\mf}{\mathfrak}
\newcommand{\mc}{\mathcal}
\newcommand{\ad}{\mathfrak{a}^\dagger}
\newcommand{\as}{\mathfrak{a}}
\newcommand{\ep}{\varepsilon}
\newcommand{\da}{\downarrow}
\newcommand{\ua}{\uparrow}
\newcommand{\no}{\hat{N}}
\newlist{steps}{enumerate}{1}
\setlist[steps, 1]{label = Step \arabic*:}
\numberwithin{equation}{section}
\title{Background for the Self Consistent Renormalisation (SCR) Theory}
\author{Bharathiganesh Devanarayanan $^{a,b}$, Akariti Sharma$^{a}$, Alpesh Sheth$^{c}$,\\ Prafulla K. Jha $^{c}$, Navinder Singh$^{a}$ $^{*}$\\
        \small $^{a}$ Theoretical Physics Division, Physical Research Laboratory, Navrangpura Ahmedabad, India - 380009 \\
        \small $^{b}$ Indian Institute of Technology, Gandhinagar, Palaj, Gujarat, India - 382355 \\
        \small $^{c}$ Department of Physics, Faculty of Science,  The Maharaja Sayajirao University of Baroda, Gujarat, India \\\\
        \small $^{*}$ Corresponding author:Navinder Singh; \tt{navinder.phy@gmail.com}
}
\date{} %leave blank

\begin{document}
%
%\maketitle
%\maketitle
\maketitle
\begin{abstract}
A detailed review is given on the evolution of the theory of itinerant magnetism.
The self-consistent renormalization (SCR) theory is quite successful in addressing the phenomenon of itinerant magnetism in weakly and nearly ferromagnetic and anti-ferromagnetic materials \cite{toru}. It goes beyond the Stoner and random phase approximation (RPA) theories in taking the correlation effects into account. The Mathematical machinery of the SCR theory is rather complicated. The aim of these notes is to provide the required background. The problems with Stoner and RPA theory are discussed and the way in which the SCR theory rectifies those problems is also discussed.
%For example in iron (a strong ferromagnet), it seems likely that the formation of local moments above the Curie temperature $T_{c}$, is responsible for spin fluctuations observed in these systems.  In weak ferromagnets, the existence of small saturation moments at absolute zero conflicts with this picture, and a best suited description is required. In this regard, the important amendments to the simple Stoner model of magnetism due to the spin wave excitations and spin fluctuations are also explained in detail. The spin correlations (fluctuations) are studied within the framework of a molecular field or random phase approximation (RPA), as resulting directly from short-range repulsion between fermions. A self-consistent renormalization (SCR) theory to magnetism and electronic structure of heavy fermion materials which takes into account dynamical many-body effects is discussed in detail. This theory combines the features of the itinerant electron theory (Stoner) of magnetic crystals with the localized-moment description (Heisenberg). The possible applicability of the itinerant electron model to the high temperature superconductor oxides and the possible importance of the effects of antiferromagnetic spin fluctuations are also discussed. 
\noindent   \end{abstract}
\tableofcontents
\newpage
\section{Introduction}

\begin{itemize}
 \item Materials can be broadly classified based on their response to an external non-homogeneous magnetic field into the following three categories \cite{mohn2006magnetism}:
 When a sample of given material is freely suspended in a non-homogeneous magnetic field as shown in fig.\ref{af1}: following cases can emerge!
 \begin{figure}[b]
    \centering
    \includegraphics[scale = 0.15]{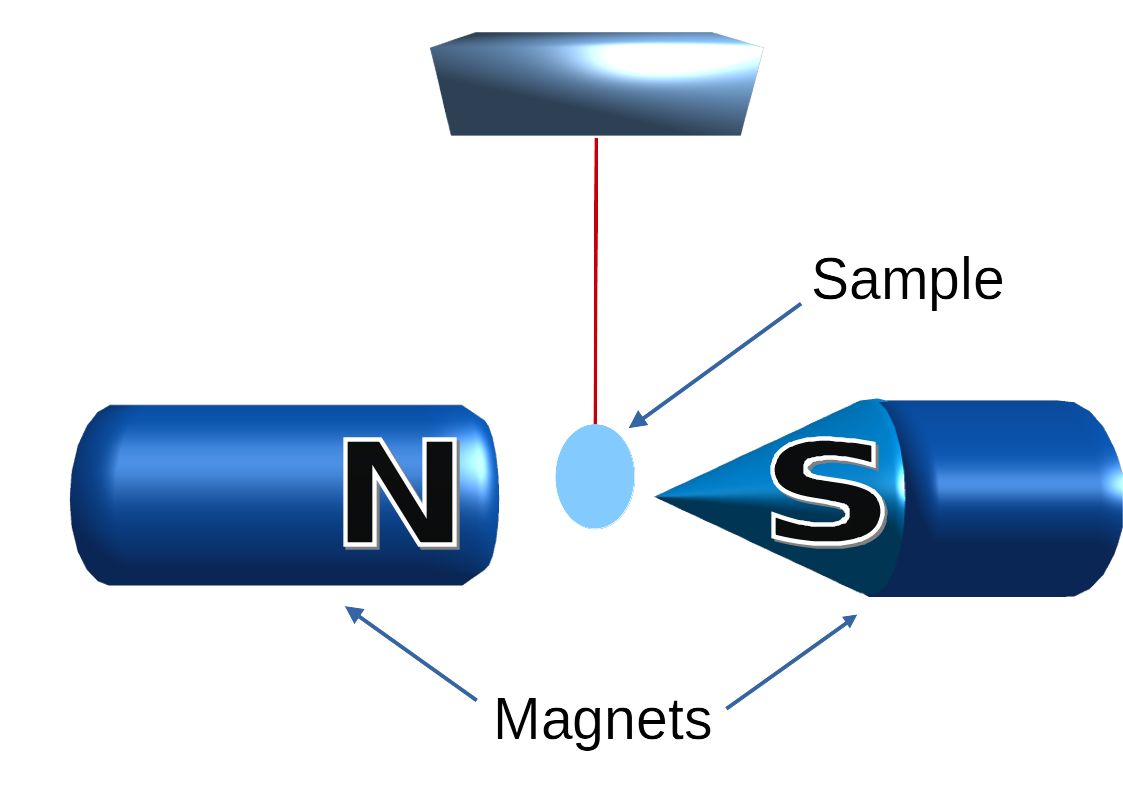}
    \caption{A sample suspended in an inhomogeneous magnetic field.}
    \label{af1}
\end{figure}
\begin{enumerate}
    \item Diamagnets: These materials are feebly repelled and move from high to low field region, Such materials are called diamagnetic materials. Examples: Noble gases, Carbon, Water etc.
    \item Paramagnets: These materials are feebly attracted and move from low to high field region, Such materials are called paramagnetic materials. Examples: Aluminium, Copper, Sodium etc.
    \item Ferromagnets: If the material is strongly attracted towards the region where the magnetic field is strong, it is called a Ferromagnetic material. These materials exhibit the phenomenon of spontaneous magnetization, Examples are: Iron, Nickel, Cobalt etc.
\end{enumerate}
Diamagnetism and Paramagnetism are weaker forms of magnetism whereas Ferromagnetism is a stronger form of magnetism.\\ 
Diamagnetism can be associated to the response of a material in accordance with ``Lenz Law". Paramagnetism can be associated with the response of a material caused due to spin and orbital motion of electrons. The essential requirement for paramagnetism is ``Permanent Magnetic Dipoles'' which can be either localized or itinerant based on which there are two types of paramagnetism.
\begin{enumerate}
    \item Curie-Langevin-Van Vleck paramagnetism: Arises due  to the localized magnetic moments in gases and salts.
    \item Pauli paramagnetism: Arises due  to quasi free (\textit{itinerant})conduction electrons carrying a permanent moment of one Bohr Magneton each.
\end{enumerate}  In Ferromagnets there are itinerant electrons that exhibit the phenomenon of spontaneous magnetization through exchange interactions. When we apply an external magnetic field $H$ to a material, the net magnetic induction in the material is given as
$$
 B = H + 4\pi M
$$
here $B$ is called the magnetic induction and $M$ is called the induced magnetisation. Assume B, H and M are all in the same direction. Dividing both the sides by $H$ we get
$$
\frac{B}{H} = 1+ 4\pi\frac{M}{H}
$$
 Magnetic permeability is defined as, $\mu =\frac{B}{H}$ and the Magnetic Susceptibility as $\chi = \frac{M}{H}$, the above equation becomes
$$
\mu = 1+ 4\pi\chi
$$
If the external field $H$ is time independent then the susceptibility $\chi$ is also time independent and is called the static susceptibility. If the external field $H$ and magnetisation $M$ are functions of time say having periodic variation with some frequency $\omega$, then the susceptibility also depends on the frequency $\omega$ and is called the dynamical susceptibility which is given as
$$
M(\omega) = \chi(\omega) H(\omega)
$$
The susceptibility $\chi$ is a dimensionless quantity. Its magnitude can vary greatly and typical susceptibility values are
\begin{enumerate}
    \item diamagnet $\sim$ $10^{-6}$
    \item paramagnet $\sim$ $10^{-4}$
    \item ferromagnet $\sim$ $10^{3}$ - $10^{6}$
\end{enumerate}
\item \textbf{Curie's law for paramagnetic substances (Experimental)}
The Pierre Curie experimentally investigated the temperature dependence of susceptibility in many paramagnetic materials. He deducted that
$$
\chi = \frac{C}{T}
$$
Here $C$ is the Curie constant. 
\item \textbf{Curie-Weiss law for ferromagnetic substances}
For ferromagnetic substances the magnetic susceptibility $\chi$ is given as 
$$
\chi = \frac{C}{T- T_{C}}
$$
here $T_{C}$ is the Curie temperature. This is known as the Curie-Weiss law \cite{weiss1907hypothese}. 
\item \textbf{The Langevin-Curie theory}

Paul Langevin in 1905 theoretically deduced that the magnetic susceptibility $\chi$ for a paramagnetic material has the form
$$
\chi = \frac{N \mu^{2}}{3K_{B}T}
$$
In his deduction he assumed that a fixed value of magnetic moment is associated with atoms and used classical statistical mechanics to reach to the above expression without using the space quantization.
\item \textbf{ The Weiss theory of ferromagnetism}
Pierre Weiss in 1907 proposed that in the case of ferromagnetism, magnetisation can be written as
$$
M = \chi H_{tot} = \frac{n\mu^{2}}{3K_{B}T}  (H + H_{0})
$$
where $ H_{0}=\lambda M$ is the internal field or molecular field. He justified this internal field by arguing that atoms in the material along with the external field also experience an internal field. This internal field is also called the molecular field. Rearranging the above Eqwe get
$$
M = \frac{\frac{n \mu^{2}}{3K_{B}T}}{1- \frac{\lambda n \mu^{2}}{3 K_{B}T}} H
$$
which gives susceptibility as 
$$
\chi = \frac{M}{H} = \frac{C}{T-T_{C}}
$$
From here we get
$$T_{c} = \frac{\lambda n \mu^{2}}{3K_{B}T}$$
the value of $\lambda$ could be calculated from the above Eq. ( by comparing it with the external data) and it turns out to be of the order $\lambda \sim 10^{5}$. But classical arguments leads to a value of $\lambda$ of the order of unity. Thus it cannot be reconciled, and only with the advent of quantum mechanics this problem was solved. Another drawback of Weiss theory is that it assumes local magnetic moments. So it cannot be applied to conduction electrons.
\item \textbf{Heisenberg model}
The Hamiltonian of the Heisenberg model is given as
$$
\mathcal{H} =- \sum_{<i,j>} J_{ij} \Vec{S_{i}}.\Vec{S_{j}} $$
The interaction in the Heisenberg model is not the dipole dipole interaction, which is very weak. $J_{ij}$ has a different origin which is attributed to the quantum mechanical exchange interaction and is known as the Heisenberg exchange interaction. From this theory, the expression for $\lambda$ is given in terms of $J$ as follows
$$
\lambda = Z\frac{J}{2n \mu ^{2}_{B}} 
$$
From the above expression, the high value of $\lambda$ required for the Weiss theory to match with experimental results could be explained. But Heisenberg model is still a localised model i.e. it considers the magnetism arising from the electrons that are localised in the atomic sites. It cannot explain magnetism arising from conduction electrons. For $LaCu_{2}O_{4}$, $J$ is negative and hence the energy is lowered because of alternative spin reversal arrangement of spins leading to anti ferromagnetism. The problem of explaining magnetism in Iron, Nickel, Cobalt is that these materials are good conductors with itinerant electrons. Hence applying the localised models to such systems will run into a logical inconsistency.
\item \textbf{Magnetism of electrons in metals}
\begin{enumerate}
 \item \textbf{Pauli paramagnetism of electrons in metals}
The free electron model of a conductor gives solutions in the form of running waves. Each wave is characterised by a particular value of momentum (k). In the momentum space the running waves form a sphere of radius $k_{F}$ as shown  fig. \ref{af2}. At any finite temperature there is a diffusion zone.
\begin{figure}[!h]
    \centering
    \includegraphics[scale = 0.20]{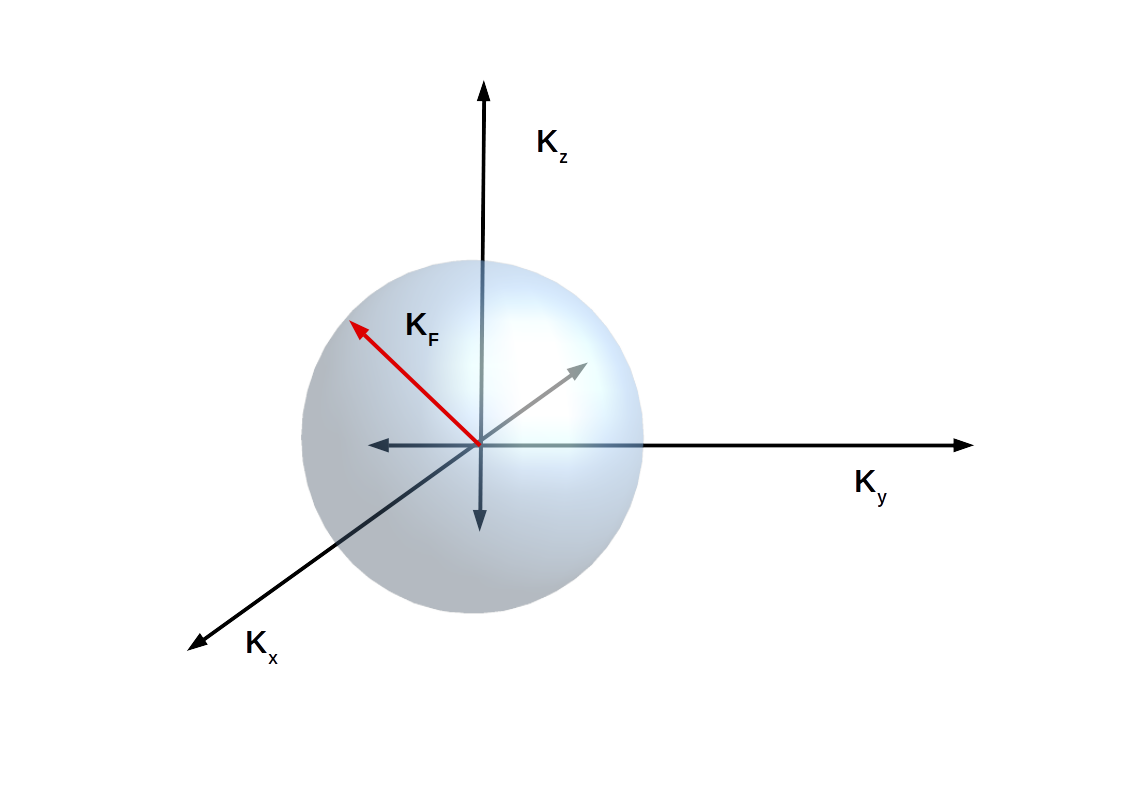}
    \caption{Fermi sphere at zero temperature.}
    \label{af2}
\end{figure}
If in the metal there are $n$ electrons  per unit volume, then in the diffusion zone there are $n_{0}$ electrons given by
$$n_{0} = n \frac{T}{T_{{F}}}.$$
The magnetic susceptibility of an electron in the diffusion zone is given as
$$\chi_{i} = \frac{\mu^{2}}{K_{B}T}.$$
Then the Pauli susceptibility of all the electrons in the diffusion zone is 
\[ \chi = \frac{n \mu^{2}}{K_{B}T_{F}} \]
This susceptibility is called the Pauli susceptibility and is temperature independent. It explains paramagnetism observed in metals like Aluminium, Copper, etc. One would want a similar theory to explain the ferromagnetism observed in metals having Fermi surface.

\item \textbf{Ferromagnetism of conduction electron: Stoner theory}
In ferromagnetic materials, there is a peculiar interaction that causes the electrons to migrate to higher energy levels by flipping their spins in a particular direction. This can be attributed to the exchange-enhanced interactions. By migrating to the higher levels, the exchange energy is lowered. But on the other hand the kinetic energy is increased and vice versa. So it is a competition between exchange energy and kinetic energy and if exchange energy wins then the system topples and goes to a Ferromagnetic state. This criterion is called the Stoner condition \cite{stoner1938collective} and is given as: $I\rho (E_{F}) > 1$. If this inequality is satisfied, then there is a ferromagnetic instability.
%
%\begin{figure}[!h]
%    \centering
%    \includegraphics[scale=0.2]{stoner_fig.png}
%    \caption{A FM configuration is shown on the right}
%    \label{fig:my_label}
%\end{figure}
\end{enumerate}
\end{itemize}

% First lecture ends here, second lecture begins after this.%
%
\section{The Semiclassical Curie-Langevin theory of magnetism} 
Let us consider an atom with $ Z $ electrons and  Hamiltonian $\mathcal{H}$
\begin{equation}\label{eq1}
  \mathcal{H}=\sum_{i=1}^{z}(\frac{\hat{\vec{p_{i}^{2}}}}{2 m_{e}}+V(r_{1},r_{2},r_{3},\cdots)) 
\end{equation}
Here the sum is taken over all the $Z$ electrons and each electron is having a kinetic energy ${p^2}/{2m} $. Using the the minimal substitution $ \vec{p} \rightarrow \vec{p}+\frac e{c}\vec{A} $ the modified Hamiltonian is given as
\begin{equation}\label{eq2}
\mathcal {H}  =\sum_{i=1}^{z} (\frac{[\hat {\vec{p}_{i}} +\frac e{c} \hat {\vec{A}} ]^{2}}{2 m_{e}}+V(r_{1},r_{2},r_{3},\cdots) ),
\end{equation}
here, the vector potential $\vec {A}$ is
\begin{equation}\label{eq3}
\vec {A} = \frac 1{2}(\vec {H} \times \vec {r}).
\end{equation}
It is called the Landau Gauge. If we consider the magnetic field along the $\hat{Z} $ direction, then 
\begin{equation*}
 A_{x}=-\frac{1}{2} y H\,\,\,;
A_{y}=-\frac{1}{2} x H\,\,\,;
A_{z}=0,
\end{equation*}
therefore one can write the Hamiltonian as 
\begin{equation} \label{eq4}
{ \mathcal {H}=\sum_{i=1}^z \frac{1}{2 m}\left\{\left(p_{x}-\frac{e}{c} y H\right)^{2}\right.}{ } \left.+\left(p_{y}-\frac{e}{c} x H\right)^{2}+p_{z}^{2}\right\} + V(r_{1},r_{2},r_{3},\cdots) %
\end{equation}
\begin{equation} \label{eq5}
\mathcal {H} =  \sum_{i=1}^z \frac{p_{i}^{2}}{2 m}+\frac{e^{2} H^{2}}{ 8 m c^{2}}\left(x_{i}^{2}+y_{i}^{2}\right)-\frac{e}{2 m c} ( p_{x_{i}} y_{i}  - p_{y_{i}} x_{i}) H + V(r_{1},r_{2},r_{3},\cdots),
\end{equation}
the above eqn.\eqref{eq5} can be written in terms of $\mathcal {H}_{0}$ (non-interacting Hamiltonian) as
\begin{equation} \label{eq6}
\mathcal {H} = \mathcal {H}_{0}  + \frac{e^2 H^2} {8 m c^2} \sum_{i=1}^z \rho_{i}^2  + \frac {e}{2 m c} \sum_{i=1}^z l_{i z} H 
\end{equation}
\begin{equation} \label{eq7}    
 \mathcal {H}=\mathcal {H}_{0} -\mu_{z} H+\frac{e^{2} H^{2}}{8 m c^{2}} \sum_{i=1}^{z} \rho_{i}^{2}
\end{equation}
 \begin{equation} \label{eq8}
 \mathcal {H}=\mathcal {H}_{0} -\mu H \cos{\theta}+\frac{e^{2} H^{2}}{8 m c^{2}} \rho_{0}^2.
 \end{equation} 
 The Helmhotz Free Energy which as a function of $\mathcal{H}$ is given as
 \begin{equation}
     F(H) = -k_{B} T \ln \mathcal{Z} (H) 
 \end{equation}
and the magnetization is calculated as
\begin{equation} \label{eq8a}
M = \frac{-1}{V} \frac {\partial F(H)}{{\partial H}}
\end{equation}
In turn, the magnetic susceptibilty is given by the well-known formula:
\begin{equation} \label{eq8a}
\chi =\frac{\partial M}{\partial H}
\end{equation}
In classical statistical mechanics
\begin{equation} \label{eq8a}
\mathcal{Z} = \int d \gamma e^{-\beta  {H}}
\end{equation}
%In Quantum Statistical Mechanics
%\begin{equation} \label{eq8b}
%\mathcal{Z} =\sum g e^{-\beta E_{n}}
%\end{equation}
%partition function takes the form as
\begin{equation} \label{eq9}
\mathcal{Z}=\int d r e^{-\beta H_{0}} \frac{1}{4 \pi}  \int_{0}^{\pi} d \theta \sin (\theta)  \int_{0}^{2 \pi} d \phi  e^{\beta \mu H \cos{(\theta)}}  e^{\frac{-\beta z e^{2} H^{2} \rho_{0} ^{2}} {8 m c^{2}}},
\end{equation}
therefore the partition function takes the following form as
\begin{equation} \label{eq10}
\mathcal{Z} =\frac{z_{0}}{2} e^{\frac{-\beta z e^{2} H^{2} \rho_{0}^{2}} {8 m c^2}}  \int_{0}^{\pi} d \theta \sin (\theta) e^{\beta \mu H \cos (\theta)} 
 \end{equation}
put
$$x =\beta \mu H \cos \theta\,\,\,;
dx =-\beta \mu H \sin \theta d \theta$$
the above eqn. then takes the form 
%\begin{equation}
%\int_{-\beta \mu H}^{+\beta \mu H} \frac{d x}{\beta M %n}=\frac{1}{\beta \mu H}\left(e^{\beta \mu H} - e^{-\beta %\mu H}\right)
%\end{equation}
%
\begin{equation} \label{eq11}
\mathcal{Z} = \frac{z_{0}}{2} e^{\frac{-\beta z e^{2} H^{2} \rho_{0}^{2}} {8 m c^2}} \frac{1}{\beta \mu H}(e^{\beta \mu H} - e^{-\beta \mu H}),
\end{equation}
therefore $F(H)$ is obtained as
\begin{equation} \label{eq12}
F(H)=-n k_{B} T\left\{-\beta \frac{z^{2} e^{2} \rho_{0}^{2}}{8 m c^{2}} H^{2}+\ln \left(\frac{e^{\beta\mu}-e^{-\beta\mu}}{\beta \mu H}\right)\right\}.
\end{equation} 
Applying the week field condition:
\begin{equation}
x=\frac{\mu H}{k_{B} T} < <  1
\end{equation} 
and Taylor expanding  $e^{\beta \mu H}$ and  $e^{-\beta \mu H}$  up to  
%$\ln \left(\frac{e^{\beta\mu}-e^{-\beta\mu}}{\beta \mu H}\right)$  
third order term and using the basic logarithmic identity : $\ln(1+x) = x $ for $x<<1$; we get
\begin{equation} \label{eq13}
F(H)=\frac{n z e^{2} \rho_{0}^{2}}{8 m c^{2}} H^{2} -\frac{n}{6} \frac{\mu^{2}}{k_{B} T} H^{2}.
\end{equation}
Further, the $F(H)$ calculated in eq. \eqref{eq13} gives the magnetization as
\begin{equation} \label{eq8a}
M=\frac{-1}{V} \frac {\partial F(H)}{{\partial H}}
\end{equation}
\begin{equation} \label{eq14}
M =  ( - \frac{n z e^{2} \rho_{0}^{2}}{4 m c^{2}} ) H + ( \frac{n}{3} \frac{\mu^{2}}{k_{B} T} ) H.
\end{equation}
Here, one should notice that the terms in the parenthesis of the above Eqare dimensionless quantities known as the magnetic susceptibilities but are having different magnitudes and signs. The term with the $-ve$ sign corresponds to a dia-magnetic susceptibility while the term with the $+ve$ sign corresponds to para-magnetic susceptibility. The diamagnetic and paramagnetic susceptiblities are thus obtained to be
\begin{equation} 
 \chi_{dia}  =  - \frac{n z e^{2} \rho_{0}^{2}}{4 m c^{2}} 
\end{equation}
\begin{equation} \label{eq14a}
\chi_{para} =  \frac{n}{3} \frac{\mu^{2}}{k_{B} T} = \frac {C} {T}
\end{equation}
\begin{equation} \label{eq15}
M =   \chi_{dia}  H + \chi_{para} H,
\end{equation}
here in eq.(\ref{eq14a}), $ C = \frac{n}{3} \frac{\mu^{2}}{k_{B}}$ is the Curie constant. It is to note that the order of $\chi_{dia} $ is $\approx 10^{-4} $ while $\chi_{para} $ is $\approx 10^{-2} $. 
%This implies the para-magnetism  always dominates diamagnetism whenever it is present in the material. 
It is important to note that this approach does not take Space Quantization into consideration and it is valid only at the ambient temperature limit. 
%
% Second lecture ends here. third lecture begins%
%
\section{Quantum Mechanical formulation of Magnetism}

We have the total Hamiltonian for a set of particles in a magnetic field is given as
\begin{equation}
    \mathcal{H} = \sum_{i=1}^{z} \frac{1}{2m} (\Vec{P} +\frac{e}{c} \Vec{A})^{2} + V(r_{1},r_{2},...)
\end{equation}
Hamiltonian can be written as shown in the previous section
\begin{equation}
    \mathcal{H} = \mathcal{H}_{0} -\mu_{z} H + \frac{e^{2}H^{2}}{8mc^{2}} \rho^{2}.
\end{equation}
Here
\begin{equation}
    \mu_{z} = \gamma \hbar J_{z},
\end{equation}
$\mu_{z}$ is the $z$- component of magnetic moment, $\gamma$ is the gyromagnetic ratio and $J_{z}$ is the $z$ - component of the total angular momentum. Therefore, the total magnetic moment is given as
\begin{equation}
    \mu = \gamma \hbar J,
\end{equation}
and
\[ \mu_{z} = \mu \frac{J_{z}}{J} \],
hereafter $J_{z}$ will be denoted by $m_{z}$ and can vary in the following range: $-J,-J+1,...,J-1,J$. We know that the total angular momentum is the sum of the spin angular momentum and the orbital magnetic momentum. Here, we will consider the spin and the orbital magnetic moment separately.  
\begin{itemize}
\item \textbf{Case A: Spin angular momentum alone (No orbital angular momentum)}
In this case
\begin{equation}
    \mu_{z} = \gamma_{s} \hbar J \,\,\,\,\,\,\,\, (\textit{For the z-component alone}),
\end{equation}
\begin{figure}[!t]
    \centering
    \includegraphics[scale=0.15]{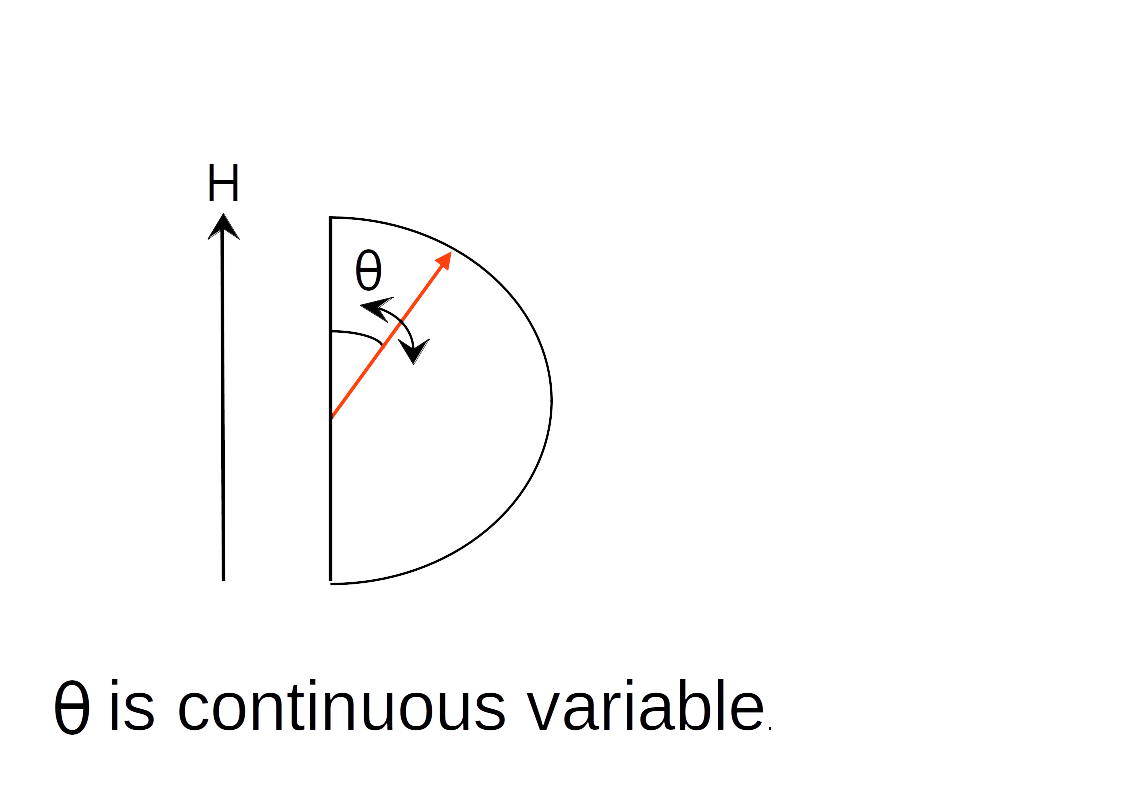}
    \includegraphics[scale=0.15]{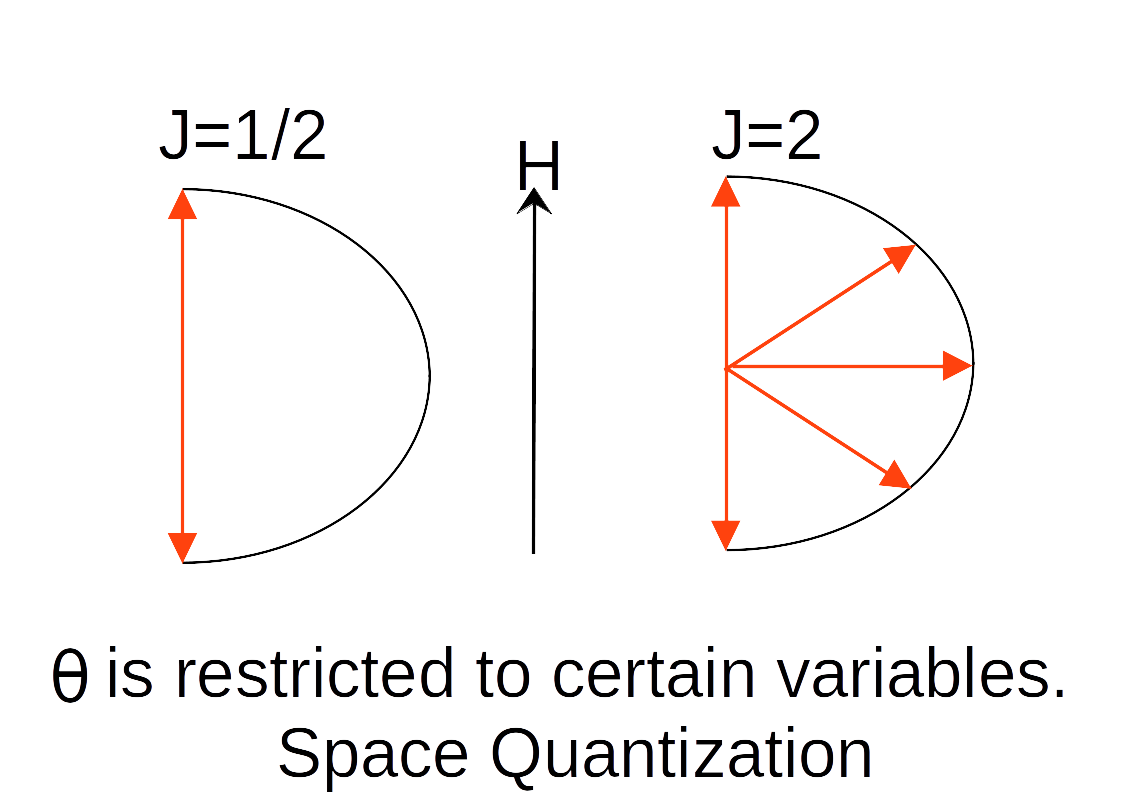}
    \caption{Space quantization of total angular momentum vector.}
    \label{af3}
\end{figure}
\begin{equation}
    \mu = \gamma \hbar S \,\,\,\,\,\,\,\, \textit{(For total angular momentum).}
\end{equation}
For an electron we have
\begin{equation}
    \gamma_{s} = -\frac{e}{mc} \,\,\, ; \,\,\, s = \frac{1}{2}
\end{equation}
which gives the total magnetic moment as
\begin{equation}
    \mu = -\frac{e \hbar}{2mc} =-\mu_{B}
\end{equation}
here $\mu_{B}$ is the Bohr magneton and its value in the CGS units is $\mu_{B} = 9.27 \times 10^{-21}$. Similarly z-component of total magnetic moment is
\begin{equation}
    \mu_{z} = \gamma_{s} \hbar S_{z} \,\,\, (\pm \mu_{B}).
\end{equation}
$ \mu_{z}$ having two possibilities corresponding to the spin quantum number that can only take the values ${+1}/{2}$ and ${-1}/{2}$ .
\item \textbf{Case B: Orbital angular momentum alone (No spin component)}
In this case when there is a contribution from only the orbital angular momentum, we have
\begin{equation}
    \mu_{l} = \gamma_{l} \hbar l,
\end{equation}
here $\it l$ is the orbital angular momentum and varies as $l=0,1,2,3,...$.Further, $\nu_{l}$ is the frequency of revolution and is defined as
\[ \nu_{l} = -\frac{e}{2mc} \].
The magnetic moment is equal to current multiplied with the area of the loop. Therefore
%magnetic moment=current $X$ area of the current loop
%
\begin{equation}
    \mu_{l} = - \left( \frac{e \hbar}{2mc}\right) l = -\mu_{B} l.
\end{equation}
Then the total magnetic moment, is given as 
\begin{equation}
    \Vec{\mu} = \gamma \hbar \Vec{J}.
\end{equation}
where $\Vec{J}$ is the total angular momentum.

\item \textbf{Example: Partition function for the spin $1/2$ particles}\\
Here we will consider the parts of Hamiltonian that contain the magnetic field as a perturbation to $\mathcal{H}_{0}$. The Hamiltonian without the magnetic field has the eigen system
\[ \mathcal{H}_{0}\phi_{0} = \varepsilon_{0}\phi_{0}; \]
and
\begin{equation}
      <{\phi_{0}|\mathcal{H}|\phi_{0}}>= \varepsilon_{0} - \mu \frac{m_{z}}{J}H + \frac{e^{2}H^{2}}{8m c^{2}} <{\rho_{0}}^{2}>
\end{equation}
because 
\[<{\phi_{0}|\hat{x}^{2} + \hat{y}^{2}|\phi_{0}}>= <{\rho}^{2}>\]
The partition function then becomes
\begin{equation}
\begin{split}
    \mathcal{Z} = \sum_{m{z} = -J}^{J} e^{-\beta \varepsilon} e^{\beta \mu \left(\frac{m_{z}}{J}\right)H}e^{-\beta\frac{ e^{2}H^{2}{\rho_{0}}^{2}}{8m c^{2}}},\\
= e^{-\beta \varepsilon}e^{-\beta\frac{ e^{2}H^{2}{\rho_{0}}^{2}}{8m c^{2}}} \sum_{m_{z},
    =-1/2}^{1/2} e^{\beta \mu \left( \frac{m_{z}}{1/2} \right)H },\\
= 2 e^{-\beta \varepsilon} e^{-\beta\frac{ e^{2}H^{2}{\rho_{0}}^{2}}{8m c^{2}}} \left( \frac{e^{-\beta \mu_{B}H} + e^{\beta \mu_{B}H} }{2}\right),\\
\mathcal{Z} = 2 e^{-\beta \varepsilon} e^{-\beta\frac{ e^{2}H^{2}{\rho_{0}}^{2}}{8m c^{2}}} \cosh(\beta \mu_{B} H).
\end{split}    
\end{equation}
From the partition function we can calculate the free energy density which is given as
\begin{equation}
    F(H) = -n K_{B} T \ln \mathcal{Z}(H),
\end{equation}
here n is the number of atoms per unit volume. From this the magnetisation can be found as
\begin{equation}\label{ae1}
M = -\frac{\partial F(H)}{\partial H}.
\end{equation}
Only those terms in the free energy which are dependent on the magnetic field are important to us. Thus we write,
\\
\begin{equation}
F(H) = n\frac{e^{2}{\rho}^{2}}{8m c^{2}}H^{2} -  \frac{n}{\beta} \ln \cosh(\beta \mu_{B}H) +(...)
\end{equation}
putting the value of $F(H)$ in Eq. \ref{ae1} we get

\begin{equation}
%\begin{split}
     M = -n \frac{e^{2}{\rho}^{2}}{4m c^{2}}H + n \mu_{B} \tanh(\beta \mu_{B}H)
%\end{split}    
\end{equation}
If we go with the high temperature or weak field case then ${\mu_{B}H}/{K_{B}T} \ll 1$, because $\mu_{B}H \sim 10^{-6} eV$ and $K_{B}T \sim 10^{-3} eV$. So $tanh(x) \sim x$, which gives
\begin{equation}
    \begin{split}
        M(H) = -n \frac{e^{2}{\rho}^{2}}{4m c^{2}}H  + \frac{n \mu_{B}^{2}}{K_{B}T}H; \\
    \end{split}
\end{equation}
$$
M(H)=(\chi_{dia} + \chi_{para})H.
$$
here
$$
\chi_{dia} = - \frac{ne^{2}{\rho}^{2}}{4m c^{2}}H 
$$
and
\[\chi_{para} = \frac{n \mu_{B}^{2}}{K_{B}T}H  \]
In the Curie-Langevin theory the magnetic moment $\mu$ is introduced in an ad-hoc way. But in the quantum mechanical treatment, one could arrive at the conclusion that this magnetic moment is indeed $\mu_{B}$ which is the Bohr magneton for the free electron case.
\\
\item \textbf{From Quantum to classical: The case of large J values}
We know that the partition function is given as

\begin{equation}
    \mathcal{Z} = \sum_{m{z} = -J}^{J} e^{-\beta \varepsilon} e^{\beta \mu \left(\frac{m_{z}}{J}\right)H}e^{-\beta\frac{ e^{2}H^{2}{\rho_{0}}^{2}}{8m c^{2}}}.
\end{equation}
If we consider a very large J, then ${m_{z}}/{J}$ is quasi-continuous and the summation can then be replaced by an integration in the above equation. Performing the integration by substituting $m_{z}/{J} = x$, we get
$$
%   \begin{split}
\mathcal{Z} = e^{-\beta \varepsilon} e^{-\beta\frac{ e^{2}H^{2}{\rho_{0}}^{2}}{8m c^{2}}} \int_{-1}^{1} dx e^{\beta \mu H x} 
 = e^{-\beta \varepsilon} e^{-\beta\frac{ e^{2}H^{2}{\rho_{0}}^{2}}{8m c^{2}}} \left( \frac{e^{\beta \mu H} - e^{-\beta \mu H} }{\beta\mu H}\right)
 $$
 \\
 \begin{equation}
= e^{-\beta \varepsilon} e^{-\beta\frac{ e^{2}H^{2}{\rho_{0}}^{2}}{8m c^{2}}}(2 +\frac{1}{2} (\beta \mu H)^{2}),
%    \end{split}
\end{equation}
then 
\[ ln\mathcal{Z} = ln (2+\frac{1}{3} (\beta \mu H)^{2})) = ln (1 + \frac{1}{6}  (\beta \mu H)^{2})\]
Therefore, the the free energy density for the paramagnetic part comes out to be
\begin{equation}
    F_{para}(H) = -\frac{n}{\beta} ln(1+ \frac{1}{6}  (\beta \mu H)^{2})),
\end{equation}
approximating  $ln(1+x)$ as $x$ for small values of  $x$, we obtain
\begin{equation}
    F_{para}(H) = -\frac{n}{6\beta} (\beta \mu H)^{2}).
\end{equation}
Then the magnetisation is 
\begin{equation}
    M=  -\frac{\partial F_{para}(H)}{\partial H} = \frac{n \mu^{2}}{3K_{B}T}H,
\end{equation}
and the paramagnetic susceptibility is given as
\begin{equation}
    \chi_{para} = \frac{n \mu^{2}}{3K_{B}T},
\end{equation}
which is eq. (\ref {eq14a}).
\end{itemize}
% Third lecture ends here. Fourth lecture begins hereafter%
\section{Magnetism in metals}
\subsection*{Pauli paramagnetism}
%\paragraph*{}
%It is well-known fact that Sommerfeld theory describes the metals considerably well, a free electron model of metal in the presence of a external Magnetic field $\vec {H} = H \hat {z}$ is framed as
%
 Consider a free electron gas in a volume $L_{x}L_{y}L_{z}$. From periodic boundary conditions we have
%In k-space
\begin{equation}\label{eq16}
\begin{array}{lcr}
L_{x}=n_{x} \lambda_{x};
n_{x}=\frac{L_{x}}{2 \pi} \frac{2 \pi}{\lambda }
=\frac{L_{x}}{2 \pi} k_{x}, \\\\
L_{y}=n_{y} \lambda_{y};
n_{y}=\frac{L_{y}}{2 \pi} \frac{2 \pi}{\lambda }
=\frac{L_{y}}{2 \pi} k_{y}, \\\\
L_{z}=n_{z} \lambda_{z};
n_{z}=\frac{L_{z}}{2 \pi} \frac{2 \pi}{\lambda }
=\frac{L_{z}}{2 \pi} k_{z}, \\\\
\end{array}
\end{equation}
where
\begin{equation}\label{eq16c}
\begin{array}{lcr}
d n_{x}=\frac{L x}{2 \pi} d k_{x};\,\,\,
d n_{y}=\frac{L y}{2 \pi} d f_{y};\,\,\,
d n_{z}=\frac{L z}{2 \pi} d k_{z}.
\end{array}
\end{equation}
%\begin{wrapfigure}{r}{0.25\linewidth}
%\centering
%\includegraphics[scale=0.40,keepaspectratio=true]{K-space.jpeg}
%\caption{k-Space}
%\end{wrapfigure}
Therefore
\begin{center}
$ d n=d n_{x} d n_{y} d n_{z}=\frac{V}{(2 \pi)^{3}} d k_{x} d k_{y} d k_{z}$,\\                                          \end{center}
\begin{equation}\label{eq17}
dn=g_{s}\frac{V}{(2 \pi)^{3}} d^{3} k=g_{s}\frac{V}{(2 \pi)^{3}} 4 \pi k^{2} d k,
\end{equation}
here, $g_{s} (=2)$ is the degeneracy factor, as each $k$-state can have maximum two numbers of electrons and the free particle energy is
\begin{equation}\label{eq18}
\varepsilon=\frac{\hbar^{2} k^{2}}{2 m}.
\end{equation}
The density of states (DOS) defined as number of states per unit energy per unit volume is given as
\begin{equation}\label{eq20}
\rho(\varepsilon)=\frac{d n}{V \varepsilon}
\end{equation}
\begin{equation}\label{eq21}
\rho (\varepsilon)=\frac{m \sqrt{2 m \varepsilon}}{\pi^{2} \hbar^{3}}.
\end{equation}
In the absence of magnetic field B, the number of electrons with spin-up and spin-down are equal ($n_{\uparrow} =n_{\downarrow}$). When B is applied, the Zeeman energy (term) shifts the bands which means that there is an imbalance of up- and down-spins electrons. This imbalance is responsible for creating magnetization in the system. 
% For simplicity, let us consider a simple model with on-site
%repulsion (i.e. an energy cost U when two electrons occupy the same %site). As per the Pauli exclusion principle, it is required that if %two electrons were to occupy the same site, they need to have opposite %spins.
In a simple model with on-site repulsion the energy states are modified in presence of magnetic field as 
\begin{equation}\label{eq22}
\begin{array}{l}
\varepsilon(\vec{k},\uparrow)=\frac{\hbar^{2} k^{2}}{2 m}+\mu_{B}H,\\\\
\varepsilon(\vec{k},\downarrow)=\frac{\hbar^{2} k^{2}}{2 m}-\mu_{B}H.\\
\end{array}
\end{equation}
Induced magnetizaton thus is given as
%\begin{equation}\label{eq23}
%M=-\frac{1}{V} \frac{d E}{d B}
%\end{equation}
%
\begin{equation}
M =-\mu_{B}(n_{\uparrow} - n_{\downarrow}),
\end{equation}
and number of electrons per unit volume is 
\begin{equation}\label{eq24}
n=\frac{N}{V}=\int_{0}^{\infty} \rho(E)f(E)dE,
\end{equation}
%\begin{wrapfigure}{L}{0.5\linewidth}
%\includegraphics[scale=0.40,keepaspectratio=true]{GRAPH.jpg}
%\caption{Density of States when Magnetic field is applied}
%\end{wrapfigure} 
here $\rho(E)$ is the density of states and $f(E)$ is the Fermi-Dirac distribution function. At $T=0K$, $f(E)$ is a step function at $E=E_{f}$. For small $H$, $\mu \ne E_{f} $ as $n_{\uparrow} \ne n_{\downarrow}$. Therefore, there is a change in DOS values in the presence of external magnetic field i.e.
\begin{equation}\label{eq25x}
\rho(\varepsilon) \rightarrow \rho_{\uparrow(\downarrow)}(\varepsilon \pm \mu_{B} H),
\end{equation}
\begin{equation}\label{eq25}
n_{\uparrow(\downarrow)}=\int_{\mp \mu_{B} H}^{\infty} \rho_{\uparrow(\downarrow)}(\varepsilon \pm \mu_{B} H) f(\varepsilon) d \varepsilon,
\end{equation}
\begin{equation}\label{eq26}
n_{\uparrow(\downarrow)}\approx \int_{0}^{\infty} \rho (\varepsilon) f(\varepsilon) d (\varepsilon \pm \mu_{B})H\int_{0}^{\infty} \rho (\varepsilon) \frac{\partial f(\varepsilon) d \varepsilon}{\partial \varepsilon},
\end{equation}
\begin{equation}\label{eq27}
\frac{\partial f(\varepsilon)}{\partial \varepsilon}=\delta\left(\varepsilon-\varepsilon_{f}\right),
\end{equation}
\begin{equation}\label{eq28}
n_{\uparrow(\downarrow)}=\frac{n}{2} \pm \frac{\mu_{B} H \rho \left(\varepsilon_{f}\right)}{2},\\
\end{equation}
then the magnetization is given as
\begin{equation}\label{eq29}
M=\mu_{B}\left(n_{\uparrow}-n_{\downarrow}\right)=\mu_{B}^{2} H \rho \left(\varepsilon_{f}\right),
\end{equation}
and the Pauli susceptibility is given as
%\begin{align*}
\begin{equation}\label{eq30}
 \chi_{Pauli} = \frac {M}{H} = \mu_{B}^2 \rho(\varepsilon_{f}).
\end{equation}
%\end{align*}
%Fourth lecture ends here. Fifth lecture starts hereafter%
%
\section{Stoner model}
In this section, we discuss a simple model for
ferromagnetism in metals.
%having partially filled d-states.
%Within this so called Stoner model, it is assumed that these orbitals form partially filled bands. These bands are expected to be quite narrow, due to the relatively small overlaps between orbitals sitting on different atoms. 
The Stoner model \cite{stoner1938collective} is basically a mean field model for itinerant electrons in magnetic metals \cite{ishikawa}.
%with partially filled d bands.
The model contains an effective exchange energy, which favors a parallel alignment of the electron spins, and the band energy, which instead favors antiparallel alignment.In the simplest version of the Stoner model, the band's assumed to be rectangular and the exchange energy between pairs of electrons is assumed to be a constant, yielding the total
interaction 
\begin{figure}[b]
    \centering
    \includegraphics[width=8cm,height=4cm]{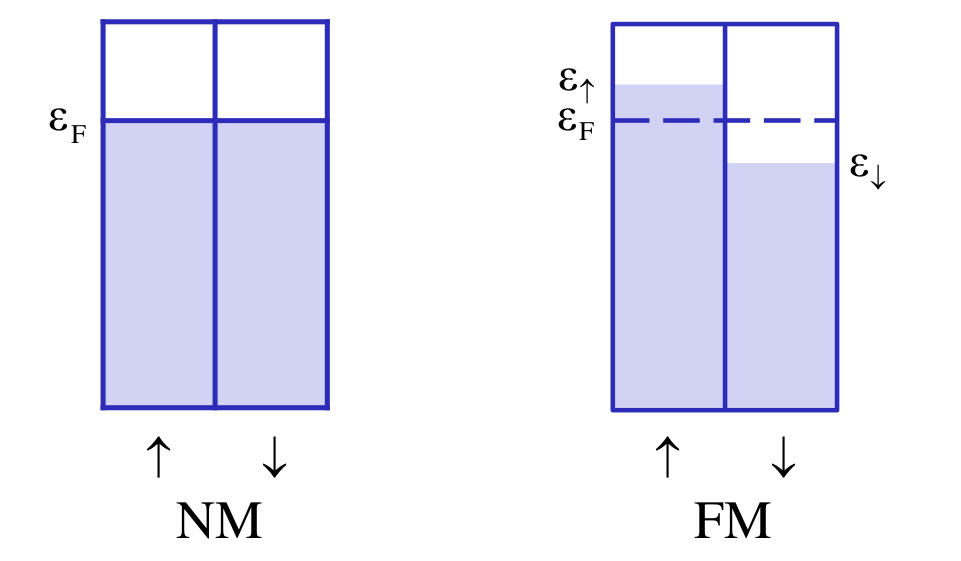}
    \caption{A FM configuration is shown on the right.}
    \label{f4}
\end{figure}
\[ \mathcal{H}_{exch} = U_{eff} n_{\uparrow} n_{\downarrow}\,\,\,\,;\,\,\,\,U_{eff}>0, \]
where $n_{\uparrow}$ and $n_{\downarrow}$ denote the spin-up and spin-down densities. The total density n of electrons is obviously given by $n= n_{\uparrow}+n_{\downarrow}$. The magnetic properties of the ground state are determined by the
competition between two tendencies i.e.
\begin{enumerate}
    \item Increase in kinetic energy 
    \item Decrease in exchange energy.
\end{enumerate}

On L.H.S of figure \ref{f4}, $n_{\uparrow}$ 
is equal to $n_{\downarrow}$ and the system show a non-magnetic configuration. On R.H.S, since there is a shifting of down-spin electrons to up-spin band which makes $U_{eff} > 0$ i.e. the exchange term alone favors a ferromagnetic configuration.
Now the exchange energy of the system is
\[ \mathcal{H}_{exch} = U_{eff} n_{\uparrow} n_{\downarrow} = U_{eff} (n - n_{\downarrow} )n_{\downarrow}. \]
We measure change in effective interaction energy from the point of maxima, i.e. we shift the axis to get
\begin{equation}
    \Delta E_{eff} = U_{eff} n_{\downarrow} n_{\uparrow} - U_{eff} \frac{n^{2}}{4}.
\end{equation}
From the above equation it is clear that 
\begin{itemize}
    \item When $\Delta E_{eff} = 0$  there is no Ferromagnetism.
    \item When $ \Delta E_{eff} < 0 $ there is Ferromagnetism.
\end{itemize}

We take an energy $\Delta$ around the Fermi level which is much smaller than the Fermi energy level and has a single occupancy, i.e. of only one spin polarisation. Let us assume that $pn$ number of electrons in down spin configuration migrates to up spin configuration. Initially there are $n/2$ electrons in both the spin up and spin down state. But this changes to $\left(\frac{n}{2} - pn \right)$ is the new number of down spin electrons and $\left(\frac{n}{2} + pn \right)$ is the new number of up spin electrons.
\[ n = n_{\uparrow} + n_{\downarrow} \]
is conserved. Therefore
%Before calculating the increase in kinetic energy, let us estimate the exchange energy in terms of $pn$, 
\begin{equation}
    \Delta E_{eff} = U_{eff} \left( \left( \frac{n}{2} -pn \right) \left( \frac{n}{2} + pn\right)\right) - U_{eff} \frac{n^{2}}{4}= -U_{eff} n^{2} (p^{2}+\frac{1}{4}).
\end{equation}
Now let us estimate the increase in kinetic energy. If we consider $\varepsilon_{F}$ as $\Delta$
\begin{equation}
    \Delta K.E. = \int_{0}^{\Delta} d\varepsilon \varepsilon \rho_{\uparrow}(\varepsilon) - \int_{-\Delta}^{0} d\varepsilon \varepsilon \rho_{\downarrow} (\varepsilon)
\end{equation}
But we know that 
\[ \rho_{\uparrow} = \rho_{\downarrow} = \frac{1}{2} \rho_{total} \]
where
$\rho_{\uparrow}$ is the density of up spin electrons, $\rho_{\downarrow}$ is the density of down spin electrons, $\rho_{total}$ is the density of total electrons. 
Integrating  above Eqwe get
\begin{equation}
     \Delta K.E. = \rho_{\uparrow}(\varepsilon_{F}) \Delta^{2} > 0.
\end{equation}
Notice the above change in kinetic energy is always positive. We now need an expression that connects $\Delta$ with p,
\[ n_{\uparrow} - n_{\downarrow} = 2 pn = \int_{-\Delta}^{\Delta} d\varepsilon \rho_{\uparrow}(\varepsilon) =  \rho_{\uparrow}(\varepsilon_{F})2 \Delta,    \]
which implies
\begin{equation}
    \Delta = \frac{pn}{\rho_{\uparrow}(\varepsilon_{F})},
\end{equation}
and the change in kinetic energy is
\begin{equation}
    \Delta K.E. = \frac{(pn)^{2}}{\rho_{\uparrow}(\varepsilon_{F})}.
\end{equation}
The total change in energy is then:
\begin{equation}
    \Delta E_{total} = -U_{eff}(pn)^{2} + \frac{(pn)^{2}}{\rho_{\downarrow}(\varepsilon_{F})} = \frac{p^{2}n^{2}}{\rho_{\downarrow}(\varepsilon_{F})}(1 - U_{eff}\rho_{\downarrow}(\varepsilon_{F})),
\end{equation}
this is negative if 
\begin{equation} \label{ae2}
    U_{eff} \rho_{\downarrow}(\varepsilon_{F}) > 1,
\end{equation}
and this is the "Stoner criterion". There is an instability if the above condition holds true. 
%This is the most fundamental result of itinerant electron magnetism.
The Stoner condition (eqn. (\ref{ae2}) demands that either $U_{eff}$ or $\rho(E_{F})$ is large near the Fermi level at Fermi surface. $\rho(E_{F})$ is generally large in d-electrons transition metals and therefore these metals exhibits the phenomenon of ferromagnetism. But there are other subtleties which are discussed in the following sections. We can also derive the change in kinetic energy in an alternative way as
\[ \Delta K.E. = K.E._{Final} - K.E._{Initial} \]\\
\[ \Delta K.E. = \int_{0}^{\varepsilon_{F} -\Delta} d\varepsilon \varepsilon \rho_{\downarrow}(\varepsilon)  + \int_{0}^{\varepsilon_{F} + \Delta} d\varepsilon \varepsilon \rho_{\uparrow}(\varepsilon) - \int_{0}^{\varepsilon_{F}} d\varepsilon \varepsilon \rho_{\downarrow}(\varepsilon)  -\int_{0}^{\varepsilon_{F}}  d\varepsilon \varepsilon \rho_{\downarrow}(\varepsilon), \] 
but
\[ \int_{0}^{\varepsilon_{F} +\Delta} d\varepsilon \varepsilon \rho_{\downarrow}(\varepsilon) = \int_{0}^{\varepsilon_{F} } d\varepsilon \varepsilon \rho_{\downarrow}(\varepsilon) + \int_{\varepsilon_{F}}^{\varepsilon_{F} +\Delta} d\varepsilon \varepsilon \rho_{\downarrow}(\varepsilon) \]
\[ \int_{0}^{\varepsilon_{F} - \Delta} d\varepsilon \varepsilon \rho_{\uparrow}(\varepsilon) = \int_{0}^{\varepsilon_{F} } d\varepsilon \varepsilon \rho_{\uparrow}(\varepsilon) + \int_{\varepsilon_{F}}^{\varepsilon_{F} -\Delta} d\varepsilon \varepsilon \rho_{\uparrow}(\varepsilon),
\]
with this change in kinetic energy becomes
\[ \Delta K.E. = \int_{\varepsilon_{F}}^{\varepsilon_{F} +\Delta} d\varepsilon \varepsilon \rho_{\downarrow}(\varepsilon) +\int_{\varepsilon_{F}}^{\varepsilon_{F} -\Delta} d\varepsilon \varepsilon \rho_{\uparrow}(\varepsilon)\]
\[ = \frac{1}{2} \rho_{\uparrow}(\varepsilon_{F}) \left((E_{f} + \Delta)^{2} - E_{F}^{2} +(E_{f} - \Delta)^{2} - E_{F}^{2} \right)\]
\[ = \rho_{\uparrow}(\varepsilon_{F}) \Delta^{2}. \]
which is in agreement with our previous result.

\section{Stoner enhancement factor in susceptibility}

For stoner model \cite{stoner1938collective}, magnetic susceptibility can be calculated in the following way. In the presence of magnetic field, the Zeeman term must be added to the magnetic energy:

\begin{eqnarray}
\Delta E_{T} = \Delta K.E. + \Delta E_{int} - 2\,p\,n\,\mu_{B}\,H \\
 \,\,\,= \frac{p^{2} n^{2}}{\rho_{\uparrow}(E_{F})} - \,U_{eff}\,n^{2}p^{2} - 2\,p\,n\,\mu_{B}\,H 
\end{eqnarray}

the equilibrium condition is:

\begin{eqnarray}
\frac{d\Delta E_{t}}{dp} = \frac{2 p n^{2}}{\rho_{\uparrow}(E_{F})} - 2\,U_{eff}\,n^{2}p - 2\,n\,\mu_{B}\,H  
\end{eqnarray}

or

\begin{eqnarray}
 p = \frac{\mu_{B}\,H }{n \left( \frac{1}{\rho_{\uparrow}(E_{F})} - U_{eff}\right)} 
\end{eqnarray}

therefore the susceptibility is given by

\begin{eqnarray}
\chi_{stoner} = \frac{2\,p\,n\,\mu_{B}}{H} = \frac{2\,\mu_{B}^{2}}{\rho_{\uparrow}(E_{F}) - U_{eff}}
\end{eqnarray}

which implies

\begin{eqnarray}\label{Schi}
\chi_{stoner} = \frac{2\,\mu_{B}^{2}\rho_{\uparrow}(E_{F})}{1 - U_{eff}\rho_{\uparrow}(E_{F})} = \frac{\chi_{pauli}}{1 - U_{eff}\rho_{\uparrow}(E_{F})}
\end{eqnarray}
The factor $\frac{1}{(1 - U_{eff}\rho_{\uparrow}(E_{F}))}$ is called the Stoner enhancement factor. At section \ref{RPAS} it will be shown that this expression is the same as that of the one derived from the RPA theory in the long wavelength and zero frequency limit.
\section{Itinerant electron magnetism within Random Phase approximation (RPA)}
%
%
%We shall now consider the microscopic theory of the electrons in metals and develop
%approximation to calculate magnetic susceptibility. 
Stoner model \cite{stoner1938collective} is the very first and the  simplest model to address the problems of ferromagnetism due to itinerant electrons. However, it does not take into account the dynamical effects. Also it completely neglects the correlation effects (it only takes into account the exchange interactions within the mean-field approximation) \cite{moriya1965ferro}. To tackle this shortcoming \textit{Bohm and Pines} put forward the notion of Random Phase Approximation.They explicitly distinguished two kinds of response of the electrons to a wave. One of these is in phase with the wave, so that the phase difference between the particle response and the wave producing it is independent of the position of the particle. This is the response which contributes to the ordering behavior of the system. The other response has a phase difference with the wave producing it which depends on the position of the particle.Because of the general random location of the particles, this second response tends to average out to zero when we consider a large number of electrons, and we shall neglect the contributions arising from this. This procedure is termed as the “Random Phase Approximation-RPA". Hence, RPA is a step ahead towards the inclusion of dynamical and correlation effects into the Stoner model.
\subsection{Dynamical susceptibility in the RPA}
In the RPA, the dynamical susceptibility is known to be given by the Fourier transform of the response function or the retarded Green function defined w.r.t spin densities. For the mathematical formulation, we shall follow the Eqof motion approach. Let us begin by considering the spin density operator defined as \\
\begin{equation}
\varsigma (\vec{r}) = \psi ^{\dagger} (\vec{r}) \,\sigma \,  \psi (\vec{r}),
\end{equation}
where $\sigma$ is Pauli matrix. $\psi $ and $\psi^{\dagger}$ are field operators given as 
 \begin{align}
 \psi = \sum_{k} \phi_{k} (\vec {r}) a_{k},\\
\psi^{\dagger} = \sum_{k} \phi_{k}^{*} (\vec {r}) a_{k}^{\dagger}. 
\end{align}
Also, the generalized dynamical susceptibility calculated in appendix A1 is given as

\begin{equation}
 \chi _{BA}(\omega) = \frac{i}{\hbar} \int_0^{\infty} d t \, e^{i \omega t - \epsilon t} \, \left< \left [ B(t) , A \right ] \right >.
\end{equation}
Let us define the operator $B(t)$ and operator $A$ as follows
\begin{equation}
\begin{split}
B(t) & = \varsigma_{\alpha}\left( \vec {r} , t \right), \\
 A& = \varsigma_{\beta} (\vec {r'}).
\end{split}
\end{equation}
Now dynamical magnetic susceptibility can be written  as
\begin{equation}
 \chi _{\alpha,\beta}(\vec {r} - \vec {r'},\omega) = \frac{i}{\hbar} \int_0^{\infty} d t \, e^{i \omega t - \epsilon t} \, \left< \left [ \varsigma_{\alpha}\left( \vec {r} , t \right) , \varsigma_{\beta} (\vec {r'}) \right ] \right >.
\end{equation}
Fourier transform of above Eqcan be  written as
\begin{equation}
\begin{split}
\chi_{\alpha,\beta}(k,\omega) & =  \frac{i}{\hbar} \int_0^{\infty} d t \, e^{i \omega t - \epsilon t} \, \int_{-\infty}^{\infty} d^3  (\vec {r}-\vec{r'}) \, e^{-i \vec{k} \cdot (\vec {r}-\vec{r'})} \, \left< \left [ \varsigma_{\alpha}\left( \vec {r} , t \right) , \varsigma_{\beta} (\vec {r'}) \right ] \right >,
\end{split}
\end{equation}
making the transformation $\vec {r}-\vec{r'} \rightarrow \vec{R} $ and $\vec {r} \rightarrow \vec{r'} + \vec{R} $ we get
\begin{equation}
\begin{split}
\chi_{\alpha,\beta}(k,\omega)& = \frac{i} {\hbar} \int_0^{\infty} d t \, {e^{i \omega t - \epsilon t}} \, \int_{-\infty}^{\infty} d^3  \vec {R} \, e^{-i \vec{k} \cdot \vec{R}} \, \left< \left [ \varsigma_{\alpha}\left( \vec {r'} + \vec{R} , t \right) , \varsigma_{\beta} (\vec {r'}) \right ] \right >.
\end{split}
\end{equation}
Further  $\left< \left [ \varsigma_{\alpha}\left( \vec {r'} + \vec{R} , t \right) , \varsigma_{\beta} (\vec {r'}) \right ] \right >$ in terms of $\vec{k}$ can be written using the following transformation:
$\varsigma_{\alpha}(\vec{r}) = \sum_{k} e^{ i \vec{k} \cdot \vec{r}} \varsigma_{\alpha}(\vec{k}) $ 
\begin{equation}
\begin{split}
\chi_{\alpha,\beta}(k,\omega) & = \frac{i} {\hbar} \int_0^{\infty} d t \, {e^{i \omega t - \epsilon t}} \, \int_{-\infty}^{\infty} d^3  \vec {R} \, e^{-i \vec{k} \cdot \vec{R}} \, \sum_{K_1}  e^{ i \vec{K_{1}} \cdot (\vec{r'} + \vec{R})}  \sum_{K_2}  e^{ i \vec{K_{2}} \cdot \vec{r'}} \left< \left [ \varsigma_{\alpha}\left( K_{1} , t \right) , \varsigma_{\beta} (\vec {K_{2}}) \right ] \right > \\
& = \frac{i} {\hbar} \int_0^{\infty} d t \, {e^{i \omega t - \epsilon t}} \, \int_{-\infty}^{\infty} d^3  \vec {R} \, e^{-i \vec{k} \cdot \vec{R}} \, \sum_{K_1, K_{2}}  e^{ i {(\vec {K_{1}} + \vec{K_{2}})} \cdot \vec{r'}}   e^{ i \vec{K_{1}} \cdot \vec{R}} \left< \left [ \varsigma_{\alpha}\left( K_{1} , t \right) , \varsigma_{\beta} (\vec {K_{2}}) \right ] \right >.
\end{split}
\end{equation}
To make the above equation independent of $\vec{r'}$ , consider 
$e^{ i {(\vec {K_{1}} + \vec{K_{2}})} \cdot \vec{r'}} = 1 $ this implies $(\vec {K_{1}} + \vec{K_{2}}) \cdot \vec{r'} = n \pi $, for $n = 0  \implies \vec {K_{1}} = -\vec{K_{2}}$. Therefore the above Eqtakes the form 
\begin{equation}
\begin{split}
\chi_{\alpha,\beta}(k,\omega) & = \frac{i} {\hbar} \int_0^{\infty} d t \, {e^{i \omega t - \epsilon t}} \, \int_{-\infty}^{\infty} d^3  \vec {R} \, e^{-i \vec{k} \cdot \vec{R}} \, e^{ i \vec{K_{1}} \cdot \vec{R}} \left< \left [ \varsigma_{\alpha}\left( \vec{K}_{1} , t \right) , \varsigma_{\beta} (\vec {-K_{1}}) \right ] \right >\\
& = \frac{i} {\hbar} \int_0^{\infty} d t \, {e^{i \omega t - \epsilon t}} \, \int_{-\infty}^{\infty} d^3  \vec {R} \, e^{i (\vec{K_{1}}-\vec{k} )\cdot \vec{R}} \, \left< \left [ \varsigma_{\alpha}\left( \vec{K}_{1} , t \right) , \varsigma_{\beta} (\vec {-K_{1}}) \right ] \right >.
\end{split}
\end{equation}
From definition of $ \delta$-function 
\begin{equation}
 \int_{-\infty}^{\infty} {d^3\vec {R}\,e^{i (\vec{K_{1}}-\vec{k} )\cdot \vec{R}}} =
\delta_{\vec{K}_1,k},   
\end{equation}
we get
\begin{equation}
\chi_{\alpha,\beta}(k,\omega) = \frac{i} {\hbar} \int_0^{\infty} d t \, {e^{i \omega t - \epsilon t}} \, \left< \left [ \varsigma_{\alpha}\left( {k} , t \right) , \varsigma_{\beta} (-{k}) \right ] \right >.
\end{equation}
%
%\subsection{The Correlation Function }
%\subsection{Stoner excitations}
Further, we will calculate $\chi_{-+}$ component of  the dynamical susceptibility, which is defined by
\begin{equation} \label{ae3}
\begin{aligned}
\chi_{-+}&(\boldsymbol{q}, \omega)=\frac{i}{\hbar} \int_{0}^{\infty} e^{-i \omega t}\left\langle\left[\varsigma_{-}(\boldsymbol{q}, t), \varsigma_{+}(-\boldsymbol{q})\right]\right\rangle d t \\
= \int_{-\infty}^{\infty} e^{-i \omega t}\langle\langle\varsigma_{-}(\boldsymbol{q}, \mathrm{t}) ; \varsigma_{+}(-\boldsymbol{q}) \rangle\rangle d t.
\end{aligned}
\end{equation}
where $\varsigma_{-+}$ is defined as
\begin{equation} \label{ae4}
 \varsigma_{-}(\boldsymbol{q}, t)=\sum_{\boldsymbol{k}} a_{\boldsymbol{k}+\boldsymbol{q} \downarrow}^{*}(t) a_{\boldsymbol{k} \uparrow}(t)  
\end{equation}
\begin{equation}
 \varsigma_{+}(\boldsymbol{-q})=\sum_{\boldsymbol{k}} a_{\boldsymbol{k} \uparrow}^{*} a_{\boldsymbol{k}+\boldsymbol{q} \downarrow}
\end{equation}
for simplicity we introduce the notation
\begin{equation}
\vartheta_{k}(\boldsymbol{q}, t) \equiv a_{k+q \downarrow}^{*}(t) a_{k \uparrow}(t),    
\end{equation}
and rewrite eqn.(\ref{ae4}) as

\begin{equation}
\varsigma_{-}(\boldsymbol{q}, t)=\sum_{\boldsymbol{k}} \vartheta_{\boldsymbol{k}}(\boldsymbol{q}, t).
\end{equation}

Define the retarded Green function as
\begin{equation}
\left\langle\langle\widetilde{\sigma}_{-}(\boldsymbol{q}, t) ; \varsigma_{+}(\boldsymbol{q})\right\rangle\rangle=\sum_{\boldsymbol{k}}\left\langle\langle\vartheta_{\boldsymbol{k}}(\boldsymbol{q}, t) ; \varsigma_{+}(-\boldsymbol{q})\right\rangle\rangle,
\end{equation}
is determined by applying the EOM approach. Therefore by using the conventional defination of retarded Green function the R.H.S of the above Eqtakes the form as
\begin{equation}\label{ae7}
%\begin{array}
\iota \hbar \frac{d}{d t}\left\langle\langle\vartheta_{k}(\boldsymbol{q}, t) ; \varsigma_{+}(-\boldsymbol{q})\right\rangle\rangle \\\\
=-\delta(t)\left\langle\left[\vartheta_{k}(\boldsymbol{q}, t), \varsigma_{+}(-\boldsymbol{q})\right]\right\rangle+\left\langle\langle\left[\vartheta_{k}(\boldsymbol{q}, t), \mathcal{H}\right] ; \varsigma_{+}(-\boldsymbol{q})\right\rangle\rangle,
%\end{array}    
\end{equation}
here $\mathcal{H}$ is the Hamiltonian of the system  and is given as
\begin{equation}
 \mathcal{H}=\mathcal{H}_{0}+\mathcal{H}_{c} \\   
\end{equation}
\begin{equation}
 \mathcal{H}_{0}=\sum_{\boldsymbol{k \mu}}\epsilon(k) \eta_{k\mu}   
\end{equation}
\begin{equation}
 \mathcal{H}_{c}=\frac{1}{2}U_{eff}\sum_{\kappa}^{'}\sum_{\boldsymbol{k}\mu}\sum_{l \nu} a_{\boldsymbol{k}+\kappa,\mu}^{*}a_{k\mu}a_{l,\nu}^{*} a_{l+\kappa,\nu}   
\end{equation}
where
\begin{equation}
\eta_{k\mu}=a_{k\mu}^{*}a_{k\mu}    
\end{equation}
Now we will calculate the commutators involved in solving the EOM in eq.\ref{ae7}:
%\begin{equation}
%
% \begin{array}{l}
\begin{equation}
\begin{aligned}
[\vartheta_{k}(\boldsymbol{q}), \varsigma_{+}(-\boldsymbol{q})]= -[ a^{\dagger}_{k+q \downarrow} a_{k\uparrow}, \sum_{p} a^{\dagger}_{p\uparrow} a_{p+q\downarrow}] 
= -\sum_{p} [ a^{\dagger}_{k+q \downarrow}a_{k\uparrow},a^{\dagger}_{p\uparrow} a_{p+q\downarrow}]\\
= \sum_{p} a^{\dagger}_{k+q \downarrow} \left[ a_{k\uparrow},a^{\dagger}_{p\uparrow} a_{p+q\downarrow}\right] - \sum_{p} \left[ a^{\dagger}_{k+q \downarrow}, a_{k\uparrow},a^{\dagger}_{p\uparrow} a_{p+q\downarrow}\right] a_{k\uparrow} -\sum_{p} a^{\dagger}_{k+q \downarrow} a^{\dagger}_{p\uparrow} \left[ a_{k\uparrow},a_{p+q\downarrow}\right] 
\\
-\sum_{p} a^{\dagger}_{k+q \downarrow} \left[ a_{k\uparrow}, a^{\dagger}_{p\uparrow} \right] a_{p+q\downarrow} -\sum_{p} a^{\dagger}_{p\uparrow}\left[a^{\dagger}_{k+q \downarrow}, a_{p+q\downarrow}\right] a_{k\uparrow} - \sum_{p} \left[a^{\dagger}_{k+q \downarrow},a^{\dagger}_{p\uparrow}\right] a_{p+q\downarrow}a_{k\uparrow} \\
\end{aligned}
\end{equation}
out of the four commutators the first and the last commutators both are zero, whereas the second and third form delta functions. Evaluating further with the properties of delta function we get,
\begin{equation}
    \begin{aligned}
        a^{\dagger}_{k+q \downarrow}a_{k+q \downarrow} + a^{\dagger}_{k\uparrow}a_{k\uparrow} =n_{k+q \downarrow}-n_{\boldsymbol{k} \uparrow}. 
    \end{aligned}
\end{equation}
Next we evaluate
%
%\begin{equation}
%\begin{aligned}
$$
{\left[\vartheta_{k}(\boldsymbol{q}), \mathcal{H}_{0}\right]= \left[ a^{\dagger}_{k+q\downarrow} a_{k\uparrow}, \sum_{p\sigma} \varepsilon(p) a^{\dagger}_{p\sigma}a_{p\sigma}\right] =  \sum_{p\sigma}\left[ a^{\dagger}_{k+q\downarrow} a_{k\uparrow}, \varepsilon(p) a^{\dagger}_{p\sigma}a_{p\sigma}\right] }
$$
%\end{aligned}
%\end{equation}
expanding the sum over the spins ($\sigma$), we get
\begin{equation}
    \begin{aligned}
\left[\vartheta_{k}(\boldsymbol{q}), \mathcal{H}_{0}\right]= \sum_{p\sigma}\varepsilon(p)\left[ a^{\dagger}_{k+q\downarrow} a_{k\uparrow},  a^{\dagger}_{p\uparrow}a_{p\uparrow}\right] + \sum_{p\sigma}\varepsilon(p)\left[ a^{\dagger}_{k+q\downarrow} a_{k\uparrow},  a^{\dagger}_{p\downarrow}a_{p\downarrow}\right]\\
=\sum_{p\sigma}\varepsilon(p) \left(a^{\dagger}_{k+q\downarrow} \left[ a_{k\uparrow}, a^{\dagger}_{p\uparrow}a_{p\uparrow}\right] + \left[ a^{\dagger}_{k+q\downarrow},a^{\dagger}_{p\uparrow}a_{p\uparrow}\right] a_{k\uparrow}\right) +  \sum_{p\sigma}\varepsilon(p) \left( a^{\dagger}_{k+q\downarrow}\left[ a_{k\uparrow}, a^{\dagger}_{p\downarrow}a_{p\downarrow}\right] + \left[ a^{\dagger}_{k+q\downarrow},  a^{\dagger}_{p\downarrow}a_{p\downarrow}\right] a_{k\uparrow} \right)\\
= \sum_{p\sigma}\varepsilon(p) \left( a^{\dagger}_{k+q\downarrow} a^{\dagger}_{p\uparrow}\left[ a_{k\uparrow},a_{p\uparrow}\right] + a^{\dagger}_{k+q\downarrow} \left[ a_{k\uparrow},a^{\dagger}_{p\uparrow}\right]a_{p\uparrow} + a^{\dagger}_{p\uparrow}\left[a^{\dagger}_{k+q\downarrow}, a_{p\uparrow} \right]a_{k\uparrow} + \left[ a^{\dagger}_{k+q\downarrow}{k\uparrow},a^{\dagger}_{p\uparrow}\right] a_{p\uparrow}a_{k\uparrow} \right)\\
+ = \sum_{p\sigma}\varepsilon(p) \left( a^{\dagger}_{k+q\downarrow} a^{\dagger}_{p\downarrow}\left[ a_{k\uparrow},a_{p\downarrow}\right] + a^{\dagger}_{k+q\downarrow} \left[ a_{k\uparrow},a^{\dagger}_{p\downarrow}\right]a_{p\uparrow} + a^{\dagger}_{p\downarrow}\left[a^{\dagger}_{k+q\downarrow}, a_{p\downarrow} \right]a_{k\uparrow} + \left[ a^{\dagger}_{k+q\downarrow}{k\uparrow},a^{\dagger}_{p\downarrow}\right] a_{p\downarrow}a_{k\uparrow} \right)
\end{aligned}
\end{equation}
Out of the eight commutators in the above Eqonly the second in the first four and the third in the next four yield a delta function. Other commutators are zero. Further evaluating the delta functions
\begin{equation}
    \begin{aligned}
        \left[\vartheta_{k}(\boldsymbol{q}), \mathcal{H}_{0}\right]= \varepsilon(k)  a^{\da
        }_{k+q\downarrow} a_{k\uparrow} - \varepsilon(k+q) a^{\da
        }_{k+q\downarrow} a_{k\uparrow} = \left(\varepsilon(\boldsymbol{k})-\varepsilon(\boldsymbol{k}+\boldsymbol{q})\right) \vartheta_{k}(\boldsymbol{q})
    \end{aligned}
\end{equation}
These commutators consist of many terms and can be expanded as follows
\begin{equation}
\begin{aligned}
\left[\vartheta_{k}(\boldsymbol{q}), \mathcal{H}_{c}\right] =  \left[a^{\dagger}_{k+q\downarrow}a_{k\uparrow}, \frac{U_{eff}}{2} \sum_{k_{1}k_{2}q'\sigma_{1}\sigma_{2}}a^{\dagger}_{k_{1} +q' \sigma} a^{\dagger}_{k_{2} -q' \sigma'} a_{k_{2}\sigma'} a_{k_{1}\sigma}\right]\\
= \frac{U_{eff}}{2} \sum_{k_{1}k_{2}q'\sigma_{1}\sigma_{2}} \left[a^{\dagger}_{k+q\downarrow}a_{k\uparrow}, a^{\dagger}_{k_{1} +q' \sigma} a^{\dagger}_{k_{2} -q' \sigma'} a_{k_{2}\sigma'} a_{k_{1}\sigma}\right]\\
= \frac{U_{eff}}{2} \sum_{k_{1}k_{2}q'\sigma_{1}\sigma_{2}}a^{\dagger}_{k+q\downarrow}\left[ a_{k\uparrow}, a^{\dagger}_{k_{1} +q' \sigma} a^{\dagger}_{k_{2} -q' \sigma'} a_{k_{2}\sigma'} a_{k_{1}\sigma}\right] + \left[a^{\dagger}_{k+q\downarrow} , a^{\dagger}_{k_{1} +q' \sigma} a^{\dagger}_{k_{2} -q' \sigma'} a_{k_{2}\sigma'} a_{k_{1}\sigma}\right] a_{k\uparrow}\\
= \frac{U_{eff}}{2} \sum_{k_{1}k_{2}q'\sigma_{1}\sigma_{2}}a^{\dagger}_{k+q\downarrow}a^{\dagger}_{k_{1} +q' \sigma}\left[  a_{k\uparrow},a^{\dagger}_{k_{2} -q' \sigma'} a_{k_{2}\sigma'} a_{k_{1}\sigma}\right] + a^{\dagger}_{k+q\downarrow} \left[a_{k\uparrow},a^{\dagger}_{k_{1} +q'}\right]a^{\dagger}_{k_{2} -q' \sigma'} a_{k_{2}\sigma'} a_{k_{1} \sigma} \\
a^{\dagger}_{k_{1} +q' \sigma}\left[ a^{\dagger}_{k+q\downarrow}, a^{\dagger}_{k_{2} -q' \sigma'} a_{k_{2}\sigma'} a_{k_{1}\sigma}\right] a_{k\uparrow} + \left[ a^{\dagger}_{k+q\downarrow}, a^{\dagger}_{k_{1} +q' \sigma}\right]a^{\dagger}_{k_{2} -q' \sigma'} a_{k_{2}\sigma'}a_{k_{1}\sigma}a_{k\uparrow}
\end{aligned}
\end{equation}
\\
The last commutator in the above Eqis zero. Evaluating further the other commutators we get 
\begin{equation}
\begin{aligned}
\left[\vartheta_{k}(\boldsymbol{q}), \mathcal{H}_{c}\right] = \frac{U_{eff}}{2} \sum_{k_{1}k_{2}q'\sigma_{1}\sigma_{2}}a^{\dagger}_{k+q\downarrow}a^{\dagger}_{k_{1} +q' \sigma}a^{\dagger}_{k_{2} -q' \sigma'}\left[a_{k\uparrow},a_{k_{2}\sigma'}a_{k_{1}\sigma}\right] + 
a^{\dagger}_{k+ q'\downarrow}a^{\dagger}_{k_{1}+q'\sigma}\left[a_{k\uparrow},a^{\dagger}_{k_{2} -q' \sigma'}\right]a_{k_{2}\sigma'}a_{k_{1}\sigma} \\
+ a^{\dagger}_{k +q \downarrow} \left[ a_{k\uparrow},a^{\dagger}_{k_{1} +q' \sigma}\right]a^{\dagger}_{k_{2} -q' \sigma'}a_{k_{2}\sigma'}a_{k_{1}\sigma} + a^{\dagger}_{k_{1} +q' \sigma}a^{\dagger}_{k_{2} -q' \sigma'} \left[ a^{\dagger}_{k+q\downarrow},a_{k_{2}\sigma'}a_{k_{1}\sigma}  \right]  a_{k\uparrow}
    \end{aligned}
\end{equation}
The first commutator in the above eqn. is zero. Evaluating the last commutator further we get,
\begin{equation}
\begin{aligned}
\left[\vartheta_{k}(\boldsymbol{q}), \mathcal{H}_{c}\right] = \frac{U_{eff}}{2} \sum_{k_{1}k_{2}q'\sigma_{1}\sigma_{2}} a^{\dagger}_{k+ q'\downarrow}a^{\dagger}_{k_{1}+q'\sigma}\left[a_{k\uparrow},a^{\dagger}_{k_{2} -q' \sigma'}\right]a_{k_{2}\sigma'}a_{k_{1}\sigma} +
a^{\dagger}_{k +q \downarrow} \left[ a_{k\uparrow},a^{\dagger}_{k_{1} +q' \sigma}\right]a^{\dagger}_{k_{2} -q' \sigma'}a_{k_{2}\sigma'}a_{k_{1}\sigma}\\
a^{\dagger}_{k_{1} +q' \sigma}a^{\dagger}_{k_{2} -q' \sigma'}a_{k_{2}\sigma'} \left[ a^{\dagger}_{k+q\downarrow},a_{k_{1}\sigma}  \right] a_{k\uparrow} +
a^{\dagger}_{k_{1} +q' \sigma}a^{\dagger}_{k_{2} -q' \sigma'} \left[ a^{\dagger}_{k+q\downarrow},a_{k_{2}\sigma'} \right] a_{k_{1}\sigma}  a_{k\uparrow}
\end{aligned}
\end{equation}
All the commutators in the above eqn. give a delta function. Evaluating this delta function we get
\begin{equation}
    \begin{aligned}
    \left[\vartheta_{k}(\boldsymbol{q}), \mathcal{H}_{c}\right] = \frac{U_{eff}}{2} \sum_{k_{1}q'\sigma_{1}} \left( - a^{\dagger}_{k+q\downarrow} a^{\dagger}_{k_{1}+q' \sigma} a_{k+q'\uparrow} a_{k_{1}\sigma}+ a^{\dagger}_{k_{1}+q' \sigma} a^{\dagger}_{k+q-q'}a_{k_{1}\sigma}a_{k\uparrow}\right) \\
    + \frac{U_{eff}}{2} \sum_{k_{2}q'\sigma_{2}}\left( -a^{\dagger}_{k+q\downarrow}a^{\dagger}_{k_{2} -q'\sigma'}a_{k_{2}\sigma'} a_{k-q'\uparrow} + a^{k+q-q' \uparrow} a^{\dagger}_{k_{2} - q'\sigma'}a_{k_{2}\sigma'}a_{k\uparrow} 
    \right)
    \end{aligned}
\end{equation}
rearranging the terms in the forms of number operators this simplifies as
\begin{equation}
    \begin{aligned}
   \left[\vartheta_{k}(\boldsymbol{q}), \mathcal{H}_{c}\right] = \frac{U_{eff}}{2} \sum_{q'} -a^{\dagger}_{k+q\downarrow}a^{\dagger}_{k+q+q'\downarrow} a_{k+q'\uparrow}a_{k+q\downarrow} + a^{\dagger}_{k+q\downarrow}a^{\dagger}_{k+q-q'\downarrow} a_{k+q-q'\downarrow}a_{k\uparrow} \\ - a^{\dagger}_{k+q\downarrow}a^{\dagger}_{k-q'\uparrow} a_{k\uparrow}a_{k-q'\uparrow} + a^{\dagger}_{k+q+q'\downarrow}a^{\dagger}_{k\uparrow} a_{k+q'\uparrow}a_{k\uparrow}
    \end{aligned}
\end{equation}
This expression can be simplified into 
\begin{equation}
% \begin{aligned}
\left[\vartheta_{k}(\boldsymbol{q}), \mathcal{H}_{c}\right]  \simeq.v \sum_{q'}\left\{\vartheta_{k+q'}(\boldsymbol{q})\left(-n_{k+\boldsymbol{q} \downarrow}+n_{\boldsymbol{k} \uparrow}\right)+\vartheta_{\boldsymbol{k}}(\boldsymbol{q})\left(-n_{\boldsymbol{k}+\kappa, \uparrow}+n_{\boldsymbol{k}+\boldsymbol{q}+\boldsymbol{k}, \downarrow}\right)\right\}
%\end{aligned}   
\end{equation}
further, in this approximation
\begin{equation}
\begin{array}{l}
\left\langle\langle n_{k, \mu} \vartheta_{i}(\boldsymbol{q}, t) ; \varsigma_{+}(-\boldsymbol{q})\right\rangle\rangle\cong\left\langle n_{k, \mu}\right\rangle\langle\left\langle\vartheta_{l}(\boldsymbol{q}, t) ; \varsigma_{+}(-\boldsymbol{q})\right\rangle\rangle \\\\
\quad=f_{k \mu}\left\langle\langle\vartheta_{\iota}(\boldsymbol{q}, t) ; \varsigma_{+}(-\boldsymbol{q})\right\rangle\rangle
\end{array}    
\end{equation}
EOM is now reduced to the simpler form as

\begin{equation}
\begin{aligned}
\iota \hbar \frac{d}{d t}\left\langle\vartheta_{k}(\boldsymbol{q}, t) ;\varsigma_{+}(-\boldsymbol{q})\right\rangle=-\delta(t)\left(f_{k+\boldsymbol{q} \downarrow}-f_{k \uparrow}\right)+\left(\tilde{\varepsilon}_{\uparrow}(\boldsymbol{k})-\tilde{\varepsilon}_{\downarrow}(\boldsymbol{k}+\boldsymbol{q})\right)\left\langle\langle\vartheta_{k}(\boldsymbol{q} \cdot t) ; \varsigma_{+}(-\boldsymbol{q})\right\rangle\rangle\\
+U_{eff}\left(f_{k \uparrow}-f_{k+q \downarrow}\right) \sum_{\kappa}\left\langle\langle\vartheta_{k+k}(\boldsymbol{q}, t) ; \varsigma_{+}^{\prime}(-\boldsymbol{q})\right\rangle\rangle \\
\end{aligned}
\end{equation}
where
\begin{equation}
\widetilde{\varepsilon}_{\mu}(\boldsymbol{k}) \equiv \varepsilon(\boldsymbol{k})- U_{eff} \sum_{\kappa}^{\prime} f_{k+\kappa, \mu}    
\end{equation}
Therefore EOM is further transformed by its Fourier transform as
\\
\begin{equation}
\begin{array}{l}
\left\{-\hbar \omega+\tilde{\varepsilon}_{\downarrow}(\boldsymbol{k}+\boldsymbol{q})-\tilde{\varepsilon}_{\uparrow}(\boldsymbol{k})\right\}\left\langle\langle\vartheta_{\boldsymbol{k}}(\boldsymbol{q}, \omega) ; \varsigma_{+}(-\boldsymbol{q})\right\rangle\rangle \\
\quad=-\left(f_{\boldsymbol{k}+\boldsymbol{q} \downarrow}-f_{\boldsymbol{k} \uparrow}\right)+U_{eff}\left(f_{\boldsymbol{k} \uparrow}-f_{\boldsymbol{k}+\boldsymbol{q} \downarrow}\right) \sum_{\boldsymbol{k}}\left\langle\langle\vartheta_{\boldsymbol{k}+\kappa}(\boldsymbol{q}, \omega) ; \varsigma_{+}(-\boldsymbol{q})\right\rangle\rangle
\end{array}    
\end{equation}
\\
dividing both sides by $(-\hbar \omega+\tilde{\varepsilon}_{\downarrow}(\boldsymbol{k}+\boldsymbol{q})-\tilde{\varepsilon}_{\uparrow}(\boldsymbol{k})$ and summing over all wave numbers we obtain the solution as 
\begin{equation}\label{ae8}
\chi_{-+}(\boldsymbol{q}, \omega)=\sum_{\boldsymbol{k}}\left\langle\vartheta_{\boldsymbol{k}}(\boldsymbol{q}, \omega) ; \varsigma(-\boldsymbol{q})\right\rangle=\frac{\Gamma_{-+}(\boldsymbol{q}, \boldsymbol{\omega})}{1-U_{eff} \Gamma_{-+}(\boldsymbol{q}, \omega)},    
\end{equation}
\\
where $\Gamma_{-+}(\boldsymbol{q}, \boldsymbol{\omega})$ is the Lindhard function and is defined as 
\begin{equation}\label{ae91}
 \Gamma_{-+}(\boldsymbol{q}, \omega)=\sum_{k} \frac{f_{k \uparrow}-f_{k+q \uparrow}}{{\varepsilon}_{\downarrow}(\boldsymbol{k}+\boldsymbol{q})-{\varepsilon}_{\uparrow}(\boldsymbol{k})-\hbar \omega}   
\end{equation}
If we put $1-U_{eff} \Gamma_{-+}(\boldsymbol{q}, \omega)=0$ in eq.(\ref{ae8}) then $\chi_{-+}^{-1}(\boldsymbol{q}, \omega)\rightarrow0$ which indicates a phase transition in the system.
\subsection{Stoner excitations }
 If we put $\tilde{\varepsilon}_{\downarrow}(\boldsymbol{k}+\boldsymbol{q})-\tilde{\varepsilon}_{\uparrow}(\boldsymbol{k}-\hbar \omega_{q}=0 $ in eq.(\ref{ae91}) then one can calculate excitations spectra of the system known as the Stoner excitations. Put
\begin{equation}
\tilde{\varepsilon}_{\downarrow}(\boldsymbol{k}+\boldsymbol{q})-\tilde{\varepsilon}_{\uparrow}(\boldsymbol{k}-\hbar \omega_{q}=0    
\end{equation}
\begin{equation}
\tilde{\varepsilon}_{\downarrow}(\boldsymbol{k}+\boldsymbol{q})- U_{eff}n_{\downarrow}-(\tilde{\varepsilon}_{\uparrow}(\boldsymbol{k})-U_{eff}n_{\uparrow})=\hbar \omega_{q}   
\end{equation}
\begin{equation}
 \tilde{\varepsilon}_{\downarrow}(\boldsymbol{k}+\boldsymbol{q})- \tilde{\varepsilon}_{\uparrow}(\boldsymbol{k})+U_{eff}(n_{\downarrow}-n_{\uparrow})=\hbar \omega_{q}   
\end{equation}
\begin{equation}
 \tilde{\varepsilon}_{\downarrow}(\boldsymbol{k}+\boldsymbol{q})- \tilde{\varepsilon}_{\uparrow}(\boldsymbol{k})+\Delta=\hbar \omega_{q}   
\end{equation}
\begin{equation}
 \hbar \omega_{q}=\frac{\hbar^{2}}{2m}((k+q)^{2}-k^{2})-\Delta  
\end{equation}
\begin{equation}
 \hbar \omega_{q}=\frac{\hbar^{2}}{2m}(k^{2}+q^{2}+2k.q-k^{2})-\Delta  
\end{equation}
\begin{equation}
 \hbar \omega_{q}=\frac{\hbar^{2}}{2m}(q^{2}+2k_{F}.q)-\Delta  
\end{equation}
\begin{figure}[t]
    \centering
    \includegraphics[scale=0.2]{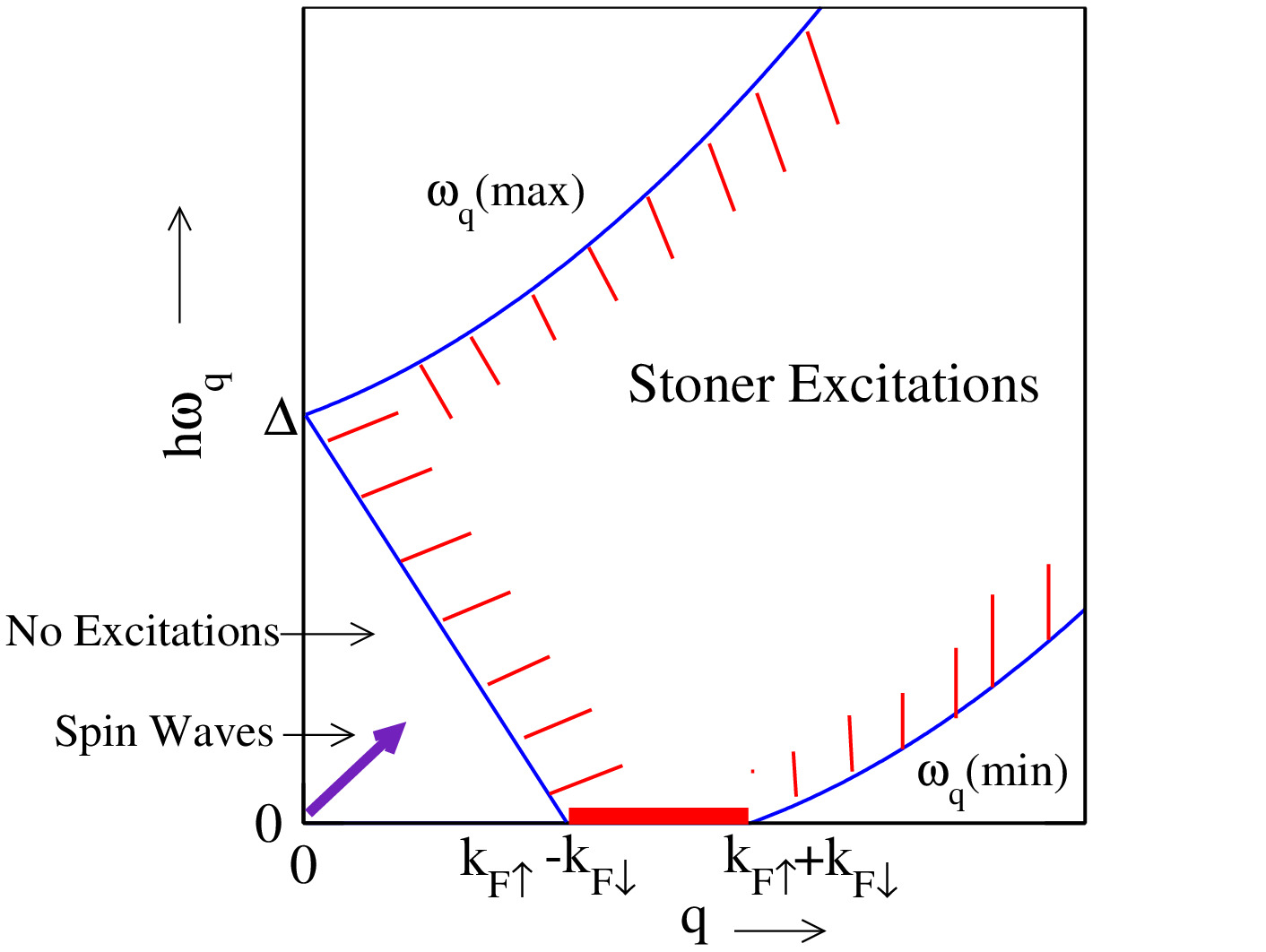}
    \caption{The Stoner excitations and the dispersion relation for spin waves which exists outside the continuum and represents therefore long living collective excitations.}
    \label{af6}
\end{figure}
\\
hence  this a quadratic Eqand roots of this Eqgives us upper and lower boundaries of Stoner excitation spectrum as
\begin{equation}
 \hbar \omega_{q}^{max}=\frac{\hbar^{2}}{2m}(q^{2}+2k_{F\uparrow}.q)-\Delta
\end{equation}
\begin{equation}
 \hbar \omega_{q}^{min}=\frac{\hbar^{2}}{2m}(q^{2}+2k_{F\downarrow}.q)-\Delta
\end{equation}
Figure \ref{af6} depicts that in the low $q$-region i.e from 0 to $k_{F\uparrow}-k_{F\downarrow}$ there are no excitations. Electron-hole pairs excitations in the Stoner theory are the spin flip excitations of electrons across the Fermi surface. The excited electrons and holes move independently in the common mean field due to other electrons. By virtue of which small spin density fluctuations are produced by thermal smearing of the Fermi level. The significance of
the interaction between the excited electrons and holes was pointed out by Slater in his theory of spin waves as excitons or bound collective modes in ferromagnetic insulators. After that, the spin wave \cite{schrieffer1968effect} theory of ferromagnetic metals has been developed by using the dynamical Hartree-Fock approximation (HFA) or the RPA. 
%Further, the dynamical HFA or the random-phase approximation
%(RPA) theory of general spin fluctuations including paramagnetic spin fluctuations has been developed.  
%If we expand denomenator of $\chi_{-+}$ to second order.
In the RPA, the $\chi_{zz}$ can be calculated in the following way, where $z$ is the direction of spontaneous magnetization. Therefore $\varsigma_{z}(\boldsymbol{q},t)$ in eq. (\ref{ae3}) is written as
\begin{equation}
 \varsigma_{z}(\boldsymbol{q},t)=\frac{1}{2}\{\sum_{k}\vartheta_{k \uparrow}(\boldsymbol{q}, t)-\sum_{k}\vartheta_{k \downarrow}(\boldsymbol{q}, t)\}
\end{equation}
where
\begin{equation}
 \vartheta_{k \sigma}(\boldsymbol{q}, t) =a_{k+q \sigma}^{*}(t)a_{k \sigma}(t) 
\end{equation}
hence Green function associated with $\chi_{zz}^{'}$ take the form as
\begin{equation}\label{ae11}
\langle\langle\varsigma_{z}(\boldsymbol{q},t);\varsigma_{z}(-\boldsymbol{q})\rangle\rangle=\frac{1}{2}\{G_{\uparrow}(\boldsymbol{q}, t)-G_{\downarrow}(\boldsymbol{q}, t)\}  
\end{equation}
with
\begin{equation}\label{ae12}
 G_{\sigma}(\boldsymbol{q}, t)= \sum_{k}\langle\langle\vartheta_{k \sigma}(\boldsymbol{q}, t);\varsigma_{z}(-\boldsymbol{q})\rangle\rangle  
\end{equation}
In RPA we put
\begin{equation}
\left[\vartheta_{k\sigma}(\boldsymbol{q}), \mathcal{H}_{c}\right]=U_{eff}(\eta_{\boldsymbol{k}+\boldsymbol{q},\sigma}-\eta_{\boldsymbol{k},\sigma})\sum_{l} \vartheta_{l,-\sigma}(\boldsymbol{q})    
\end{equation}
here $-\sigma$ corresponds to anti-parallel spin state to $\sigma$. Further the EOM is obtained as
\begin{equation}
\begin{aligned}
i \hbar \frac{d}{d t}\left\langle\langle\vartheta_{k \sigma}(\boldsymbol{q}, t) ;\varsigma_{z}(-\boldsymbol{q})\right\rangle\rangle=-\frac{1}{2} \delta(t)\left(f_{k+q, \sigma}-f_{k, \sigma}\right) \eta_{\sigma}\\
+\{\varepsilon(\boldsymbol{k})-\varepsilon(\boldsymbol{k}+\boldsymbol{q})\}\left\langle\langle\vartheta_{\boldsymbol{k}, \sigma}(\boldsymbol{q}, t) ;\varsigma_{z}(-\boldsymbol{q})\right\rangle\rangle+U_{eff}\left(f_{\boldsymbol{k}+\boldsymbol{q}, \sigma}-f_{\boldsymbol{k}, \sigma}\right) \sum_{\boldsymbol{l}}\left\langle\langle\vartheta_{\boldsymbol{l},-\sigma}(\boldsymbol{q}, t) ;\varsigma_{z}(-\boldsymbol{q})\right\rangle\rangle
\end{aligned}
\end{equation}
by doing Fourier transform of above equation we get
\begin{equation}
\begin{array}{l}
\{-\hbar \omega+\varepsilon(\boldsymbol{k}+\boldsymbol{q})-\varepsilon(\boldsymbol{k})\}\langle\langle\vartheta_{k \sigma}(\boldsymbol{q}, \omega) ; {\varsigma_{z}(-\boldsymbol{q})\rangle\rangle}\\\\
\quad=-\frac{1}{2}\left(f_{k+\boldsymbol{q}, \sigma}-f_{\boldsymbol{k}, \sigma}\right) \eta_{\sigma}+U_{eff}\left(f_{\boldsymbol{k}+\boldsymbol{q}, \sigma}-f_{\boldsymbol{k}, \sigma}\right) \sum_{\boldsymbol{l}}\left\langle\langle\vartheta_{\ell,-\sigma}(\boldsymbol{q}, \omega) ; \varsigma_{z}(-\boldsymbol{q})\right\rangle\rangle
\end{array}    
\end{equation}
\\
this is a coupled equation for up spin-and down-spin electrons. Dividing both sides by the factor appearing in L.H.S of this equation and summing over $k^{,}s$, one easily arrive at
\begin{equation}\label{ae101}
\begin{array}{l} 
\sum_{\boldsymbol{k}}\left\langle\langle\vartheta_{\boldsymbol{k}}(\boldsymbol{q}, \omega) ; \varsigma(-\boldsymbol{q})\right\rangle\rangle=\frac{1}{2}\sum_{k} \frac{f_{k \uparrow}-f_{k+q \uparrow}}{{\varepsilon}_{\downarrow}(\boldsymbol{k}+\boldsymbol{q})-{\varepsilon}_{\uparrow}(\boldsymbol{k})-\hbar\omega}\\\\
+U_{eff}\sum_{k} \frac{f_{k \uparrow}-f_{k+q \uparrow}}{{\varepsilon}_{\downarrow}(\boldsymbol{k}+\boldsymbol{q})-{\varepsilon}_{\uparrow}(\boldsymbol{k})-\hbar \omega}\times\sum_{\boldsymbol{l}}\left\langle\langle\vartheta_{\ell,-\sigma}(\boldsymbol{q}, \omega) ; \varsigma_{z}(-\boldsymbol{q})\right\rangle\rangle
\end{array} 
\end{equation}
\\
this implies
\begin{equation}
 G_{\sigma}(\boldsymbol{q}, \omega)=\Gamma_{\sigma}(q,\omega)\{\frac{1}{2}\eta_{\sigma}-U_{eff} G_{-\sigma}(q,\omega)\}   
\end{equation}
where $G_{\sigma}(q,\omega)$ is the Fourier transform of $G_{\sigma}(q,t)$ given in eq.(\ref{ae12})
%now we use already defined parameters $\Gamma_{\sigma}(q,\omega)$ and $G_{\sigma}(q,\omega)$ in eq.(\ref{}) and eq.(\ref{}),respectively and above Eqbecomes
%\begin{equation}\label{ae10}
%\chi_{-+}(\boldsymbol{q}, \omega)=\sum_{\boldsymbol{k}}\left\langle\vartheta_{\boldsymbol{k}}(\boldsymbol{q}, \omega) ; \varsigma(-\boldsymbol{q})\right\rangle=\frac{\Gamma_{-+}(\boldsymbol{q}, \boldsymbol{\omega})}{1-U_{eff} \Gamma_{-+}(\boldsymbol{q}, \omega)},    
%\end{equation}
%Adopting the same approximation $\chi_{zz}(\boldsymbol{q}, \omega)$ can be obtained with spontaneous magnetization in z-direction.
and 
\begin{equation}\label{gamma}
\Gamma_{\sigma}(\boldsymbol{q}, \omega)=\sum_{k} \frac{f_{k \sigma}-f_{k+q \sigma}}{\varepsilon(\boldsymbol{k}+\boldsymbol{q})-\varepsilon(\boldsymbol{k})-\hbar \omega}.
\end{equation}
Therefore $G_{\sigma}(q,\omega)$ is written in terms of $\Gamma_{\sigma}(q,\omega)$ as
\begin{equation}
G_{\sigma}(\boldsymbol{q}, \omega)=\frac{1}{2}\eta_{\sigma}\frac{\Gamma_{\sigma}(\boldsymbol{q}, \omega)\{1+U_{eff} \Gamma_{-\sigma}(q,\omega)\}}{1-U_{eff}^{2} \Gamma_{-\sigma}(q,\omega)\Gamma_{\sigma}(q,\omega)}.
\end{equation}
%
%From (4.22) we obtain
$$
%\mathrm{G}_{\sigma}(\boldsymbol{q}, \omega)=\frac{1}{2} \eta_{\sigma} \frac{\Gamma_{\sigma}(\boldsymbol{q}, \omega)\left\{1+U_{eff}\Gamma_{-\sigma}(\boldsymbol{q}, \omega)\right\}}{1-U_{eff}^{2} \Gamma_{-\sigma}(\boldsymbol{q}, \omega) \Gamma_{\sigma}(\boldsymbol{q}, \omega)}
$$
Now, from eq.(\ref{ae11}) and eq. (\ref{ae3}) $\chi_{z z}(\boldsymbol{q}, \omega)$ can be obtained as
\begin{equation}\label{chizz}
\chi_{z z}(\boldsymbol{q}, \omega)=\frac{1}{4} \frac{\Gamma_{\mathrm{\downarrow}}(\boldsymbol{q}, \boldsymbol{\omega})+\Gamma_{\uparrow}(\boldsymbol{q}, \boldsymbol{\omega})+2 U_{eff} \Gamma_{\downarrow}(\boldsymbol{q}, \omega) \Gamma_{\mathrm{\uparrow}}(\boldsymbol{q}, \boldsymbol{\omega})}{1-U_{eff}^{2} \Gamma_{\mathrm{\uparrow}}(\boldsymbol{q}, \omega) \Gamma_{\mathrm{\downarrow}}(\boldsymbol{q}, \omega)}
\end{equation}

\subsection{Deduction of Stoner susceptibility from RPA susceptivility}\label{RPAS}

In this section we will derive the Stoner susceptibility from Eq.\eqref{chizz} in the limit: $q \rightarrow 0$ and $\omega = 0$. We have from Eq. \eqref{gamma}:
\begin{equation}
    \Gamma_{\sigma}(\boldsymbol{q}, \omega)=\sum_{k} \frac{f_{k \sigma}-f_{k+q \sigma}}{\mathcal{E}(k + q)-\mathcal{E}(k)-\hbar \omega}.
\end{equation}
In the limit $q \rightarrow 0$ and putting $\omega = 0$, we can do a Taylor expansion to get:
\begin{equation}
    f_{k + q \sigma} = f_{k \sigma} + q \frac{\partial f_{k \sigma}}{\partial k} 
\end{equation}
and 
\begin{equation}
    \mathcal{E}(k + q) = \mathcal{E}(k) + q \frac{\partial \mathcal{E}(k)}{\partial k} 
\end{equation}
substituting the above in Eq. \eqref{gamma} we get:
\begin{equation}
    \Gamma_{\sigma}(\boldsymbol{q}, 0)=\sum_{k} \frac{\partial f_{k \sigma}/\partial k}{\partial \mathcal{E}(k)/\partial k} = \rho_{\sigma}(\mathcal{E}_{F})
\end{equation}
Substituting the above in the Eqfor RPA susceptibility we get:
\begin{equation}\label{c1}
    \chi_{zz}(\boldsymbol{q}, 0) = \frac{1}{4} \frac{\rho_{\downarrow}(\mathcal{E}_{F})+\rho_{\uparrow}(\mathcal{E}_{F})+2 U_{eff}\rho_{\downarrow}(\mathcal{E}_{F})\rho_{\uparrow}(\mathcal{E}_{F})}{1 - U_{eff}^{2}\rho_{\downarrow}(\mathcal{E}_{F})\rho_{\uparrow}(\mathcal{E}_{F})}
\end{equation}
Near the Ferromagnetic instability we have:
\begin{equation}
    \rho_{\downarrow}(\mathcal{E}_{F}) \sim \rho_{\uparrow}(\mathcal{E}_{F})
\end{equation}
Substituting the above in \eqref{c1}, we get
\begin{equation}
     \chi_{zz}(\boldsymbol{q}, 0) = \frac{1}{4} \frac{2 \rho(\mathcal{E}_{F})(1 + U_{eff} \rho(\mathcal{E}_{F}))}{ 1 - U_{eff}^{2} \rho(\mathcal{E}_{F})^{2}}  = \frac{\frac{\rho(\mathcal{E}_{F})}{2}}{1 - U_{eff}\rho(\mathcal{E}_{F})} = \frac{\chi_{0}}{1 - U_{eff}\rho(\mathcal{E}_{F})}   = \chi_{stoner}
\end{equation}
which is the same as Eq. \eqref{Schi}.
\section{Self Consistent Renormalization (SCR) Theory}

\begin{itemize}
\item \textit{Stoner Theory} is based on the \textit{Hartree-Fock} ``Mean Field'' approximation (Only the exchange term is taken into account and the Correlation terms are neglected).
\item Random Phase Approximation (RPA) goes ahead of the \textit{Stoner Theory} by including the correlation effects. It gives an aid to study the dynamical effects like stoner-excitation, spin-wave etc. using the dynamical magnetic susceptibility $\chi{(q,\omega)}$.
\item There is an additional free energy due to the correlation effects which is not taken into account in the RPA theory. This extra free energy will be expressed in terms of the transverse  dynamical susceptibility using the fluctuation dissipation theorem.  
\item The SCR theory takes into account this neglected free energy and in doing so it goes beyond the RPA theory.
\item Thus the SCR theory leads to a serious modification of the results obtained through RPA theory.
\end{itemize}
\begin{itemize}
 \item[1.] The calculated value of $T_{c}$ (using SCR Theory) agrees well with the experiments for weakly ferromagnetic and anti ferromagnetic systems.
 \item[2.] It reproduces the \textit{Curie-Weiss} (CW) law.
\end{itemize}
\par
Various authors have applied the Stoner theory \cite{stoner1938collective} to the study of magnetism in d-band materials without much success. It was not able to explain the \textit{Curie-Weiss} behaviour observed in these materials. On the other hand, the experiments clearly indicate the existence of a new mechanism for the CW law. Therefore, one has to go beyond the HFA and RPA theory by taking in to account the influence of exchange-enhanced spin fluctuations on the thermodynamical quantities. Such a renormalization effect was studied by various authors  \cite{lonzarich1985effect,moriya1965ferro,moriya1973effect,moriya2006developments,toru,moriya2012} for nearly ferromagnetic (paramagnetic) metals. Their study includes the enhancements of low-T specific heat and the T-dependence of magnetic susceptibility due to strongly exchange-enhanced long-wavelength spin fluctuations. The RPA was used for the
spin fluctuations in calculating the renormalized free energy and the results were successfully
applied in the low-T limit. As T- increases, the RPA becomes less applicable, and one has to
calculate the renormalized thermal equilibrium state and the spin fluctuations at the same time
in a self-consistent manner \cite{toru}. Furthermore, this self-consistency has been proved to be essential for the
theory of ferromagnetic metals at any T. In other words, one has to deal with the mutually
coupled modes of spin fluctuations self- consistently. Along this line, a coupling theory of
spin fluctuations in weakly ferromagnetic metals was developed later\cite{toru,moriya2012}.  This theory provided a new mechanism for the
CW susceptibility, which explains not only the disagreement between the effective moment
deduced from the Curie constant and the spontaneous moment, but also the CW susceptibility.
The results of the SCR theory were derived later by various different methods and the theory
was extended to cover antiferromagnetic metals too. Comparison between the theory and
experimental results confirm that the SCR theory is quantitatively correct. 
%\end\

\subsection{A General Formula for the Magnetic Susceptibilty}
The starting single band Hamiltonian is expressed in terms of creation and annihilation operators (\textit{Hubbard Hamiltonian}) as
\begin{equation}
    \mathcal{H} = \mathcal{H}_0 +\mathcal{H}'(\mathcal{I})
\end{equation}
\begin{equation}
    \mc{H}_{0} = \sum_{\mf{k}, \sigma} \ep_{\mf{k}} \ad_{\mf{k}\sigma} \as_{\mf{k} \sigma}
\end{equation}
\begin{equation}\label{8.3}
\mc{H}'(I) = I \sum_{\mf{k,k',q}} \ad_{\mf{k+q}\ua}\ad_{\mf{k'-q}\da}\as_{\mf{k'}\da}\as_{\mf{k}\ua}
\end{equation}
\begin{equation}\label{8.4}
\mc{H}'(I) = \frac{1}{2} \mc{U} \no - \frac{1}{2} I \sum_{\mf{q}} \left\{ \mc{S}_+(\mf{q}) \mc{S}_-(\mf{-q}) \right\}
\end{equation}
where $\mc{N}$ is the total number of electrons, $\mc{U} = \mc{N}_{0} I$ is the intra-atomic exchange energy, $\mc{N}_{0}$ is the number of atoms in the crystal and the $\{\,\,\}$ anti-commutator. Now starting from Eq. \eqref{8.4} and writing spin raising $\mc{S}_+(\mf{q})$ and lowering $\mc{S}_-(\mf{q})$ operator in terms of creation and annihilation operators one can reach at Eq. \eqref{8.3} in the following way. Define
\begin{equation*}
%\begin{split}
\mc{S}_+(\mf{q}) = \sum_{\mf{k}} \ad_{\mf{k+q}\ua}\as_{\mf{k}\da} \,\,;\, 
\mc{S}_-(\mf{q}) = \sum_{\mf{k}} \ad_{\mf{k+q}\da}\as_{\mf{k}\ua}
%\end{split}
\end{equation*}
Proof of the equivalence of Eq. \eqref{8.3} and Eq. \eqref{8.4} is derived as: we begin with the Eq. \eqref{8.4} as shown below 

\begin{equation}
\begin{split}
\mc{H}'(I) = \frac{1}{2} \mc{U} \sum_{\mf{k\sigma}} \ad_{\mf{k\sigma}\ua}\as_{\mf{k\sigma}\da} - \frac{1}{2} I \sum_{\mf{q}} \{ \sum_{\mf{k}} \ad_{\mf{k+q}\ua}\as_{\mf{k}\da} \sum_{\mf{k'}} \ad_{\mf{k'-q}\da}\as_{\mf{k'}\ua} \\ 
 +  \sum_{\mf{k'}} \ad_{\mf{k'-q}\da}\as_{\mf{k'}\ua} \sum_{\mf{k}} \ad_{\mf{k+q}\ua}\as_{\mf{k}\da} \}
\end{split}
\end{equation}

\begin{eqnarray}\nonumber
\mc{H}'(I) = \frac{1}{2} \mc{U} \sum_{\mf{k\sigma}} \ad_{\mf{k\sigma}\ua}\ad_{\mf{k\sigma}\da} - \frac{1}{2} I \sum_{\mf{q k k'}} \{ \ad_{\mf{k+q}\ua}\underbrace{\as_{\mf{k}\da}  \ad_{\mf{k'-q}\da}} \as_{\mf{k'}\ua} \\  + \ad_{\mf{k'-q}\da} \underbrace{\as_{\mf{k'}\ua} \ad_{\mf{k+q}\ua }} \as_{\mf{k}\da} 
\end{eqnarray}

\begin{eqnarray}\nonumber
\mc{H}'(I) = \frac{1}{2} \mc{U} \sum_{\mf{k\sigma}} \ad_{\mf{k\sigma}\ua}\ad_{\mf{k\sigma}\da} - \frac{1}{2} I \sum_{\mf{q k k'}} \{ \ad_{\mf{k+q}\ua}\underbrace{(\delta_{k,k'-q} - \ad_{\mf{k'-q}\da} \as_{\mf{k}\da}  )} \as_{\mf{k'}\ua} \\  + \ad_{\mf{k'-q}\da} \underbrace{ ( \delta_{k',k+q} - \ad_{\mf{k+q}\ua} \as_{\mf{k'}\ua}  )} \as_{\mf{k}\da} \}
\end{eqnarray}

\begin{eqnarray}\nonumber
\mc{H}'(I) = \frac{1}{2} \mc{U} \sum_{\mf{k\sigma}} \ad_{\mf{k\sigma}\ua}\ad_{\mf{k\sigma}\da}  - \frac{1}{2} I \sum_{\mf{q k k'}} \{ \ad_{\mf{k+q}\ua}\delta_{k,k'-q} \as_{\mf{k'}\ua} - \ad_{\mf{k+q}\ua}\ad_{\mf{k'-q}\da} \as_{\mf{k}\da} \as_{\mf{k'}\ua} \\  + \ad_{\mf{k'-q}\da} \delta_{k',k+q} \as_{\mf{k}\da} - \ad_{\mf{k'-q}\da} \ad_{\mf{k+q}\ua} \as_{\mf{k'}\ua} \as_{\mf{k}\da} )  \}
\end{eqnarray}

\begin{eqnarray}\nonumber
\mc{H}'(I) = \frac{1}{2} \mc{U} \sum_{\mf{k\sigma}} \ad_{\mf{k\sigma}\ua}\ad_{\mf{k\sigma}\da}  - \frac{1}{2} I \sum_{\mf{q k k'}} \{ \ad_{\mf{k+q}\ua} \as_{\mf{k+q}\ua} - \ad_{\mf{k+q}\ua}\ad_{\mf{k'-q}\da} \as_{\mf{k}\da} \as_{\mf{k'}\ua} \\  + \ad_{\mf{k'-q}\da}  \as_{\mf{k'-q}\da} - \ad_{\mf{k'-q}\da} \ad_{\mf{k+q}\ua} \as_{\mf{k'}\ua} \as_{\mf{k}\da} )  \}
\end{eqnarray}

\begin{eqnarray}\nonumber
\mc{H}'(I) = \frac{1}{2} \mc{U} \sum_{\mf{k\sigma}} \ad_{\mf{k\sigma}\ua}\ad_{\mf{k\sigma}\da}  - \frac{1}{2} I \sum_{\mf{q k k'}} \{ \mc{N}_{0} \no_{\ua} - \ad_{\mf{k+q}\ua}\ad_{\mf{k'-q}\da} \as_{\mf{k}\da} \as_{\mf{k'}\ua} \\  + \mc{N}_{0} \no_{\da} - \ad_{\mf{k'-q}\da} \ad_{\mf{k+q}\ua} \as_{\mf{k'}\ua} \as_{\mf{k}\da} )  \}
\end{eqnarray}
rearranging terms in above equation we get
\begin{eqnarray}\nonumber
\mc{H}'(I) = \frac{1}{2} \mc{U} \sum_{\mf{k\sigma}} \ad_{\mf{k\sigma}\ua}\ad_{\mf{k\sigma}\da}  - \frac{1}{2} I \sum_{\mf{q k k'}} \{ \mc{N}_{0} (\no_{\ua} + \no_{\da} ) - \ad_{\mf{k+q}\ua}\ad_{\mf{k'-q}\da} \as_{\mf{k}\da} \as_{\mf{k'}\ua} \\   - \ad_{\mf{k'-q}\da} \ad_{\mf{k+q}\ua} \as_{\mf{k'}\ua} \as_{\mf{k}\da} )  \}
\end{eqnarray}
\begin{eqnarray}\nonumber
\mc{H}'(I) = \frac{1}{2} \mc{U} \sum_{\mf{k\sigma}} \ad_{\mf{k\sigma}\ua}\ad_{\mf{k\sigma}\da}  - \frac{1}{2} I \sum_{\mf{q k k'}} \{ \mc{N}_{0} \no  - \ad_{\mf{k+q}\ua}\ad_{\mf{k'-q}\da} \as_{\mf{k}\da} \as_{\mf{k'}\ua} \\ \ - \ad_{\mf{k'-q}\da} \ad_{\mf{k+q}\ua} \as_{\mf{k'}\ua} \as_{\mf{k}\da} )  \}
\end{eqnarray}

\begin{eqnarray}\nonumber
\mc{H}'(I) = \frac{1}{2} \mc{U} \underbrace {\sum_{\mf{k\sigma}} \ad_{\mf{k\sigma}\ua}\ad_{\mf{k\sigma}\da}}  - \frac{1}{2} I \mc{N}_{0} \no - \frac{1}{2} I \sum_{\mf{q k k'}} \{ - \ad_{\mf{k+q}\ua}\ad_{\mf{k'-q}\da} \as_{\mf{k}\da} \as_{\mf{k'}\ua} \\   - \ad_{\mf{k'-q}\da} \ad_{\mf{k+q}\ua} \as_{\mf{k'}\ua} \as_{\mf{k}\da} )  \}
\end{eqnarray}

\begin{eqnarray}\nonumber
\mc{H}'(I) = \underbrace {\frac{1}{2} \mc{U}  \no }  - \frac{1}{2} I \mc{N}_{0} \no - \frac{1}{2} I \sum_{\mf{q k k'}} \{ - \ad_{\mf{k+q}\ua}\ad_{\mf{k'-q}\da} \as_{\mf{k}\da} \as_{\mf{k'}\ua} \\   - \ad_{\mf{k'-q}\da} \ad_{\mf{k+q}\ua} \as_{\mf{k'}\ua} \as_{\mf{k}\da} )  \}
\end{eqnarray}
using definition of $\mc{U}= \mc{N}_{0} I$, we get

\begin{eqnarray}
\mc{H}'(I) =\frac{1}{2} I \sum_{\mf{q k k'}} \left\{ + \ad_{\mf{k+q}\ua}\ad_{\mf{k'-q}\da} \as_{\mf{k}\da} \as_{\mf{k'}\ua} + \ad_{\mf{k'-q}\da} \ad_{\mf{k+q}\ua} \as_{\mf{k'}\ua} \as_{\mf{k}\da} )  \right\}
\end{eqnarray}
interchanging $k\leftrightarrow k'$ wherever required we reach to  Eq. \eqref{8.3} as \\
\begin{equation}\label{8.17}
\underbrace{\mc{H}'(I) = I \sum_{\mf{k,k',q}} \ad_{\mf{k+q}\ua}\ad_{\mf{k'-q}\da}\as_{\mf{k'}\da}\as_{\mf{k}\ua}}
\end{equation}
%hence proved.
%\par
Next, the magnetic susceptibility in the unit of $\mu_{B}^2$ is given as
\begin{equation}\label{8.18}
\chi=\left[\left(\frac{\delta^2F(M)}{\delta M^2}\right)^{-1}\right]_{M=M^*}
\end{equation}
here, $M$ is the magnetization, $M^*$ is its saturation value and $F(M)$ is the total free energy as a function of $M$.
\begin{equation*}
M=N_{\downarrow}-N_{\uparrow}\,\,\,;\,\,\,
N=N_{\downarrow}+N_{\uparrow}    
\end{equation*}
The partition function of the system in the presence of magnetic field is given as
\begin{equation}
Z(H)=Tr \left[e^{\frac{-[\mc{H}+ H M_{z}]}{K_{B}T}}\right]
\end{equation}
here $H$ is the magnetic field aligned along the $z$-axis, $\mc{H}$ is the Hamiltonian of the system and $M_{z}$ is the component of magnetization along $H$. Therefore, the free energy of the system is given as
\begin{equation}
F(H)=-T ln Z(H)    
\end{equation}
Further, the free energy can be expressed in terms of $M$ by using the Laplace transformation in the following way
\begin{equation}\label{8.19}
Z(M)=\frac{1}{2\pi i}\int_{-i\infty-\epsilon}^{i\infty+\epsilon} d\left(\frac{H}{T}\right) e^{\left(-\frac{H}{T}\right)M}Z(H)   
\end{equation}
define $F(M)=-T ln Z(M)$ is free energy for a given value of $M$ and we have
\begin{equation}
\frac{F(M)}{T}=ln Z(M)    
\end{equation}
\begin{equation}\label{8.118}
Z(M)=e^{-\frac{F(M)}{T}}   
\end{equation}
from Eq. \eqref{8.19} and Eq. \eqref{8.118}  we get

\begin{equation}
e^{-\frac{F(M)}{T}}=\frac{1}{2\pi i T}\int_{-i\infty-\epsilon}^{i\infty+\epsilon} d(H) e^{\left(-\frac{1}{T}\right) \left[(HM+F(H)\right]} 
\end{equation}
Therefore, using the saddle point approximation we get
\begin{equation}\label{8.24}
M=M^{*}=-\frac{\delta F(H)}{\delta H}    
\end{equation}
$M^{*}$ is the saturation value of $M$. From Eq. \eqref{8.24} (saddle point integral) we get
\begin{equation}
F(M^{*})=F(H)+HM^{*}    
\end{equation}
In physical sense, the free energy $F(M)$ can be estimated by calculating the free energy under an external magnetic field $H$ which gives rise to the magnetization $M$ and then subtracting the energy due to external field i.e. $-HM^{*}$. Further we express $F(M)$ as follows
\begin{equation}
  F(M) = F_{0}(M) + \int_{0}^{I}  \frac{dI}{I} \left< \mc{H}' (I)\right>_{M,I}.
\end{equation}
%A.sharma end
Here, the total free energy of the magnetic system in terms of magnetization can be expressed as the sum of free energy terms due to free electrons, free energy contributed by the HFA (Mean Field Contribution) and the term solely contributed by the correlation effects.
The HFA term is of linear order while the correlation term is found to be of second order in nature.\\\\
\begin{eqnarray}
 F(M) = F_{0}(M) +\underbrace{F_{\textit{HF}}^{I}(M) + \varDelta F^{I}(M)}
\end{eqnarray}
\begin{center}
~~~~~~~~~~~~~~~~~~~~$F^{I}(M)$
\end{center}
%\begin{eqnarray}
% {F}({M}) = {F}_{0}({M}) + %{F}_{\textit{HF}}^{{I}}({M}) + \varDelta %{F}^{{I}}({M}).
%\end{eqnarray}
We can see in Eq. \eqref{8.18} that the second order differentiation of $F(M)$ w.r.t. magnetization gives us $\chi^{-1}$. The term ($F_{\textit{HF}}^{{I}}(M) + \varDelta F^{{I}}(M)$) is the interaction term and is defined as
%\begin{eqnarray}
% F(M) = F_{0}(M) %+\underbrace{F_{\textit{HF}}^{I}(M) + %\varDelta F^{I}(M)}
%\end{eqnarray}
%\begin{center}
%~~~~~~~~~~~~~~~~~~~~$F^{I}(M)$
%\end{center}
\begin{eqnarray} \nonumber
\begin{split}
F^{{I}}(M) & = \int_{0}^{{I}} d {I}' \frac{1}{{I}'}\left< \mc{H}' ({I}')\right>  \\
\mc{H}'({I}) & = {I} \sum_{\mf{k,k',q}} \ad_{\mf{k+q}\ua}\ad_{\mf{k'-q}\da}\as_{\mf{k'}\da}\as_{\mf{k}\ua} \\
\mc{H}'({I})& = {I} \, \mc{H}_{\text{Int.}}
\end{split}
\end{eqnarray}
The Hamiltonian corresponding to the magnetic system is given as
\begin{eqnarray}
\mc{H} = \mc{H}_{0}+\mc{H}'(I)- H M 
\end{eqnarray}
%where $H$ is the magnetic field.
Assuming that the eigensystem of the total Hamiltonian $\mc{H}$ has the eigen energies $\mf{E}_{n}(I,M)$ and eigen state $\left| \phi(I,M) \right> $, so one can write $\mf{E}_{n}(I,M) = \left. \left< \mc{H} \right>  \right| _{I,M} $ at constant $ I $ and $ M $. Therefore
\begin{eqnarray}
< \mc{H} >  = \sum_{n} e^{-\beta \mf{E}_{n}(I,M)} <\phi_{m}(I,M)| \mc{H} |\phi_{n}(I,M) >
\end{eqnarray}
here $\left< \mc{H} \right>$ is the ensemble average. Putting $\mc{H} = \mc{H}_{0} - H M + \mc{H}'(I)$ and $\mc{H}'(I) = I \,\mc{H}_{\text{Int}}$ in above equation we get
\begin{eqnarray}
\begin{split}
 \frac{\partial \mf{E}_{n}(I,M)}{\partial I}&= \sum_{n} e^{-\beta \mf{E}_{n}(I,M)} \left< \phi_{m}(I,M) \left| \mc{H}_{\text{Int.}} \right| \phi_{n}(I,M) \right> \\ 
 & = \left< \mc{H}_{\text{Int.}}\right>_{I,M}
\end{split}
\end{eqnarray}
\begin{eqnarray}
\begin{split}
 \mf{E}_{n}(I,M) -  \mf{E}_{n}(0,M) & = \int_{0}^{I} \frac {d I'} {I'} \left< I' \mc{H}_{\text{Int.}}\right>_{I,M} \\
F^{I}(M) & = \int_{0}^{I} \frac {d I'} {I'} \left<  \mc{H}^{'}(I')\right>_{I,M} 
\end{split}
\end{eqnarray}
by doing this we include the contribution of free energy coming from the electronic interaction. Therefore
\begin{eqnarray}
\begin{split}
 F(M) & = F_{0}(M) + F^{I}(M) \\
 & = F_{0}(M) + \underbrace{ \int_{0}^{I} \frac {d I'} {I'}\left<  \mc{I'}  \sum_{\mf{k,k',q}} \ad_{\mf{k+q}\ua}\ad_{\mf{k'-q}\da}\as_{\mf{k'}\da}\as_{\mf{k}\ua}  \right>_{I,M} }
\end{split}
\end{eqnarray}
and
\begin{equation}
\mc{H}'(I) = I \sum_{\mf{k,k',q}} \ad_{\mf{k+q}\ua}\ad_{\mf{k'-q}\da}\as_{\mf{k'}\da}\as_{\mf{k}\ua} = \frac{1}{2} \mc{U} \no - \frac{1}{2} I \sum_{\mf{q}} \left\{ \mc{S}_+(\mf{q}) \mc{S}_-(\mf{-q}) \right\}
\end{equation}
thus free energy is obtained as
\begin{eqnarray} \label{8.34}
\begin{split}
 F(M) & = F_{0}(M) + \underbrace{ \int_{0}^{I} \frac {d I'} {I'}\left<  \frac{1}{2} \mc{U} \no - \frac{1}{2} I' \sum_{\mf{q}} \left\{ \mc{S}_+(\mf{q}) \mc{S}_-(\mf{-q}) \right\} \right>_{I,M} } \\ 
 & = F_{0}(M) + \frac{1}{2} \mc{U} \mc{N} - \frac{1}{2}   \underbrace{ \int_{0}^{I}  {d I'} \left<  \sum_{\mf{q}} \left\{ \mc{S}_+(\mf{q}) \mc{S}_-(\mf{-q}) \right\} \right>_{I,M} }\\
& = F_{0}(M) + \frac{1}{2} \mc{U} \mc{N} -  \frac{1}{2}  I \left< \left\{ \mc{S}_+(\mf{q}) \mc{S}_-(\mf{-q}) \right\} \right>_{0,M} \\
&- \frac{1}{2}  I \sum_{\mf{q}}  \left(  {\underbrace{ \left< \left\{ \mc{S}_+(\mf{q}) \mc{S}_-(\mf{-q}) \right\} \right>_{I,M} - \left< \left\{ \mc{S}_+(\mf{q}) \mc{S}_-(\mf{-q}) \right\} \right>_{0,M} }} \right) 
\end{split}
\end{eqnarray}
%We know that $\mc{H}'(I) = I \sum_{\mf{k,k',q}} \ad_{\mf{k+q}\ua}\ad_{\mf{k'-q}\da}\as_{\mf{k'}\da}\as_{\mf{k}\ua} = \frac{1}{2} \mc{U} \no - \frac{1}{2} I \sum_{\mf{q}} \left\{ \mc{S}_+(\mf{q}) \mc{S}_-(\mf{-q}) \right\} $
%With this let us evaluate the third term,
Next, the last term in above equation is evaluated as
\begin{equation}
\begin{aligned}
\sum_{q} \left\langle \sum_{k} a^{\dagger}_{k+q\uparrow}a_{k\downarrow}\sum_{k'}a^{\dagger}_{k'-q\downarrow}a_{k'\uparrow} + \sum_{k'}a^{\dagger}_{k'-q\downarrow}a_{k'\uparrow} \sum_{k} a^{\dagger}_{k+q\uparrow}a_{k\downarrow} \right\rangle \\
= \sum_{kk'q} \left\langle  a^{\dagger}_{k+q\uparrow}a_{k\downarrow}a^{\dagger}_{k'-q\downarrow}a_{k'\uparrow} + a^{\dagger}_{k'-q\downarrow}a_{k'\uparrow}  a^{\dagger}_{k+q\uparrow}a_{k\downarrow} \right\rangle
\end{aligned}
\end{equation}
using the commutation relations for the Fermionic operators, we get
\begin{equation}
    \begin{aligned}
    \sum_{kk'q} \langle a^{k+q\uparrow} \left(\delta_{k,k'-q} - a^{\dagger}_{k'-q\downarrow}a_{k\uparrow}\right)a_{k'\uparrow} +  a^{\dagger}_{k'- q\downarrow} \left(\delta_{k',k+q} - a^{\dagger}_{k+q\uparrow}a_{k\downarrow}\right)a_{k\downarrow)\rangle}
    \end{aligned}
\end{equation}
simplifying the first and third term in the form of number densities we get,
\begin{equation}
    \begin{aligned}
    N_{0}N_{\uparrow} + N_{0}N_{\downarrow} = N_{0}N
    \end{aligned}
\end{equation}
while, the second and fourth term becomes
\begin{equation}
    \begin{aligned}
    - \sum_{kk'q} \left \langle a^{\dagger}_{k+q\uparrow} a^{\dagger}_{k'-q\downarrow} a_{k\downarrow}a_{k'\uparrow} +
    a^{\dagger}_{k'-q\downarrow}
    a^{\dagger}_{k+q\uparrow}  a_{k'\uparrow}a_{k\downarrow}\right\rangle
    \end{aligned}
\end{equation}
put $k'-q =k$ in above equation, we get the equation simplified in terms of number operators as
\begin{equation}
    \begin{aligned}
    -\sum_{qk} \left( n_{k+q\uparrow} n_{k\downarrow} + n_{k+q\uparrow} n_{k\downarrow} \right)
    \end{aligned}
\end{equation}
further, put $q=-k+q'$ in above equation and get
\begin{equation}
    -\sum_{kq'} \left(n_{q'}n_{k} + n_{q'}n_{k}\right)  = \left( N_{\uparrow} N_{\downarrow} +  N_{\uparrow} N_{\downarrow} \right)
\end{equation}
To get back our original second and fourth term we multiply the above equation by $-\frac{I}{2}$ to get
\begin{equation}
    \begin{aligned}
    \frac{1}{2}  I \left< \left\{ \mc{S}_+(\mf{q}) \mc{S}_-(\mf{-q}) \right\} \right>_{0,M} = -\frac{1}{2}I N_{0}N + IN_\uparrow N_{\downarrow}\\
    = -\frac{1}{2}I N_{0}N + \frac{1}{4} \left(M^{2} -N^{2}\right)
    \end{aligned}
\end{equation}
This equation is nothing but the term that we had obtained in the Stoner model corresponding to the Hartree-Fock term. Therefore the other terms except this are the terms beyond the Hartree-Fock approximation. Now, we express these terms in the form of transverse dynamical susceptibility \cite{izuyama1963band}. Consider
$\left< \left\{ \mc{S}_+(\mf{q}) \mc{S}_-(\mf{-q}) \right\} \right>_{I,M}$ is ensemble average of $\mc{S}_+(\mf{q}) \mc{S}_-(\mf{-q})$ and is given as
\begin{eqnarray}
\begin{split}
\left< \left\{ \mc{S}_+(\mf{q}) \mc{S}_-(\mf{-q}) \right\} \right>_{I,M} = \frac{1}{\mc{Z}} \sum_{m} e^{-\beta \mf{E}_{m}(I,M)}\\
%\end{eqnarray*}
%\begin{eqnarray}
\times\left< \phi_{m}(I,M) \left| \left\{  \mc{S}_+(\mf{q}) \mc{S}_-(\mf{-q}) \right\} \right| \phi_{m}(I,M)\right>_{I,M}
\end{split}
\end{eqnarray}
For simplicity take $\left| \phi_{m}\right> = \left| \mf{m} \right>  $ in above equation and we get 
\begin{eqnarray}
\begin{split}
\left< \left\{ \mc{S}_+(\mf{q}) \mc{S}_-(\mf{-q}) \right\} \right>_{I,M} = \frac{1}{\mc{Z}} \sum_{m} e^{-\beta \mf{E}_{m}(I,M)} \\
\times\left( \left< \mf{m} \left|   \mc{S}_+(\mf{q}) \mc{S}_-(\mf{-q}) \right| \mf{m} \right>_{I,M} + \left< \mf{m} \left|   \mc{S}_-(\mf{-q}) \mc{S}_+(\mf{q}) \right| \mf{m} \right>_{I,M} \right)
\end{split}
\end{eqnarray}
\begin{equation}\label{8.44}
    \begin{aligned}
    = \frac{1}{\mc{Z}} \sum_{m,n} e^{-\beta E_{m}} \left( \left\langle m|S_{+}(q)|n \right\rangle \left\langle n|S_{-}(-q)|m \right\rangle +  \left( \left\langle m|S_{-}(-q)|n \right\rangle \left\langle n|S_{+}(q)|m \right\rangle    \right) \right)\\
    \end{aligned}
\end{equation}
using the following property
\begin{equation}
    \left(\left\langle n| S_{-}(-q)| m \right\rangle\right)^{\dagger} = \left\langle m|S^{\dagger}_{-}(-q)| n \right\rangle = \left\langle m|S_{+}(q)| n \right\rangle
\end{equation}
Eq. \eqref{8.44} becomes
\begin{equation}\label{8.45}
    \begin{aligned}
    = \frac{1}{\mc{Z}} \sum_{mn} e^{-\beta E_{m}} \left( |\left\langle m|S_{+}(q)|n \right\rangle|^{2} + |\left\langle m|S_{-}(-q)|n \right\rangle|^{2} \right)
    \end{aligned}
\end{equation}
Further, using the following relation
\[ \int_{-\infty}^{+\infty} d\omega \delta(\omega - \omega_{nm}) = 1 \]
Eq. \eqref{8.45} yields
\begin{equation}
    \begin{aligned}
   \left< \left\{ \mc{S}_+(\mf{q}) \mc{S}_-(\mf{-q}) \right\} \right>_{I,M} = \int_{-\infty}^{+\infty} d\omega ( \frac{1}{\mc{Z}} \sum_{mn} e^{-\beta E_{m}}|\left\langle m|S_{+}(q)|n\right\rangle|^{2}\delta(\omega-\omega_{nm})\\ 
   + \frac{1}{\mc{Z}} \sum_{mn} e^{-\beta E_{m}}|\left\langle m|S_{-}(-q)|n\right\rangle|^{2}\delta(\omega-\omega_{nm}) )
    \end{aligned}
\end{equation}
expressing the R.H.S of the above equation in terms of new compact variables $\tilde{S}_{+}(q,\omega)$ and $\tilde{S}_{-}(-q,\omega)$ is shown as
\begin{equation}
     \left< \left\{ \mc{S}_+(\mf{q}) \mc{S}_-(\mf{-q}) \right\} \right>_{I,M} = \int_{-\infty}^{+\infty} d\omega \left(\tilde{S}_{+}(q,\omega) + \tilde{S}_{-}(-q,\omega) \right)
\end{equation}
We now try to simplify $\tilde{S}_{-}(-q,\omega)$ in to a much convenient form as
\begin{equation}
    \begin{aligned}
        \tilde{S}_{-}(-q,\omega) = \frac{1}{\mc{Z}} \sum_{mn} e^{-\beta E_{m}}|\left\langle m|S_{-}(-q)|n\right\rangle|^{2}\delta(\omega-\omega_{nm}) \\
        = \frac{1}{\mc{Z}} \sum_{mn} e^{-\beta E_{m}}|\left\langle n|S_{+}(q)|m\right\rangle|^{2}\delta(\omega+\omega_{nm}) \\
        =  \frac{1}{\mc{Z}} \sum_{mn} e^{-\beta \omega}|\left\langle m|S_{+}(q)|n\right\rangle|^{2}\delta(\omega-\omega_{nm}) \\
        =  \frac{1}{\mc{Z}}  e^{-\beta \omega}\sum_{mn}|\left\langle m|S_{+}(q)|n\right\rangle|^{2}\delta(\omega-\omega_{nm})
    \end{aligned}
\end{equation}
Therefore the final expression becomes
\begin{equation}
\begin{aligned}
    \left< \left\{ \mc{S}_+(\mf{q}) \mc{S}_-(\mf{-q}) \right\} \right>_{I,M} = \int_{-\infty}^{+\infty} d\omega \left(\tilde{S}_{+}(q,\omega) + e^{-\beta \omega}\tilde{S}_{+}(q,\omega) \right) \\
    =  \int_{-\infty}^{\infty} d\omega \left( 1 + e^{-\beta \omega}\right)\tilde{S}_{+}(q,\omega) 
    \end{aligned}
\end{equation}
our next aim is to express $\tilde{S}_{+}(q,\omega)$ in terms of $\chi_{M,I}^{+-}(q,\omega)$,
\begin{equation}\label{8.48}
    \chi_{M,I}^{+-}(q,\omega) = i \int_{0}^{\infty} d\omega e^{i\omega t } \left\langle \left[ S_{+}(q,t),S_{-}(-q)\right]\right\rangle_{M,I}
\end{equation}
let us introduce a two sided Fourier transform in the following way
\begin{equation}\label{anno}
    f_{+-}(q,\omega) = i \int_{-\infty}^{\infty} dt e^{i\omega t} \left\langle \left[ S_{+}(q,t)S_{-}(-q)\right]\right\rangle_{M,I}
\end{equation}
 note this is very similar to the expression of susceptibility but the limit in the integral goes from $-\infty$ to $+\infty$. First we will break the limits of integration in the R.H.S. of Eq.\eqref{anno}:

 \begin{equation}
 \begin{aligned}
  i \int_{-\infty}^{0} d t e^{i \omega t}\langle[S_{+}(q, t), S_{-}(-q, 0)]\rangle_{M, I}
 +i \int_{0}^{\infty} d t e^{j \omega t}\langle[S_{+}(q, t), S_{-}(-q, 0)]\rangle_{M, I}
 \end{aligned}
 \end{equation}
\begin{equation}
 = i\int_{0}^{\infty} d t e^{-i \omega t}\left\langle\left[S_{+}(q,-t), S_{-}(-q, 0)\right]\right\rangle_{M, I}+\chi_{M, I}^{+-}(q, \omega).
\end{equation}
\begin{equation}
 = i \int_{0}^{\infty} d t e^{-i \omega t}(-1)\left\langle\left[S_{-}(-q,0), S_{+}(q,-t)\right]\right\rangle_{M, I}+\chi_{M, I}^{+-}(q, \omega)
\end{equation}
\begin{equation}
 = -i \int_{n}^{\infty} d t e^{-i \omega t} \left\langle\left[S_{-}(-q, t), S+(q,0)\right]\right\rangle_{M,I}+\chi_{M, I}^{+-}(q, \omega)
\end{equation}
%\begin{array}{l}
\begin{equation}
 = -\chi_{M, I}^{-+}\left(-q,\omega\right)+\chi_{M I}^{+-}(q, \omega)
\end{equation}
\begin{equation}
 = \chi_{M, I}^{+-}(q, w)-\chi_{M,I}^{-+}(-q,-\omega)
\end{equation}
We have,
\begin{equation}
\chi_{M, I}^{-+}(-q, \omega)=i \int_{0}^{\infty} d t e^{i(-\omega) t}\left\langle\left[S_{-}(-q, t), S_{+}(q,0)\right]\right\rangle_{M,I}.
%\end{array}
\end{equation}
Perform the Hermitian conjugate operation on both sides, 
\begin{equation}\label{8.59}
\begin{aligned} \chi_{M, I}^{-+}(-q,-\omega)^{*} &=-i \int_{0}^{\infty} e^{i \omega t}\left\langle\left[S_{+}^{\dagger}(q,0), S_{-}^{\dagger}(-q, t)\right]\right\rangle_{M, I} \\ &=i \int_{0}^{\infty} e^{i \omega t}\left\langle\left\{S^{\dagger}(-q, t), S_{+}^{\dagger}(q, 0)\right]\right\rangle_{M,I} \end{aligned}
\end{equation}
using the property
$$\quad S_{-}^{\dagger}(-q)=S_{+}(q)$$
we get,
\begin{equation}
\chi_{M, I}^{-+}(-q,-\omega)=i \int_{0}^{\infty} d t e^{i \omega t}\left\langle\left[S_{+}(q, t), S_{-}(q,0)\right]\right\rangle_{M, I}.
\end{equation}
now
\begin{equation}
\left(\chi_{M,I}^{-+}(-q,-\omega)\right)=\left(\chi_{M, I}^{+-}(q, \omega)\right)^{*}
\end{equation}
by using these properties, $f_{+-}(q,\omega)$ takes the form:
\begin{equation}
f_{+-}\left(q, \omega)=\chi_{M T}^{+-}(q, \omega)-\chi_{M, I}^{+,-}\left(q_\omega\right)^{*}\right.
\end{equation}
therefore by using the properties of the complex conjugate we have
\begin{equation}
%\begin{array}{l}
f_{+-}(q, \omega)=2 i \operatorname{Im}\chi_{M, I}^{+-}(q, \omega)
\end{equation}
now we will use the following identity to calculate $f_{+-}(q,\omega)$ as
\begin{equation}
\int_{-\infty}^{\infty} d t e^{i \omega t}\langle A B(t)\rangle=e^{-\beta \hbar \omega} \int_{-\infty}^{\infty} d t e^{i \omega t}\langle B(t) A\rangle
%\end{array}
\end{equation}
therefore
\begin{equation}
\begin{aligned}
f_{+-}(q,\omega)=i \int_{-\infty}^{+\infty} d t e^{i \omega t}\langle S_{-}(-q, 0), S_{+}(q, t)\rangle_{M, I}\\
-i \int_{-\infty}^{+\infty} d t e^{j \omega t}\langle S_{-}(-q, 0)S_{+}(q, t)\rangle_{M, I}.
\end{aligned}
\end{equation}
now the second term in the above equation takes the form:
\begin{equation}
%\begin{array}{l}
%\therefore 2^{\mathrm{n} d} \text { term in the abeve %Eqtaker the form an: }\\
-i e^{-\beta \omega} \int_{-\infty}^{+\infty} d t e^{i \omega t}\left\langle S_{+}(q, t) S_{-}(-q,0)\right.
\end{equation}
thus
\begin{equation}
f_{+,-}(q, \omega)=i\left(1-e^{-\beta \omega}\right) \int_{-\infty}^{+\infty} d t e^{i \omega t}\left\langle S_{+}(q, t) S_{-}(-q,0)\right\rangle
%\end{array}
\end{equation}
\begin{equation}\label{8.69}
 2 i \operatorname{Im}\chi_{M, I}^{+-}(q, \omega)=i\left(1-e^{-\beta \omega}\right) \int_{-\infty}^{+\infty} d t e^{i \omega t}\left\langle S_{+}(q, t) S_{-}(-q,0)\right\rangle
\end{equation}
The L.H.S of the above equation is $\tilde{S}_{+-}(q, \omega)$ with some Constant factors. Thus by substinuting $\tilde{S}_{+-}(q, \omega)$ into equation \eqref{8.59}, we can express the free energy $\Delta_{2} F_{M}$ in terms of the dynamical spin susceptibilities. Before doing that we express the L.H.S  from the above equation in terms of $\tilde{S}_{+-}(q, \omega)$ as
\begin{equation}
\tilde{S}_{+-}(q, \omega)=\frac{1}{z} \sum_{m, n} e^{-\beta E_{n}} \mid\left\langle\psi_{m}\left|S_{+}(q)\right| \psi_{n}\right\rangle^{2} \delta\left(\omega-\omega_{nm}\right)
\end{equation}
where,
\begin{equation}
\delta\left(\omega-\omega_{n m}\right)=\frac{1}{2 \pi} \int_{-\infty}^{+\infty} d t e^{i t\left(\omega-\omega_{n m}\right)}
\end{equation}
therefore,
\begin{equation}
\begin{aligned}
\tilde{S}_{+-}(q,\omega)=\frac{1}{z} \sum_{m, n} e^{-g E_{m}}\left\langle\psi_{m}\right| S_{+}(q)\left|\psi_{n}\right\rangle\left\langle\psi_{n}\left|\delta_{-}(-\varepsilon)\right| \psi_{m}\right\rangle\\
\times \frac{1}{2 \pi} \int_{-\infty}^{+\infty} d t e^{ {i\omega t }} e^{-i\omega_{nm}t}
\end{aligned}
\end{equation}
\begin{equation}
\begin{aligned}
=\frac{1}{2 \pi} \int_{-\infty}^{+\infty} d t \bar{e}^{i \omega t} \cdot \frac{1}{2} \sum_{m, n} e^{-\beta E_{m}}\left\langle\psi_{m}\left|e^{i E_{m} t} S_{+}(q) e^{-i \varepsilon_{n} t}\right| \psi_{n}\right\rangle \\
\times \left\langle\psi_{m}\left|S_{-}(-q)\right| \psi_{m}\right\rangle
\end{aligned}
\end{equation}
\begin{equation}
=\frac{1}{2 \pi} \int_{-\infty}^{+\infty} d t e^{i \omega t} \frac{1}{z} \sum_{m,n}e^{-\beta E_{m}}\left\langle\psi_{m}\left|S_{+}(q,t)\right| \psi_{n}\right\rangle\left\langle\psi_{n}\right| S_{-}(-q) \mid \psi_{m}\rangle
\end{equation}
\begin{equation}
\tilde{S}_{+-}(q, \omega)=\frac{1}{2 \pi} \int_{-\infty}^{+\infty} d t e^{i \omega t}\left\langle S_{+}(q,t) S_{-}(-q)\right\rangle
\end{equation}
From equation \eqref{8.69} we obtain
\begin{equation}
\tilde{S}_{+-}(q, \omega)=\frac{1}{\pi} \frac{\operatorname{Im} \chi_{M I}^{+-}(q, \omega)}{1-e^{-\beta \omega}}
\end{equation}
substituting this in equation \eqref{8.48} we get
\begin{equation}
\langle\left[S_{+}(q), S_{-}(-q)\right]\rangle_{M, I}=\frac{1}{\pi} \int_{-\infty}^{+\infty} d \omega \frac{1+e^{-\beta\omega}}{1-e^{-\beta \omega}} \operatorname{Im} \chi_{m, I}^{+-}(q, \omega).
\end{equation}
\begin{equation}\label{8.77}
\langle\left[S_{+}(q)_{i} S_{-}(-2)\right\rangle_{M, I}=\frac{1}{\pi} \int_{-\infty}^{+\infty} d \omega \operatorname{coth}\left(\frac{1}{2}\beta \omega\right) \operatorname{Im} \chi_{M, I}^{+-}(q, \omega).
\end{equation}
On substituting equation \eqref{8.77} in equation \eqref{8.34} we obtain
\begin{equation}
\begin{aligned}
\Delta F^{I}(M)=-\frac{1}{2 \pi} \sum_{q} \int_{0}^{I} d I\left\{\int_{-\infty}^{+\infty} d \omega \operatorname{coth}\left(\frac{1}{2} \beta \omega\right) \operatorname{Im} \chi_{M, I}^{+-}(q, \omega)\right.\\
-\left.\int_{-\infty}^{+\infty} d \omega \operatorname{coth}\left(\frac{1}{2} \beta \omega\right) \operatorname{Im} \chi_{M, 0}^{+-}(q, \omega)\right\}
\end{aligned}
\end{equation}
where
\begin{equation}
\chi_{M,0}^{+-}(q, \omega)=i \int_{0}^{\infty} d t e^{i \omega t}\langle\left[S_{+}(q, t), S_{-}(-q)]\right\rangle_{M, 0}.
\end{equation}
now equation \eqref{8.34} takes the form as
\begin{equation}
\Delta F^{I}(M)=-\frac{1}{2 \pi} \int_{-\infty}^{t \infty} d \omega \operatorname{coth}\left(\frac{1}{2} \beta \omega\right) \operatorname{Im} \int_{0}^{I} d t \sum_{q}\left[\chi_{M,I}^{+-}(q,\omega)-\chi_{M, 0}^{+-}(q, \omega)\right]
\end{equation}
\begin{equation}
%\begin{aligned}
\Delta F^{I}(M)= -(1 / 2) \sum_{q} \int_{0}^{I} \mathrm{~d} I\left\{\left\langle\left[S_{+}(q), S_{-}(-q)\right]_{+}\right\rangle_{M, I}\right.\left.\\
-\left\langle\left[S_{+}(q), S_{-}(-q)\right]_{+}\right\rangle_{M, 0}\right\}
%\end{aligned}
\end{equation}
this is one of the main results of the SCR theory \cite{moriya2012}. Therefore the exact value of the susceptibility is given by
%$$
%\therefore$ From $\varepsilon_{2}\left(A_{1}\right)$ and %$\varepsilon_{q}\left(A_{4}\right)$
%$$
%, the exact value of the susceptibiliy in gien by.
\begin{equation}
\chi=\frac{\chi_{0}}{1-\frac{1}{2} I \chi_{0}+\lambda(T)}
\end{equation}
where
\begin{equation}
\lambda(T)=\frac{1}{2 \pi} \int_{-\infty}^{+\infty} d\omega \operatorname{coth}\left(\frac{1}{2} \beta \omega\right)G(\omega)
\end{equation}
and
\begin{equation}\label{akku8.4}
G(\omega)=-\chi_{0} \operatorname{Im} \frac{\partial^{2}}{\partial M^{2}}\left(\int_{0}^{\infty} d I \sum_{q}\left[\chi_{M, I}^{+-}(q, \omega)-\chi_{M, 0}^{+-}(q, \omega)\right]\right).
\end{equation}
 \subsection{Use of modified Random Phase Approximation for spin fluctuations}
 If we use RPA susceptibilities in Eq. (\ref{akku8.4}), then one runs into a problem. The problem is that the longer wavelength and zero frequency value of the dynamical susceptibility does not agree with the uniform susceptibility that we are calculating. This inconsistency can be rectified using modified random phase approximation \cite{moriya1973effect} for the calculation of transversal dynamical susceptibilities and Curie temperature for  heavy fermion like materials \cite{coleman2001fermi,gegenwart2008quantum}. A convenient feature of this approximation is that the value of magnetization $M$ is kept constant i.e $\chi_{0} B$, if we keep the longitudinal molecular field  $B$ to be constant for varying values of $I$. 
%According to statistical physics  thus have
%$$
%\partial^{2} \Delta F_{M} / \partial M^{2}=\chi_{0}^{-2}\left(\partial^{2} \Delta F_{\mathrm{B}} / \partial B^{2}\right) .
%$$
The transversal dynamical susceptibilities under a fixed longitudinal molecular field $B$ by using a random phase approximation is obtained as
\begin{equation}
%\left.\begin{array}{l}
\chi^{+-}(q, \omega)=\frac{\chi_{0}^{+-}(q, \omega)} {\left[1-I \chi_{0}^{+-}(q, \omega)\right]}\\
\end{equation}
\begin{equation}
\chi_{0}^{+-}(q, \omega)=\frac{\sum_{k}\left[f\left(\varepsilon_{k+q}-B\right)-f\left(\varepsilon_{k}+B\right)\right] }
{\left(\varepsilon_{k}-\varepsilon_{k+q}+2 B-\omega\right)},
%\end{array}\right\}
\end{equation}
where $f(\varepsilon)=\left[\mathrm{e}^{\beta(\varepsilon-\mu)}+1\right]^{-1}, \mu$ being the chemical
potential, and $\chi_{0}^{+-}(q, \omega)$ is the dynamical susceptibility for $I=0 . \quad G(\omega)$ accounts for the spin fluctuations and is calculated as
\begin{equation}
\begin{aligned}
G(\omega)=&-\xi \alpha^{2}(4 \pi)^{-1} \operatorname{Im} \int \mathrm{d} q\left\{f_{0}\left(\partial^{2} f_{0} / \partial B^{2}\right)\left(1-\alpha f_{0}\right)^{-1}\right.\\
&\left.+\left(\partial f_{0} / \partial B\right)^{2}\left(1-\alpha f_{0}\right)^{-2}\right\}
\end{aligned}
\end{equation}
with
\begin{equation}
\left.\begin{array}{l}
f_{0}=f_{0}(q, \omega+\mathrm{is})=\chi_{0}^{+-}(q, \omega+\mathrm{is}) /\left(\chi_{0} / 2\right) \\
\xi=\left(k_{\mathrm{F}}^{3} / 2 \pi^{2} \varepsilon_{\mathrm{F}} \chi_{0}\right), \quad \alpha=I \chi_{0} / 2
\end{array}\right\}
\end{equation}
  \begin{figure}[h]
    \centering
    \includegraphics[scale = 0.30]{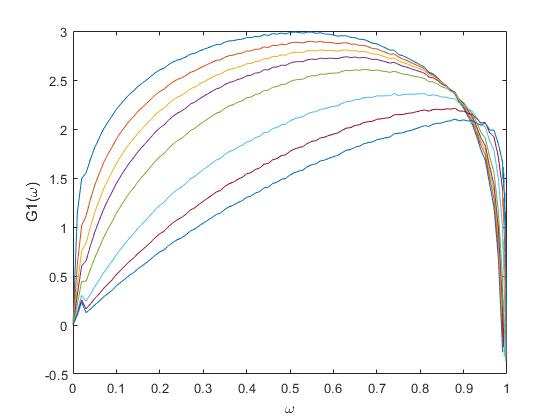}
     \includegraphics[scale = 0.30]{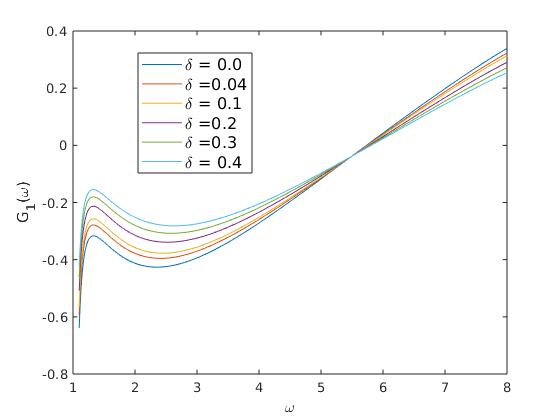}
    \caption{$G_{1}(\omega)$ for $\omega<1$ at $\delta=0.01,0.02,0.04,0.06,0.1,0.2,0.3$ and $0.4$ (from top to bottom: reference point $\omega=0.5$) in [panel (a)] and $G_{1}(\omega)$ for $\omega>1$ at $\delta=0.01,0.04,0.1,0.2$ and $0.4$ (from bottom to top: reference point $\omega\approx 1.5$) in [panel(b)].}
    \label{f1}
\end{figure} 
%
%  \begin{figure}[t]
%    \centering
%    \includegraphics[scale = 0.50]{G1.jpg}
%     \includegraphics[scale = 0.50]{G1_geq_1.jpg}
%    \caption{$G1(\omega)$ for $\omega<1$ at $\delta=0.01,0.02,0.04,0.06,0.1,0.2,0.3$ and $0.4$ (from top to bottom: reference point $\omega=0.5$) in [panel (a)] and $G1(\omega)$ for $\omega>1$ at $\delta=0.01,0.04,0.1,0.2$ and $0.4$ (from bottom to top: reference point $\omega\approx 1.5$) in [panel(b)]}
%    \label{f1}
%\end{figure} 
%
\\
where $\zeta$ equals to 1 for an electron gas model at $T=0\mathrm{~K}$. In modified RPA \cite{toru}, we replace $(1-\alpha) / \alpha$ in the conventional RPA expression by
\begin{equation}
\delta=\chi_{0} / \alpha \chi=2 / I \chi=(1-\alpha+\lambda) / \alpha.
\end{equation}
we thus get
\begin{equation}
\begin{aligned}
G(\omega)=-\xi(4 \pi)^{-1} Im \int \mathrm{d} q\{\alpha f_{0}\left(\partial^{2} f_{0} / \partial B^{2}\right)\left(\delta+1-f_{0}\right)^{-1}\\+(\partial f_{0} / \partial B)^{2}(\delta+1-f_{0})^{-2}\}.
\end{aligned}
\end{equation}
and 
\begin{equation}
\begin{aligned}
\lambda &=\chi_{0}\left(\partial^{2} \Delta_{2} F_{M} / \partial M^{2}\right) \\
&=(2 \pi)^{-1} \int_{-\infty}^{\infty} \mathrm{d} \omega \operatorname{coth}(\beta \omega / 2) G(\omega)
\end{aligned}
\end{equation}
The magnetic susceptibility can be obtained self-consistently by solving above three equations for $\delta$, $\lambda$ and $G(\omega)$ simultaneously.
\begin{figure}[b]
    \centering
    \includegraphics[scale = 0.30]{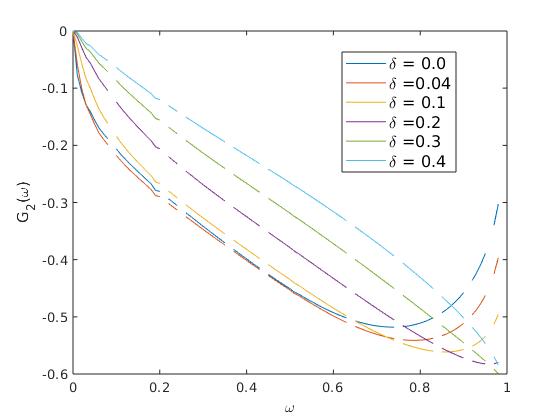}
    \includegraphics[scale = 0.30]{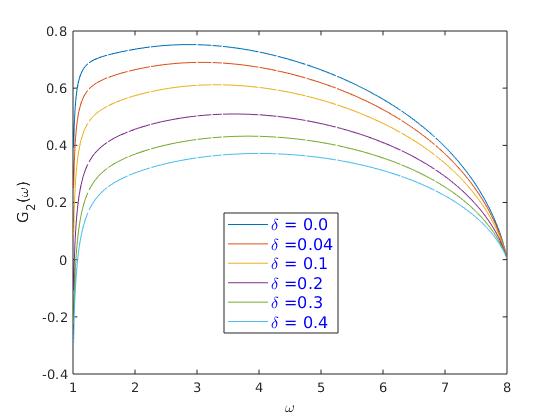}
    \caption{$G_{2}(\omega)$ for $\omega<1$ at $\delta=0.0,0.04,0.1,0.2,0.3$ and $0.4$ in [panel (a)] and $G_{2}(\omega)$ for $\omega>1$ at $\delta=0.0,0.04,0.1,0.2,0.3$ and $0.4$ in [panel(b)].}
    \label{f2}
\end{figure}
%\begin{figure}[t]
%    \centering
%    \includegraphics[scale = 0.50]{G2_geq_1.jpg}
%     \includegraphics[scale = 0.50]{G2_leq_1.jpg}
%    \caption{$G2(\omega)$ for $\omega<1$ at $\delta=0,0.04,0.1,0.2,0.3$ and $0.4$ in [panel (a)] and $G2(\omega)$ for $\omega>1$ at $\delta=0,0.04,0.1,0.2, 0.3$ and $0.4$ in [panel(b)].}
%    \label{f2}
%\end{figure}
%
Further, the Curie temperature is determined by putting $\delta=0$ and we have
\begin{equation}
1-\alpha\left(T_{\mathrm{c}}\right)+\lambda\left(0, T_{\mathrm{c}}\right)=0  
\end{equation}
where $\delta$-dependence of $\lambda(0, T_{c})$ comes from $G(\omega)$. $G(\omega)$ is the quantity of central importance and its detailed calculations are as follows
$$
f_{0}(q, \omega)=f_{0}^{\prime}(q, \omega)+i f_{0}^{\prime \prime}(q, \omega)
$$
\begin{equation}
\begin{aligned}
G(\omega)=&-\int \mathrm{d} q q^{2}\left[E ( q , \omega ) \left\{\alpha(1+\delta)\left[f_{0}^{\prime \prime}\left(\partial^{2} f_{0}^{\prime} / \partial B^{2}\right)+f_{0}^{\prime}\left(\partial^{2} f_{0}^{\prime \prime} / \partial B^{2}\right)\right]\right.\right.\\
&\left.-\alpha\left[\left(f_{0}^{\prime}\right)^{2}+\left(f_{0}^{\prime \prime}\right)^{2}\right]\left(\partial^{2} f_{0}^{\prime \prime} / \partial B^{2}\right)\right\}
+2[E(q, \omega)]^{2}\left\{\left[\left(\delta+1-f_{0}^{\prime}\right)^{2}-\left(f_{0}^{\prime \prime}\right)^{2}\right]
\left(\partial f_{0}^{\prime} / \partial B\right)\left(\partial f_{0}^{\prime \prime} / \partial B\right)\right.\\
&\left.\left.+\left(\delta+1-f_{0}^{\prime}\right) f_{0}^{\prime \prime}\left[\left(\partial f_{0}^{\prime} / \partial B\right)^{2}-\left(\partial f_{0}^{\prime \prime} / \partial B\right)^{2}\right]\right\}\right]
\end{aligned}
\end{equation}
where
\begin{eqnarray}
E(q, \omega)=\left[\left(\delta+1-f_{0}^{\prime}\right)^{2}+\left(f_{0}^{\prime \prime}\right)^{2}\right]^{-1}
\end{eqnarray}
Important expressions for the real and imaginary parts of the free electron model are as follows
%\footnotemark \footnotetext{These are the %correct formulas after rectifying typing %mistakes in the formulas given in  %}:
\\\\
Real parts
\begin{equation}
%\begin{eqnarray}\label{A1}
\begin{aligned}
f_{0}^{\prime}(q, \omega)=(1 / 2)-\left\{\left[q^{4}-(4-2 \omega) q^{2}+\omega^{2}\right] / 16 q^{3}\right\} \log \left|\left(q+q_{1}\right)\left(q+q_{2}\right) /\left(q-q_{1}\right)\left(q-q_{2}\right)\right| \\
-\left\{\left[q^{4}-(4+2 \omega) q^{2}+\omega^{2}\right] / 16 q^{3}\right\} \log \left|\left(q+q_{0}\right)\left(q+q_{3}\right) /\left(q-q_{0}\right)\left(q-q_{3}\right)\right|, \\
{\left[\partial f_{0}^{\prime}(q, \omega) / \partial B\right]_{B=0}=\left(\omega / 4 q^{3}\right) \sum_{m=0}^{3} \log \left|\left(q+q_{m}\right) /\left(q-q_{m}\right)\right|,} \\
{\left[\partial^{2} f_{0}(q, \omega) / \partial B^{2}\right]_{B=0}=\left(1 / 2 q^{2}\right)\{\left[q^{4}-(4-\omega) q^{2}+4 \omega\right]\left(q^{2}-q_{1}{ }^{2}\right)^{-1}\left(q^{2}-q_{2}^{2}\right)^{-1}+\left[q^{4}-(4+\omega) q^{2}\right.} \\
\quad-4 \omega]\left(q^{2}-q_{0}^{2}\right)^{-1}\left(q^{2}-q_{3}{ }^{2}\right)^{-1}-q^{-1}\left[1+\left(q^{2} / 4\right)\right] \sum_{m=0}^{3} \log \left|\left(q+q_{m}\right) /\left(q-q_{m}\right)\right| \}
\end{aligned}
\end{equation}
%\end{eqnarray}
\\
Imaginary parts
\begin{eqnarray}\label{A2}
\begin{aligned}
f_{0}^{\prime \prime}(q, \omega)=(\pi / 4)(\omega / q) \theta\left(q-q_{1}\right) \theta\left(q_{2}-q\right) 
\quad+\left(\pi / 16 q^{3}\right)\left(q^{2}-q_{0}^{2}\right)\left(q_{3}{ }^{2}-q^{2}\right)\\\left[\theta\left(q-q_{0}\right) \theta\left(q_{3}-q\right)-\theta\left(q-q_{1}\right) \theta\left(q_{2}-q\right)\right]
=\left(\pi / 16 q^{3}\right)\left[\left(q^{2}-q_{0}^{2}\right)\left(q_{3}^{2}-q^{2}\right) \theta\left(q-q_{0}\right) \theta\left(q_{3}-q\right)\right. \\
\left.-\left(q^{2}-q_{1}{ }^{2}\right)\left(q_{2}{ }^{2}-q^{2}\right) \theta\left(q-q_{1}\right) \theta\left(q_{2}-q\right)\right], \\
{\left[\partial f_{0}{ }^{\prime \prime}(q, \omega) / \partial B\right]_{B=0}=(\pi / 4)\left(\omega / q^{3}\right)\left[\theta\left(q-q_{0}\right)\left(q_{3}-q\right)-\theta\left(q-q_{1}\right) \theta\left(q_{2}-q\right)\right],} \\
{\left[\partial^{2} f_{0}{ }^{\prime \prime}(q, \omega) / \partial B^{2}\right]_{B=0}=-\left(\pi / 2 q^{3}\right)\left[1+\left(q^{2} / 4\right)\right]\left[\theta\left(q-q_{0}\right) \theta\left(q_{3}-q\right)-\theta\left(q-q_{1}\right) \theta\left(q_{2}-q\right)\right]} \\
\quad+\left(\pi / 8 q^{2}\right)\left\{(1+\omega)^{-1 / 2}\left[q_{0}{ }^{2} \delta\left(q-q_{3}\right)+q_{3}{ }^{2} \delta\left(q+q_{0}\right)\right]\right. 
\left.\quad-(1-\omega)^{-1 / 2}\left[q_{2}{ }^{2} \delta\left(q-q_{1}\right)+q_{1}{ }^{2} \delta\left(q-q_{2}\right)\right]\right\}
\end{aligned}
\end{eqnarray}

where
$$
\left.\begin{array}{l}
q_{1} \\
q_{2}
\end{array}\right\}=1 \mp(1-\omega)^{1 / 2}
$$

$$
\left.\begin{array}{l}
q_{0} \\
q_{3}
\end{array}\right\}=1 \mp(1+\omega)^{1 / 2}
$$

$$
\begin{array}{l}
s=\omega / q \\\\
F(x)=(1 / 2)\left\{1+\left[\left(1-x^{2}\right) / 2 x\right] \log |(1+x) /(1-x)|\right\}
\end{array}
$$
This $G(\omega)$ is the correction to the conventional RPA theory. It is found that $G(\omega)$ constituted of $\alpha \, G_{1}(\omega)$ and $G_{2}(\omega)$. For obtaining  $G(\omega)$ we require both $G_{1}(\omega)$ and $G_{2}(\omega)$ in long wavelength limit. Mathematical concepts involved in the numerical integration of both $G_{1}(\omega)$ and $G_{2}(\omega)$ plays an important role in analytical calculations of spin fluctuations \cite{moriya1965ferro}. The numerical integration of $G_{1}(\omega)$ involves functions like : $f_{0}^{\prime}(q,\omega)$, $f_{0}^{\prime \prime}(q,\omega)$, ${\partial^{2} f_{0}^{\prime}(q,\omega)}/{\partial^{2} B}$ and ${\partial^{2} f_{0}^{\prime \prime}(q,\omega)}/{\partial^{2} B}$.  These functions and their derivatives consists of various $\theta-$ and $\delta-$ functions. The parts of calculations involving the $\delta-$ functions is quite straight forward except for the singularity corrections. It is observed here that the roots of the functions inside the $\theta-$ and $\delta-$ functions plays an important role. In our single band model the $q$-value is restricted up to $q_{c} = 2$, the integration is performed form $q = 0$ to $q = q_{c} =2$. For $\omega > 1$, the contribution comes from only one root which lies inside the region of integration. For $\omega > 8$, there is no contribution from any root in the calculations of $G_{1}(\omega)$. In $\omega < 1$ region,  the singularity correction conditions are used to estimate the contributions of roots of functions inside the $\theta-$ and $\delta-$  functions. Unlike $G_{1}(\omega)$, the numerical integration of $G_{2}(\omega)$ is straightforward as it involves only $\theta-$ functions and does not need any singularity corrections. For $\omega > 1$, the theta function involving the contributions of at least one root upto $\omega = 8$ and there is no contribution thereafter. For $\omega < 1$, it is found that the contribution of all the roots are quiet straightforward and can be calculated by using pre-defined $\theta-$ functions that are readily available which is not true for $G_{1}(\omega)$.
Hence, the calculations $G(\omega)$ are crucial and play an important role in the settling of algorithm for the calculations of other properties \cite{toru}. So far, numerical calculation of $G(\omega)$ has been carried out for the free electron gas model and the results are shown in Figures (\ref{f1}) and (\ref{f2}). 
%$G(\omega)$ consists of two terms which are separated as
%
%$$
%G(w)=\alpha G1(w) +G2(w)
%$$
here $G_{1}(\omega)$ and $G_{2}(\omega)$ include only $\delta$ as a varying parameter. Numerical results at $q=2$ are shown for entire $\omega$-range. For $\omega>1$,  $G_{1}(\omega)$ becomes negative and reflects the attractive nature of spin correlations in such systems. 
\par
Next, We may rewrite $\lambda$ as follows
$$
\begin{array}{rl}
\lambda & =\lambda_{0}+\lambda_{1}, \\\\
\lambda_{0}(\delta, T) & =\pi^{-1} \int_{0}^{\infty} \mathrm{d} \omega G(\omega), \\\\
\lambda_{1}(\delta, T) & =(2 / \pi) \int_{0}^{\infty} \mathrm{d} \omega\left(\mathrm{e}^{\omega / T}-1\right)^{-1} G(\omega)
\end{array}
$$
here $\lambda_{0}$ is weakly temperature dependent through
$\alpha$, while $\lambda_{1}$ is strongly temperature dependent and vanishes at $T=0 .$ The condition for the appearance of ferromagnetism is therefore given by
\begin{equation}
a(0)-1-\lambda_{0}(0,0)>0
\end{equation}
which is generally more accurate than the Stoner condition: $a(0)-1>0$.
Therefore, the susceptibility is rewritten as
\begin{equation}
\begin{aligned}
\chi_{0} / \chi=& \alpha \delta=\left[\Delta \alpha(T)-\Delta \alpha\left(T_{\mathrm{C}}\right)\right]+[\Delta \lambda_{0}(\delta, T)-\Delta \lambda_{0}\left(0, T_{\mathrm{C}}\right)]+\Delta \lambda_{1}(\delta, T)
\end{aligned}
\end{equation}
where
\begin{equation}
\begin{array}{rl}
\Delta \alpha(T) & =\alpha(0)-\alpha(T) \\\\
\Delta \lambda_{0}(\delta, T) & =\lambda_{0}(\delta, T)-\lambda_{0}(0,0) \\\\
\Delta \lambda_{1}(\delta, T) & =\lambda_{1}(\delta, T)-\lambda_{1}\left(0, T_{\mathrm{c}}\right)
\end{array}
\end{equation}
Now the following equation must be satisfied for the calculation of Curie temperature as
\begin{equation}
\begin{array}{c}
\Delta \alpha\left(T_{\mathrm{c}}\right)+\Delta \lambda_{0}\left(0, T_{\mathrm{c}}\right)+\lambda_{1}\left(0, T_{\mathrm{c}}\right)=\alpha(0)-1-\lambda_{0}(0,0)
\end{array}
\end{equation}
%\subsection{Results obtained so far}
%
\begin{figure}[h]
    \centering
    \includegraphics[scale = 0.30]{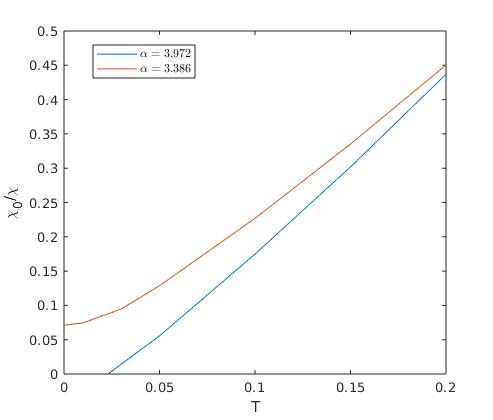}
    \includegraphics[scale = 0.30]{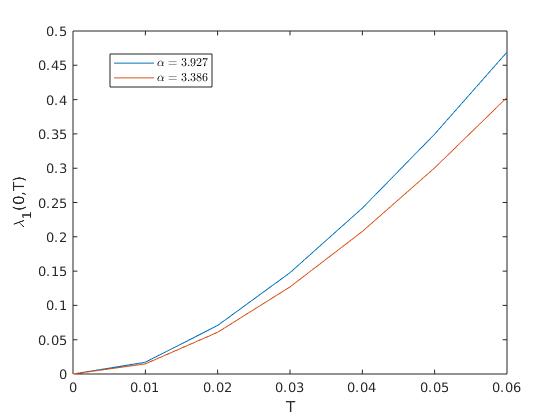}
    \caption{$\chi_{0}/\chi$ vs $T$ plots at $q_{c}=2$ for $\alpha=3.973$ and $\alpha=1.733$ in the SCR theory [left panel] and $\lambda_{1}(0,T)$ vs $T$ plots at $q_{c}=2$ for indicated $\alpha=3.973$ and $\alpha=1.733$ values [right panel].}
    \label{f3}
\end{figure}
Considering a system of itinerant electrons where the spin fluctuations are long and slow, a low frequency and long wavelength limit was studied. Calculations of $F(M, T)$ and $\chi(q, \omega)$ in static and long wavelength limit $\chi(q=0, \omega=0)$ agrees well with the one calculated from the thermodynamic relation, i.e. $\chi=-{\partial^{2} F(H, T)}/{\partial H^{2}}$. This method developed by Moriya is known as the SCR theory of spin fluctuations \cite{toru,moriya1965ferro,moriya1973effect}. The static $\chi(T)$ above $T_{\mathrm{C}}$ for itinerant magnets can be derived from the SCR theory, and the results matches well with the Curie-Weiss fit. The inverse Curie-Weiss-like magnetic susceptibility $\chi^{-1}(T)$ is calculated as
\begin{equation}
\chi_{0}(T) / \chi(T)=1-\alpha+\lambda(T)
\end{equation}
here $\chi_{0}$ is the Stoner susceptibility with $I=0.\, \alpha$ is equal to $\alpha_{0}$ in the Stoner criterion having weak $T$-dependence and $\lambda(T)$ is proportional to the mean-square local amplitude of the spin fluctuations, i.e. the $q$-dependent spin correlation function. Here $\lambda(T)$ is the T-dependent correction quantity.
\begin{figure}[h!]
    \centering
     \includegraphics[scale = 0.10]{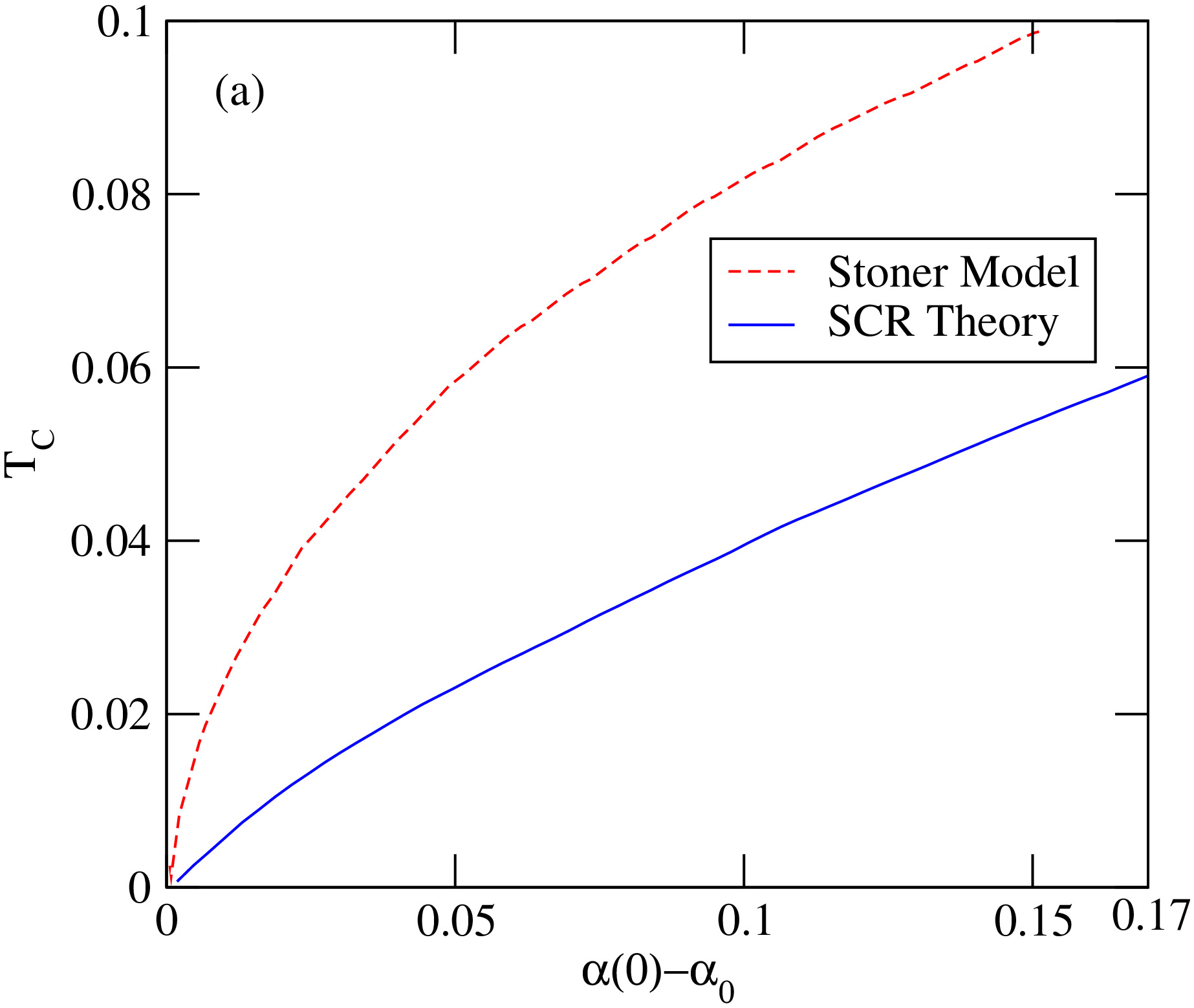}
      \includegraphics[scale = 0.10]{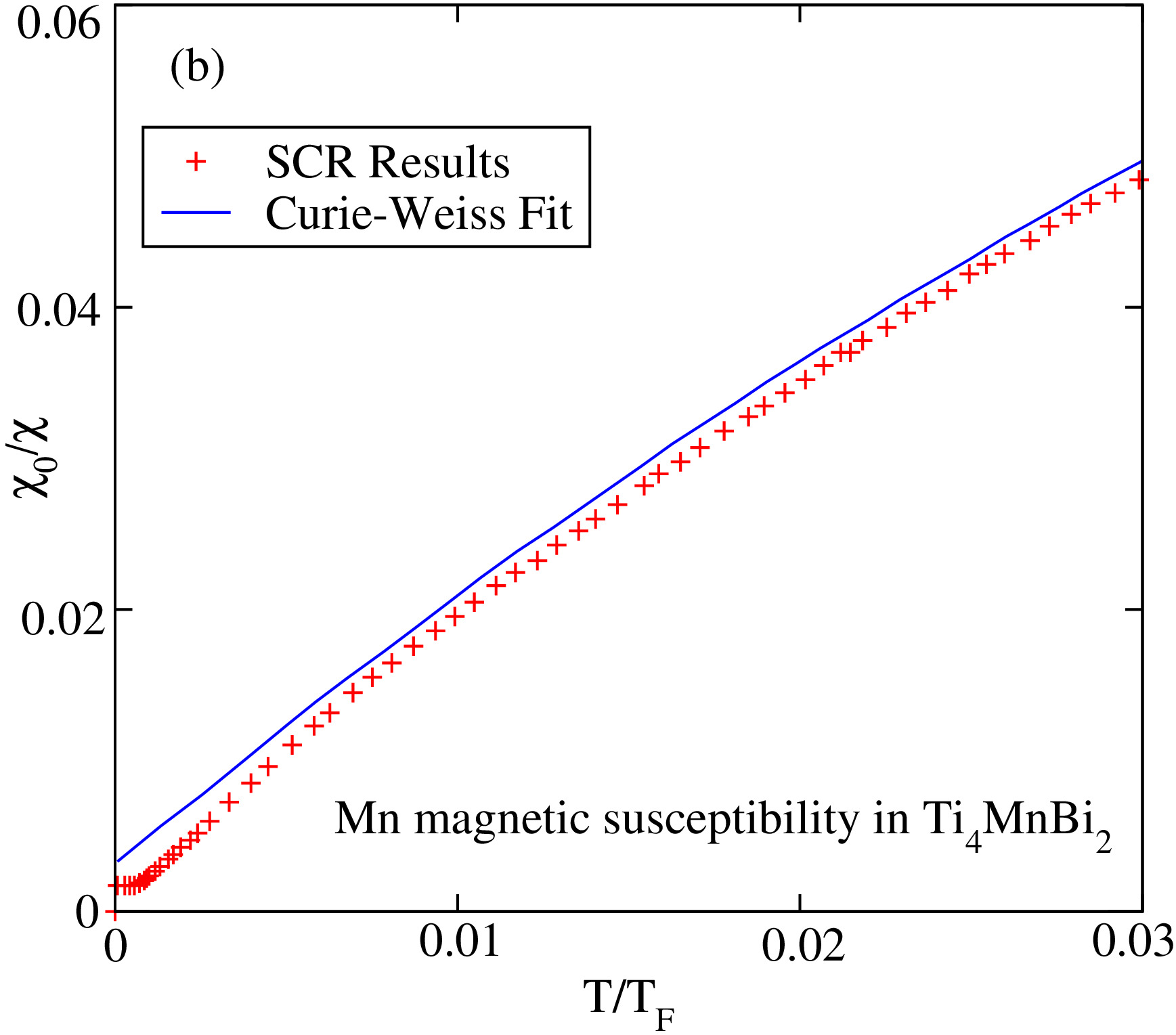} 
    \caption{$T_{C}$ vs $\alpha(0)–\alpha_{0}$ plots in comparison with Stoner model (dashed lines) [panel(a)] and $\chi_{0}/\chi$ vs $T/T_{F}$ plots at $q_{c}=2$ in the SCR theory in comparison with Curie-Weiss fit (obtained from ref.\cite{pandey2020correlations}) [panel(b)].}
    \label{f3}
\end{figure}
 
Next, the calculations of $\lambda_{1}(0,T)$ involves $G(\omega)$ at $\delta=0$ and the results are shown in figure (8). $\chi_{0}/\chi$ (renormalized Susceptibility) vs. T plots are found to be T-linear and follow Curie-Weiss Law above $T_{c}$. Correlation effects and hence spin fluctuations are found to be increased with increase in $\alpha$. It is also observed in the calculations that Stoner Model underestimate the value of renormalized susceptibility due to the over estimation of spin fluctuations at particular value of $\alpha$ so as $T_{c}$. The correction factor part $\lambda_{1}(0,T)$ is strongly T-dependent and found to increases with increase in temperature. Increase in $\lambda_{1}(0,T)$ with $T$  indicating the reduction in effective spin fluctuations. The study of T-dependence of $\lambda_{1}(0,T)$ is most important in SCR theory as it overcome the Stoner contribution and $\chi_{0}/\chi$ is much closer to Curie-Weiss susceptibility above $T_C$. Greater are the interactions in the system greater will be the amplitude of these spin fluctuations \cite{lonzarich1985effect}. 
In figure (9a), $T_{C}$ vs $\alpha(0)–\alpha_{0}$ plots are shown in comparison with Stoner model (dashed lines). $T_{C}$ vs $\alpha(0)–\alpha_{0}$ curves shows that the calculated $T_{C}$ values based on the SCR theory are small, while overestimated values are obtained by the Stoner model. It is seen that this lowering in $T_{C}$ is due to the inclusion of spin-fluctuations by adding the correction factor. In figure (9b), $\chi_{0}/\chi$ vs $T/T_{F}$ plots at $q_{c}=2$ in the SCR theory are depicted in comparison with Curie-Weiss fit. In order to check the validly of our calculations, we applied the SCR theory for realistic system i.e $Ti_{4}MnBi_{2}$ compound \cite{pandey2020correlations} in order to calculate the renormalized susceptibility for $Mn$ and results are compared with the Curie-Weiss Fit.
%
%It is observed that in low-T limit, $\chi_{0}/\chi$ decreases with decrease in $\alpha$. In next step, the self-consistently calculated $G(\omega)$, $\lambda(T)$ and $\chi_{0}/\chi$  are used to calculate $T_{C}$.
%
\subsection{Conclusion}
In conclusion, it is clear that one of the the most successful development in the field of itinerant magnetism is the SCR
theory. This theory goes beyond the Stoner theory (Hartree-Fock approximation) and the random phase approximation in treating the correlation effects. In addition, this theory takes into account the effects of temperature on spin fluctuations that is the effect of re-normalized equilibrium state. 
This theory shows that the T-dependence of the magnetic susceptibility is governed by the spin fluctuation effects. This is true even for weakly ferromagnetic metals and can be true for nearly ferromagnetic metals. It is also found that due to the inclusion of corrections in spin fluctuations the calculated T-dependence of the susceptibility becomes more linear than in the Stoner theory and the Curie temperature is lower than the Stoner value \cite{moriya2012}.  

%Using this theory, one can find that the temperature dependence of the susceptibility is stronger than in the Stoner theory and the Curie temperature is lower than the Stoner value.
\section*{Appendix A1}
\subsection*{Background of linear response theory}
Let us assume that our system is described by the Hamiltonian $\widehat{H}_0$
and let it be perturbed by some external probe.
The interaction between the system and 
external probe is described be  an interaction operator. We separate the time dependence of the perturbation and 
write it as
\begin{equation}
\widehat{H}^{'}=\widehat{A}F(t)
\label{c1e16}
\end{equation}
Here $\widehat{A}$ is the 
interaction operator and $F(t)$ is a scalar function which
contains the time dependence of the perturbation. If $F(t)$ is the electric field then $\widehat{A}$ is the electric dipole moment or if $F(t)$ is the magnetic field then $\widehat{A}$ is the electric dipole moment. The total Hamiltonian is

\begin{equation}
\widehat{H}= \widehat{H}_0+A^0 F(t)
\label{c1e17}
\end{equation}
For $\widehat{H}_0$ let assume the density matrix of the system is $\rho_{0}$.\\
At equilibrium \\
\begin{equation}
 [\widehat{H}_0,\widehat{\rho_{0}}]_{(-)}=0  
\end{equation}
The total Hamiltonian is
\begin{equation}
\widehat{H}= \widehat{H}_0+\widehat{H'}
\label{c1e17}
\end{equation}
\begin{equation}
\widehat{\rho}= \widehat{\rho}_0+\Delta\widehat{\rho}
\label{c1e17}
\end{equation}
therefore
\begin{equation}
\iota\hbar\frac{d\rho}{dt}=[\widehat{H},\widehat{\rho}]_{(-)}   
\end{equation}
This is the famous Liouville-von-Neumemn Equation and can we written as
\begin{equation}
\iota\hbar\frac{d\rho}{dt}=[\widehat{H}+\widehat{H}_{0},\widehat{\rho}+\Delta \widehat{\rho}]_{(-)}   
\end{equation}
While expanding the above Equation we keep only the linear terms and neglect the higher order terms.
Therefore
\begin{equation}
\iota\hbar\frac{d\rho}{dt}=[\widehat{H}_{0},\Delta\widehat{\rho}]+[\widehat{H'},\widehat{\rho_{0}}]  
\end{equation}
\begin{equation}
\frac{d\rho}{dt}=\frac{1}{\iota\hbar}[\widehat{H}_{0},\Delta\widehat{\rho}]-\frac{1}{\iota\hbar}[\widehat{A},\widehat{\rho_{0}}]F(t)  
\end{equation}
where $\widehat{H}^{'}=-\widehat{A}F(t)$.
\begin{equation}
\frac{d\rho}{dt}-\frac{1}{\iota\hbar}[\widehat{H}_{0},\Delta\widehat{\rho}]=-\frac{1}{\iota\hbar}[\widehat{A},\widehat{\rho_{0}}]F(t)  
\end{equation}
let us define the time evolution operator
\begin{equation}
\mathcal{L}\mathcal{O}=\frac{1}{\iota\hbar}[\widehat{H}_{0},\mathcal{O}]
\end{equation}
in this $e^{-\mathcal{L}t}$ is the operator integrating factor.
\begin{equation}
e^{-\mathcal{L}t}\frac{d\rho}{dt}-\frac{1}{\iota\hbar}e^{-\mathcal{L}t}[\widehat{H}_{0},\Delta\widehat{\rho}]=-\frac{1}{\iota\hbar}e^{-\mathcal{L}t}[\widehat{A},\widehat{\rho_{0}}]F(t)  
\end{equation}
\begin{equation}
e^{-\mathcal{L}t}\frac{d\rho}{dt}-e^{-\mathcal{L}t}
\mathcal{L}(\Delta\widehat{\rho})=-\frac{1}{\iota\hbar}e^{-\mathcal{L}t}[\widehat{A},\widehat{\rho_{0}}]F(t)  
\end{equation}
the above Eqcan be written as
\begin{equation}
\frac{d}{dt}[e^{-\mathcal{L}t}{\Delta\rho}]=-\frac{1}{\iota\hbar}e^{-\mathcal{L}t}[\widehat{A},\widehat{\rho_{0}}]F(t)  
\end{equation}
\begin{equation}
e^{-\mathcal{L}t}{\Delta\rho(t)}=-\frac{1}{\iota\hbar}\int_{-\infty}^{t} dt' e^{-\mathcal{L}t}[\widehat{A},\widehat{\rho_{0}}]F(t')  
\end{equation}
\begin{equation}{\label{jaggo}}
{\Delta\rho(t)}=-\frac{1}{\iota\hbar}\int_{-\infty}^{t} dt' e^{-\mathcal{L}(t-t')}[\widehat{A},\widehat{\rho_{0}}]F(t')  
\end{equation}
now we do Taylor expansion of $e^{-\mathcal{L}t}\mathcal{O}$ as
\begin{equation}
e^{-\mathcal{L}t}\mathcal{O}=\left(1+\mathcal{L}t+\frac{t^2}{2\iota}\mathcal{L}\mathcal{L}+\frac{t^3}{3\iota}\mathcal{L}\mathcal{L}\mathcal{L}\right) \mathcal{O}  
\end{equation}
\begin{equation}
e^{-\mathcal{L}t}\mathcal{O}=\left(\mathcal{O} +t\mathcal{L}\mathcal{O} +\frac{t^2}{2\iota}\mathcal{L}\mathcal{L}\mathcal{O} +\frac{t^3}{3\iota}\mathcal{L}\mathcal{L}\mathcal{L}\mathcal{O}\right) 
\end{equation}
\begin{equation}
e^{-\mathcal{L}t}\mathcal{O}=\left(\mathcal{O} +t\frac{1}{\iota \hbar}[\widehat{H}_{0},\mathcal{O}] +\frac{t^2}{2\iota}\mathcal{L}\left(\frac{1}{\iota \hbar}[\widehat{H}_{0},\mathcal{O}]\right)+......\right) 
\end{equation}
\begin{equation}
e^{-\mathcal{L}t}\mathcal{O}=\left(\mathcal{O} +t\frac{1}{\iota \hbar}[\widehat{H}_{0},\mathcal{O}] +\frac{t^2}{2\iota}\left(\frac{1}{\iota \hbar} \right)^{2}[\widehat{H}_{0},[\widehat{H}_{0},\mathcal{O}]+......\right) 
\end{equation}
\begin{equation}
\frac{d}{dt}\mathcal{O}=\dot{\mathcal{O}}=\frac{1}{\iota \hbar}[\mathcal{O},\widehat{H}_{0}]
\end{equation}
\begin{equation}
\mathcal{O}(-t) = \mathcal{O} + t\dot{\mathcal{O}}(O) +\frac{t^{2}}{2 i}\dot{\dot{\mathcal{O}}}(O) +... 
\end{equation}
therefore Eq. (\ref{jaggo}) takes the form as
\begin{equation}
{\Delta\rho(t)}=-\frac{1}{\iota\hbar}\int_{-\infty}^{t} dt' e^{\frac{i}{\hbar}{H_{0}(t'-t)}}\mathcal{O}e^{\frac{-i}{\hbar}{H_{0}(t'-t)}}F(t')  
\end{equation}
\begin{equation}
{\Delta\rho(t)}=-\frac{1}{\iota\hbar}\int_{-\infty}^{t} dt' e^{\frac{i}{\hbar}{H_{0}(t'-t)}}[A,\rho_{0}]e^{\frac{-i}{\hbar}{H_{0}(t'-t)}}F(t')  
\end{equation}
Now, we wish to calculate the change in the property of the system due to  the external perturbation. Hence
\begin{equation}
B=\Delta B;
\Delta B(t)=tr(B\Delta \rho)
\end{equation}
therefore
\begin{equation}
{\Delta B(t)}=-\frac{1}{\iota\hbar}\int_{-\infty}^{t} dt' tr \left(B e^{\frac{i}{\hbar}{H_{0}(t'-t)}}[A,\rho_{0}]e^{\frac{-i}{\hbar}{H_{0}(t'-t)}}\right)F(t')  
\end{equation}
now we will use the cyclic properties of trace $i.e.$ $tr(ABC)=tr(BCA$ and get
\begin{equation}
{\Delta B(t)}=-\frac{1}{\iota\hbar}\int_{-\infty}^{t} dt' tr \left([A,\rho_{0}]B(t'-t)\right)   
\end{equation}
where
\begin{equation}
B(t-t')=e^{\frac{i}{\hbar}{H_{0}(t-t')}}B
e^{\frac{-i}{\hbar}H_{0}(t-t')}F(t')
\end{equation}
we also assumed that there was no external perturbation at $t=-\infty$(in very remote past) and let us consider a simple case when the external perturbation has been
applied in the form of sudden impulse, i.e.
\begin{equation}
F(t)=\delta(t)
\nonumber
\end{equation}
then we have
\begin{equation}
{\Delta B(t)}=-\frac{1}{\iota\hbar} tr \left([A,\rho_{0}]B(t)\right)   
\end{equation}
\begin{align}
\Delta B(t) & = 0\,\,\,\,    
for \,\,\,\, t<0\, \nonumber\\
& = -\frac{1}{\iota\hbar} tr \left([A,\rho_{0}]B(t)\right)\,\,\,\,for\,\,\,\, t>0
\label{c1e30}
\end{align}
This is the response function of the system.
\begin{equation}
{\Delta B(t)}=-\frac{1}{\iota\hbar} tr \left([A\rho_{0}B(t)-\rho_{0}AB(t)\right)   
\end{equation}
\begin{equation}
{\Delta B(t)}=-\frac{1}{\iota\hbar} tr \left(\rho_{0}[B(t),A]\right)   
\end{equation}
\begin{equation}
\Delta B(t) =\frac{-1}{\iota\hbar}\langle\,[\widehat{A},{B}
  (t)]\,\rangle
\label{c1e31}
\end{equation}
This way we can define response function for any external perturbation operator
$\widehat{A}$ and any dynamical variable  $\widehat{B}$ of the system.
The above expression is completely general.
Now if instead of a $\delta$-function perturbation we have the
general time-dependent perturbation, then
\begin{equation}
\Delta B(t)=\frac{-1}{\iota\hbar}
\int_{-\infty}^{+\infty}\langle[B(t-t^{'}),A]\rangle\,F(t^{'})\,dt^{'}
\label{c1e32}
\end{equation}
now we will see if
\begin{align}
F(t')=F_{0}\cos(\omega t) \nonumber\\
F(t)=Re F_{0}e^{-\iota \omega t}e^{\epsilon t}
\end{align}
At remote  past $F(t)=Re F_{0}e^{-\iota \omega t}e^{\epsilon t}$ and expectation values are always real, therefore putting $t-t'=\tau$ we get
\begin{equation}
\Delta B(t)=\frac{-1}{\iota\hbar}
Re\int_{0}^{+\infty}\,d\tau \langle[B(\tau),A]\rangle\,F_{0}e^{-\iota \omega (t-\tau)}e^{\epsilon (t'-\tau)}
\label{c1e32}
\end{equation}
\begin{equation}
\Delta B(t)=\frac{-1}{\iota\hbar}
Re\int_{0}^{+\infty}\,d\tau e^{\iota \omega t -\epsilon t} \langle[B(\tau),A]\rangle\,F_{0}e^{-\iota \omega t}e^{\epsilon t}
\label{c1e32}
\end{equation}
Let us define $\chi_{BA}(\omega)$ as
\begin{equation}
\chi_{BA}(\omega)=\frac{\iota}{\hbar}\int_{0}^{+\infty}dt
\,e^{(\iota\omega-\epsilon) t}\,\langle[B(\tau),A]\rangle
\label{c1e38}
\end{equation}
Because the response of the system is real, hence the Fourier transform of
response function will satisfy the reality condition as well, i.e.,
\begin{equation}
\chi^*_{BA}(-\omega)=\chi_{BA} (\omega)
\label{c1e39}
\end{equation}
Now we have 
\begin{equation}
\Delta B(t) =\chi_{BA}(\omega)\, F_{0}\,
e^{(-\iota\omega+\epsilon)t}
\label{c1e41}
\end{equation}
Here, $\chi_{BA} (\omega)$ is generalized susceptibility of the system.
\section*{Appendix A2}
\subsection*{Correlation function and fluctuation dissipation theorm}
The correlation function of spin density of electrons is  given as \\
\begin{equation}
\mathcal{S}_{\alpha,\beta} (\vec{r} - \vec{r'} , t) = \left < \varsigma_{\alpha}(\vec{r}, t) \varsigma_{\beta}(\vec{r'})\right >
\end{equation}
where the $\left < \cdots \right >$ represents the canonical ensemble average in $r-$space . Again using the Fourier transform for $\vec {r} \rightarrow  \vec{k} $ and $ t  \rightarrow \omega $ we get
\begin{equation}
{S}_{\alpha,\beta} (k,\omega) = \frac {1}{2\pi}\int_{-\infty}^{\infty}dt \, e^{-i \omega t} \left < \varsigma_{\alpha}(k,t) \varsigma_{\beta}(-k)\right >
\end{equation}
%The magnetic scattering of neutrons by electron's spins in any system 
As an example of the correlation function,
%is described by 
the differential cross section formula given as \\
\begin{equation}
\begin{split}
\frac{d^{2} \sigma}{d \Omega d \omega} & =\left(\frac{2 g e^{2}}{m_{0} c^{2}}\right)^{2} \frac{k^{\prime}}{k} \sum_{\alpha, \beta}\left(\delta_{\alpha, \beta}-\hat{\kappa}_{\alpha} \hat{\kappa}_{\beta}\right) {S}_{\alpha \beta}(\boldsymbol{\kappa}, 
\omega) \\
& =\left(\frac{2 g e^{2}}{m_{0} c^{2}}\right)^{2} \frac{k^{\prime}}{k} \sum_{\alpha, \beta}\left(\delta_{\alpha, \beta}-\hat{\kappa}_{\alpha} \hat{\kappa}_{\beta}\right) \bar{{S}}_{\alpha \beta}(\boldsymbol{\kappa}, \omega) 
\end{split}
\end{equation}
where \textbf{$\vec{k}$} is the wave vector of an incident neutron and \textbf{$\vec{k'}$} that of a scattered wave \textbf{$ \vec{\boldsymbol\kappa} = \vec{k'} - \vec{k}$} the change in the wave vector. $ \bar{{S}}_{\alpha \beta}(\boldsymbol{\kappa}, \omega) $ is the summarized correlation function and is calculated experimentally by neutron scattering experiments \cite{ishikawa}. Also, in $compound Y_{0.97}Sc_{0.03}Mn_{2}$ Shiga et al \cite{shiga} have observed the huge scattering due to zero-point fluctuations
at the lowest-$T$. Further, we know that the dynamical susceptibility %
%\subsection{The Fluctuation Dissipation Theorem}
$\chi_{\alpha ,\beta} (k, \omega) $ is a complex quantity having real and imaginary parts. The correlation function $\bar{{S}}_{\alpha\beta}(\boldsymbol{\kappa}, \omega)$ is related to the imaginary part of dynamical susceptibility $\chi_{\alpha ,\beta} (k, \omega)$ as
%\begin{center}
\begin{equation}
 \bar{{S}}_{\alpha \beta}(\boldsymbol{\kappa}, \omega) = \frac{-2 \hbar}{1 - e^{-\beta \hbar \omega}} \mathcal{I}m \chi_{\alpha,\beta}^{{(S)}}(k,\omega)
\end{equation}
%\end{center}
The quantity on the L.H.S is the average value of the density fluctuations,
while the quantity on the R.H.S is related to the dissipation in the system. This
equation may be regarded as a form of the fluctuation-dissipation theorem. 
%The proof of this theorem is given in Appendix. 
%\section*{Appendix A3}
%\subsection*{Response function and generalized susceptibility}
\section*{Appendix A3}
\subsection*{Computation details}
In this section we give some remarks on the computational aspects that should be kept in mind while devising the algorithm when for $G(\omega)$ numerical calculations. It is important to note that we are not giving the complete algorithm but only limited to making some comments which are useful in designing the algorithm for computation. $G(\omega)$ is the sum of $\alpha \, G_{1}(\omega)$ and $G_{2}(\omega)$. Thus for obtaining  $G(\omega)$ we need to obtain both $G_{1}(\omega)$ and $G_{2}(\omega)$. We will next discuss separately the concepts involved in the numerical integration of both $G_{1}(\omega)$ and $G_{2}(\omega)$ one by one. \\\\
{\bf Numerical integration of $G_{1}(\omega)$}\\\\
The numerical integration of $G_{1}(\omega)$ is the most involved. The first thing one should note is that for obtaining $G_{1}(\omega)$, we will have to integrate an integrand which is a function of $q$, $f_{0}^{\prime}(q,\omega)$, $f_{0}^{\prime \prime}(q,\omega)$, $\frac{\partial^{2} f_{0}^{\prime}(q,\omega)}{\partial^{2} B} |_{B = 0}$ and $\frac{\partial^{2} f_{0}^{\prime \prime}(q,\omega)}{\partial^{2} B} |_{B = 0}$. Since $\frac{\partial^{2} f_{0}^{\prime \prime}(q,\omega)}{\partial^{2} B} |_{B = 0}$ consists of various theta and delta functions one would need to evaluate the integration with the part involving the theta functions and the delta functions separately and then take the sum. The part involving the delta functions is quite straight forward except for something that we would like to call "Singularity Corrections". Before getting in to the details let us recall the definitions of some special points $q_{0}$,$q_{1}$,$q_{2}$,$q_{3}$,
$$
\left.\begin{array}{l}
q_{1} \\
q_{2}
\end{array}\right\}=1 \mp(1-\omega)^{1 / 2}
$$
$$
\left.\begin{array}{l}
q_{0} \\
q_{3}
\end{array}\right\}=1 \mp(1+\omega)^{1 / 2}
$$
These points are special because they are the roots of the functions inside the theta and delta functions. Since the integration of a delta function gives the functional value of the integrand without the delta function at the point where the delta function peaks provided it is inside the region of integration and is zero otherwise. For the purpose of understanding, if we consider $q_{c} = 2$, the integration is performed form $q = 0$ to $q = q_{c} =2$. When we consider the domain $\omega > 1$, then only $|q_{0}|$ lies inside the region of integration for $1<\omega<8$ and the only contribution comes from this term. For $\omega > 8$, there is no peak inside the region of integration thus the contribution is zero. If we consider the domain $\omega < 1$, then there are three peaks corresponding to $|q_{0}|$, $q_{1}$ and $q_{2}$ inside the region of integration. Note that if we put the values of $|q_{0}|$, $q_{1}$ or $q_{2}$ directly in $f_{0}^{\prime}$ or $\frac{\partial^{2} f_{0}^{\prime}(q,\omega)}{\partial^{2} B} |_{B = 0}$ it diverges. To avoid this one has to add a small value inside the logarithm to obtain a meaningful value and this is exactly what we called the "Singularity Corrections". The contributions from each of these peaks should be calculated separately. Next while integrating the parts involving the theta functions, as in the above case for $\omega > 1$, the only non zero contribution comes from the theta function involving $|q_{0}|$ for $1<\omega<8$ and there are no non zero contributions for $\omega > 8$. It is important to note that $\frac{\partial^{2} f_{0}^{\prime}(q,\omega)}{\partial^{2} B} |_{B = 0}$ quadratically diverges for $|q_{0}|$ and one should use some logical conditions to remove the points where the integrand diverges. For $\omega < 1$, there are non zero contributions from the theta functions involving  $|q_{0}|$, $q_{1}$ and $q_{2}$. Here It is important to note that the function $\frac{\partial^{2} f_{0}^{\prime}(q,\omega)}{\partial^{2} B} |_{B = 0}$ quadratically diverges for $|q_{0}|$, $q_{1}$ and $q_{2}$ and similar to the 
$\omega > 1$ case we will have to remove the points where the functional value of the integrand is not finite. One should also keep in my mind that since the integrand drastically peaks near these points, one will have to take more points near these regions for integration to get a satisfactory accuracy up on numerical integration. 
\\\\
{\bf Numerical integration of $G_{2}(\omega)$}\\\\
The numerical integration of $G_{2}(\omega)$ is straightforward when compared to $G_{1}(\omega)$, as it involves only theta functions and does not need any "Singularity corrections". As before for $\omega > 1$ only the theta function involving $|q_{0}|$ contributes until $\omega = 8$ and there is no contribution thereafter. For $\omega < 1$, there are contributions from $|q_{0}|$, $q_{1}$ and $q_{2}$ and can be calculated quiet straightforwardly either by designing a code for the theta function or using pre-defined theta functions that are readily available.
\bibliographystyle{unsrt}
\bibliography{main}
\end{document}